%% file: main.tex
\documentclass[review]{elsarticle}
\usepackage[english]{babel} 
\usepackage[T1]{fontenc}
\usepackage{lmodern} 
\usepackage[ansinew]{inputenc}
\usepackage{amsmath}
\usepackage{amssymb}
\usepackage{pgfplotstable}
\usepackage{xfrac}
\usepackage{multirow}
\usepackage{tikz}
\usepackage{tikz-dimline}
\usepackage{lineno}
\usepackage{datatool}
\usepackage[ruled,linesnumbered]{algorithm2e}
\usepackage{pgfplots}
\pgfplotsset{
  compat=1.5,
  legend image code/.code={
    \draw[mark repeat=2,mark phase=2]
    plot coordinates {
      (0cm,0cm)
      (0.15cm,0cm)        %% default is (0.3cm,0cm)
      (0.3cm,0cm)         %% default is (0.6cm,0cm)
    };
  }
}
\usepackage{geometry}
\let\print\pgfmathprintnumber
\usepackage{blindtext}
\usepackage{float}
\usepackage{graphicx}
\usepackage{xcolor}
\usepackage{caption}
\usepackage{hyperref}

\usepackage{array}
\usepackage{siunitx}
\usepackage{multirow}
\usepackage{adjustbox} % to rescale tables only if they exceed textwidth
\usepackage{subcaption}
\usepackage{bm}
\usepackage{mathtools}
\usepackage{pgfplotstable}
\usepackage{setspace}

\definecolor{TUMblue}{RGB}{0,101,189}
\definecolor{TUMblack}{RGB}{0,0,0}
\definecolor{TUMorange}{RGB}{227,114,34}
\definecolor{TUMgreen}{RGB}{162,173,0}
\definecolor{TUMgray}{RGB}{71,80,88}
\definecolor{TUMpink}{RGB}{181,92,165}

\usepgfplotslibrary{groupplots,dateplot}
\usetikzlibrary{patterns,shapes.arrows,calc,external,decorations,shapes,positioning,arrows.meta}
\usetikzlibrary{angles,quotes}
\usepgfplotslibrary{colormaps}
\usetikzlibrary{pgfplots.colormaps}
\usepgfplotslibrary{fillbetween}
\tikzset{>=latex}
\pgfplotsset{compat=newest}

\begin{document}

\title{Data-driven shape inference in  
three-dimensional steady-state supersonic flows using ODIL and JAX-Fluids}

\author[a]{Aaron B. Buhendwa\corref{cor}}
\ead{aaron.buhendwa@tum.de}

\author[a]{Deniz A. Bezgin}
\ead{deniz.bezgin@tum.de}

% \corref{cor}\fnref{fn}

\author[b]{Petr Karnakov}
\ead{pkarnakov@gmail.com}

\cortext[cor]{Corresponding authors}
% \cortext[alph]{In alphabetical order}
% \fntext[fn]{Authors with equal contribution.}

\author[a,c]{Nikolaus A. Adams}
\ead{nikolaus.adams@tum.de}

\author[b]{Petros Koumoutsakos}
\ead{petros@seas.harvard.edu}

\address[a]{Technical University of Munich, School of Engineering and Design, Chair of Aerodynamics and Fluid Mechanics, Boltzmannstra{\ss}e 15, 85748 Garching bei M\"unchen, Germany}
\address[b]{Computational Science and Engineering Laboratory, Harvard University, Cambridge, MA 02138, USA}
\address[c]{Technical University of Munich, Munich Institute of Integrated Materials, Energy and Process Engineering, Lichtenbergstra{\ss}e 4a, 85748 Garching bei M\"unchen, Germany}

\begin{frontmatter}
    \begin{abstract}
      \input{sections/abstract.tex}      
    \end{abstract}

\end{frontmatter}
\journal{APS}
% \journal{Computer Physics Communications}

% \linenumbers
\input{sections/introduction}
\input{sections/methodology}
\input{sections/results}
\input{sections/conclusion}

\input{sections/acknowledgements}

\input{sections/appendix}

% \bibliographystyle{ieeetr}
% \newpage
\bibliography{bibliography}

\end{document}

%% file: sections/abstract.tex
We present a novel data- and first-principles-driven method for inferring the shape of a solid obstacle and its flow field in three-dimensional steady-state supersonic flows.
The proposed method combines the Optimizing a Discrete Loss (ODIL) technique with the automatically differentiable JAX-Fluids computational
fluid dynamics (CFD) solver to study the joint reconstruction of flow fields and obstacle shapes.
ODIL minimizes the discrete residual of the governing partial differential equation (PDE) by gradient descent-based algorithms.
The ODIL framework inherits the characteristics of the chosen numerical discretization of the underlying PDE,
including its consistency and stability. Discrete residuals and their automatic differentiation gradients are
computed by the JAX-Fluids solver which provides nonlinear shock-capturing schemes and level-set-based immersed solid boundaries.
We use synthetic data to validate this approach on challenging inverse problems including the shape inference of a solid obstacle in three-dimensional
steady-state supersonic flow.
Specifically, we study flows around a cylinder, a sphere, and an ellipse.
We investigate two distinct approaches for the obstacle shape representation: (1) \textit{parametric shape representation}, where the obstacle is described by a small set of parameters (e.g., the radius of the cylinder and the sphere) that are optimized together with the flow field, and (2) \textit{free shape representation}, where the level-set function is directly optimized at each point of the computational mesh, without relying on predefined shapes.
For the former, a thorough comparison with Physics-informed Neural Networks is provided.
We show that the nonlinear shock-capturing discretization in combination with the level-set-based interface representation allows for accurate inference of the obstacle shape and its flow field for the ODIL method.
The proposed approach opens new avenues for solving complex inverse problems in supersonic aerodynamics.

%% file: sections/introduction.tex
\section{Introduction}
\label{sec:introduction}

Recent advances in computing and artificial intelligence have sparked
significant interest in addressing challenging inverse problems in
aerodynamics~\cite{Li2022}.
Effective algorithms for inverse problems in aerodynamics
are critical for a plethora of applications ranging from shape
optimization to operational envelopes of supersonic vehicles.
This often involves inferring unknown parts of the flow field,
material properties of the fluid or unknown terms and boundary
conditions of the governing partial differential equations (PDE).

In experimental settings, direct measurements of
density, flow velocity, and pressure fields are sparse, as they
are expensive to obtain, often involve intrusive measurement techniques,
or may not be feasible at all~\cite{Chanetz2020}.
However, in high-speed flows, schlieren photography is a readily available,
nonintrusive method for flow field visualization.
Schlieren images visualize gradients of the refractive index of the fluid,
which, for a single-component flow, correspond to gradients
of the density field~\cite{Settles2001}.
Recent work has explored the reconstruction of flow fields
from a combination of schlieren images and sparse measurements of primitive variables~\cite{Cakir2023,Hu2024}.

A priori knowledge of fundamental physical principles,
often expressed in the form of PDEs, can be incorporated to regularize the ill-posed nature of inverse problems.
In recent applications of ML methods to inverse problems, 
Physics-informed Neural Networks (PINNs)~\cite{Raissi2019}
have been used for flow reconstruction from measurements~\cite{Zhang2023,Sliwinski2023}. 
PINNs express the solution of the flow field with a multilayer perceptron
that maps spatio-temporal coordinates to flow quantities. The underlying
physical evolution laws are enforced by including the residual of the
corresponding PDEs in the loss function.
The PDE loss serves as a physical regularization for optimization.
Raissi et al.~\cite{Raissi2019a} applied
this technique to reconstruct the incompressible two-dimensional viscous flow
around a cylinder at Reynolds number Re =100. In their work, measurements included the concentration
of a passive scalar, i.e., smoke or dye, and 
combined with the physical laws, the authors reconstructed the flow field
and inferred the induced drag and lift force on the cylinder.
Buhendwa et al.~\cite{Buhendwa2021c} applied PINNs to incompressible two-dimensional
two-phase flows in the presence of capillary and viscous effects.
Measurements of the motion of the fluid-fluid interface were used to reconstruct
the entire flow field.
Mao et al.~\cite{Mao2020} and Jagtap et al. ~\cite{Jagtap2022} were the first
to use PINNs for the investigation of inviscid supersonic two-dimensional flows. 
The measurements of primitive variables were enriched by numerically obtained Schlieren data
to infer the flow field.
Recent works have also used experimental BOS measurements to
reconstruct missing flow-field data. For example, Molnar et al.~\cite{Molnar2023} have
investigated supersonic flow around axisymmetric cones,
and Rohlfs et al.~\cite{Rohlfs2023} reconstructed the quasi-inviscid section
of a shock wave/boundary layer interaction.

We note that inverse problems can be interpreted from a Bayesian perspective~\cite{Stuart2010}.
Although computationally more involved, the Bayesian approach allows one to rank possible solutions according to their relative probabilities
and to quantify uncertainty.
The Bayesian approach has been used in the context of flow reconstruction.
For example, in~\cite{Kontogiannis2024}, this approach has been applied to
joint flow field reconstruction and parameter learning of 3D steady laminar flows given magnetic resonance velocimetry measurements.

Recently, optimization of discrete loss (ODIL)~\cite{Karnakov2023a,Karnakov2023b}
has been introduced as a framework for inverse problems for PDEs. ODIL utilizes automatic differentiation (AD) to minimize a loss function for the discrete approximation of the PDE residual, where the discrete solution itself is the set of parameters being optimized.
Because ODIL minimizes the discrete residual, it inherits the characteristics of the underlying numerical discretizations. In ODIL, gradient descent-based methods are typically used for optimization, where the gradients are computed by AD.
Karnakov et al.~\cite{Karnakov2023b} applied ODIL to inverse problems in
incompressible viscous flows. Examples included a two-dimensional
lid-driven cavity, where a comparison with PINNs showed
a significant advantage of ODIL both in computational cost and accuracy. Furthermore, the potential of the framework was illustrated
by jointly reconstructing the flow field and obstacle shape for three-dimensional incompressible flows.

In this paper, we extend the ODIL framework for the joint reconstruction of the
flow field and obstacle shape for steady-state supersonic flows governed by the compressible Euler equations.  As ODIL minimizes the discrete PDE residual with gradient descent-based algorithms, the numerical discretization of the underlying PDE plays a central role in the optimization process. We use the differentiable computational fluid dynamics solver JAX-Fluids~\cite{Bezgin2022,Bezgin2024} to evaluate the discrete PDE residual. JAX-Fluids implements nonlinear shock-capturing schemes \cite{Toro2009a} and a sharp-interface immersed boundary method \cite{Hu2006}. Especially for flows containing discontinuities such as shocks, we show that the present approach offers the following advantages over PINNs: (1) the conservative finite-volume formulation ensures convergence towards physically correct weak solutions, and (2) high-order shock-capturing reconstruction schemes improve accuracy around shocks.

We demonstrate the capabilities of the method with two optimization scenarios:
(1) \textit{Parametric shape representation}, where we learn
a small set of parameters defining the obstacle shape, along with the flow field.
Here, a comprehensive comparison between PINN and ODIL is provided.
(2) \textit{Free shape representation}, which involves optimizing the
level set field in each cell of the underlying mesh directly.
For both scenarios, numerically generated data is used as a proxy for sparse measurements of primitive variables and schlieren data.
We find that ODIL, combined with JAX-Fluids, provides accurate results for both optimization scenarios.
% We remark the importance of the ODIL formulation that operates on the discrete form of the equations: first, the nonlinear
% shock-capturing scheme enables accurate flow field reconstruction near discontinuities,
% while, second, the level-set-based sharp interface representation is crucial for accurate and efficient inference of obstacle shapes.

The manuscript is structured as follows.
In Section \ref{sec:methodology}, we present the governing equations
and briefly describe the underlying methodology for ODIL and PINNs.
% Further detail about the methodology is provided in \ref{appendix:numerical_methods}, \ref{appendix:odil}, and \ref{appendix:pinn}.
We discuss the obstacle shape representation and the loss function.
In Section \ref{sec:results}, we present results for the flow field and obstacle
shape reconstruction.
Section \ref{sec:conclusion} summarizes the present work and gives an overview.

%% file: sections/methodology.tex
\section{Methodology}
\label{sec:methodology}

\subsection{Governing equations}
\label{subsec:governing_equations}
We describe the state of the fluid at any location
$\mathbf{x} = \left[ x, y, z \right]^T = \left[ x_1, x_2, x_3 \right]^T$
in the flow field using either the vector of primitive variables $\mathbf{W} = \left[ \rho, u, v, w, p \right]^T$ or
the vector of conservative variables $\mathbf{U} = \left[ \rho, \rho u, \rho v, \rho w, E \right]^T$.
%
% \begin{align}
%     \mathbf{W} = \begin{bmatrix}
%         \rho \\ u \\ v \\ w \\ p
%     \end{bmatrix}, \quad
%     \mathbf{U} = \begin{bmatrix}
%         \rho \\ \rho u \\ \rho v \\ \rho w \\ E
%     \end{bmatrix},
%     \label{eq:prime_cons_vector}
% \end{align}
%
The primitive variables include the fluid density $\rho$, 
the velocity components $u$, $v$, and $w$ (in $x$-,$y$-, and $z$-direction, respectively),
and the pressure $p$.
$\mathbf{u} = \left[ u, v, w \right]^T = \left[ u_1, u_2, u_3 \right]^T$ is the velocity vector.
$E = \rho e + \frac{1}{2} \rho \mathbf{u} \cdot \mathbf{u} $ denotes the
total (volume-specific) energy of the fluid and $e$ is the specific internal energy.

Inviscid, compressible flows are governed by the compressible Euler equations
which consist of conservation equations for mass (continuity equation),
momenta, and total energy, while neglecting  viscous effects.
For the vector of conservative variables, 
the steady-state compressible Euler equations are expressed as
\begin{align}
    \boldsymbol{\mathcal{R}}_E (\mathbf{U}) = 
    \frac{\partial \mathbf{F}(\mathbf{U})}{\partial x}
    + \frac{\partial \mathbf{G}(\mathbf{U})}{\partial y}
    + \frac{\partial \mathbf{H}(\mathbf{U})}{\partial z} = 0,
    \label{eq:DiffConsLaw1}
\end{align}
where $\boldsymbol{\mathcal{R}}_E (\mathbf{U})$ represents the residual operator
of the steady-state Euler equations. $\mathbf{F}, \mathbf{G},$ and $\mathbf{H}$ denote the convective fluxes in the $x$-, $y$- and $z$-direction, respectively.
\begin{align}
    \mathbf{F}(\mathbf{U}) = \begin{bmatrix}
        \rho u \\
        \rho u^2 + p \\
        \rho u v \\
        \rho u w \\
        u (E + p)
    \end{bmatrix} \quad
    \mathbf{G}(\mathbf{U}) = \begin{bmatrix}
        \rho v \\
        \rho v u\\
        \rho v^2 + p \\
        \rho v w \\
        v (E + p)
    \end{bmatrix} \quad
    \mathbf{H}(\mathbf{U}) = \begin{bmatrix}
        \rho w\\
        \rho w u\\
        \rho w v \\
        \rho w^2 + p \\
        w (E + p)
    \end{bmatrix}
    \label{eq:convective_fluxes}
\end{align}

The system of equations \eqref{eq:DiffConsLaw1} is closed by an equation of state (EOS).
As EOS, we employ the ideal gas law
\begin{align}
    p &= \left( \gamma - 1 \right) \rho e = \rho R T, \\
    c &= \sqrt{\gamma \frac{p}{\rho}} = \sqrt{\gamma R T}  \ .
    \label{eq:ideal_gas_eos}
\end{align}
Here, $\gamma$ is the ratio of specific heats, $R$ is the non-dimensional specific gas constant,
and $c$ is the speed of sound, respectively.
We use $\gamma = 1.4$ and $R = 1$ throughout this work.
The specific heat capacity at constant pressure is $c_p = \gamma/(\gamma - 1) R$.

The Mach number of the flow is defined as the ratio of velocity magnitude
$\left\| \mathbf{u} \right\|= \sqrt{\mathbf{u} \cdot \mathbf{u}}$ to the speed of sound,
$M = \left\| \mathbf{u} \right\|/c$.
The schlieren are defined as the absolute density gradient $S=\left\| \nabla \rho \right\|$.

For steady and inviscid flows, energy conservation becomes 
\begin{equation}
    \mathbf{u} \cdot \nabla \left( h + \frac{\mathbf{u} \cdot \mathbf{u}}{2} \right) = 0,
    \label{eq:constant_enthalpy}
\end{equation}
where $h = e + p / \rho$ is the specific enthalpy.
Equation \eqref{eq:constant_enthalpy} states that $h + (\mathbf{u} \cdot \mathbf{u})/{2}$
is constant along any streamline.
With the total temperature (or the temperature of the stagnation point) $T_0$ defined as
\begin{equation}
    T_0 = T \left(1 + \frac{\gamma - 1}{M^2}\right) 
        = T + \frac{\mathbf{u} \cdot \mathbf{u}}{2 c_p},
\end{equation}
we can recast Eq. \eqref{eq:constant_enthalpy} as 
\begin{equation}
    \mathbf{u} \cdot \nabla T_0 = 0.
    \label{eq:constant_total_temperature}
\end{equation}
which implies that the total temperature is constant along any streamline
but can vary between different streamlines.
In this work, we focus on flows with uniform inflow
conditions so that $h + (\mathbf{u} \cdot \mathbf{u}) / {2} = \text{const}$ 
and $T_0 = \text{const}$ throughout the domain.
Flows with $\left\| \nabla T_0 \right\| = 0$ are homenthalpic.

\subsection{PDE-constrained Optimization using ODIL and PINN}
\label{subsection:PDE_costrained_optimization}

Our goal is to infer the flow field, represented by the vector of primitive variables $\mathbf{W}$, and the obstacle shape,
parameterized by $\boldsymbol{\theta}_s$, given schlieren measurements $S^\text{ref}$ 
and (sparse) measurements of the primitive variables $\mathbf{W}^\text{ref}$.
We require that the solution satisfies the underlying partial differential equations (PDEs), i.e., the steady-state Euler equations, as defined in Eqs.~\eqref{eq:DiffConsLaw1} and \eqref{eq:convective_fluxes}.
This task is formulated as a PDE-constrained optimization problem that minimizes the loss function $\mathcal{L}(\mathbf{W}, \boldsymbol{\theta}_s)$.
\begin{equation}
    (\mathbf{W}, \boldsymbol{\theta}_s) = \arg\min_{\mathbf{W}, \boldsymbol{\theta}_s} \mathcal{L}(\mathbf{W}, \boldsymbol{\theta}_s)
    \label{eq:optimization_problem}
\end{equation}
The loss function generally has the following contributions.
\begin{equation}
    \begin{aligned}
        % (\mathbf{W}, \boldsymbol{\theta}_s) &= \arg\min_{\mathbf{W}, \boldsymbol{\theta}_s} \mathcal{L}(\mathbf{W}, \boldsymbol{\theta}_s) \\
        % \mathcal{L}(\mathbf{W}, \boldsymbol{\theta}_s) &=
        % \lambda_E \underbrace{\left\| \boldsymbol{\mathcal{R}}_E(\mathbf{W}, \boldsymbol{\theta}_s) \right\|^2}_{\text{PDE residual loss} \ \mathcal{L}_\text{PDE}} +
        % \lambda_W \underbrace{\left\| \mathbf{W}^\text{ref} - \mathbf{W} \right\|^2}_{\text{Primitive data loss} \ \mathcal{L}_{W}} +
        % \lambda_S \underbrace{\left\| S^\text{ref} - S \right\|^2}_{\text{Schlieren data loss} \ \mathcal{L}_S}  \\
        % &+ \sum_i \lambda_i \underbrace{\left\| \mathcal{R}_i(\mathbf{W}, \boldsymbol{\theta}_s) \right\|^2}_{\text{Additional regularizations} \ \mathcal{L}_i}
        \mathcal{L}(\mathbf{W}, \boldsymbol{\theta}_s) &=
        \underbrace{\left\| \boldsymbol{\mathcal{R}}_E(\mathbf{W}, \boldsymbol{\theta}_s) \right\|^2}_{\text{PDE residual loss} } +
        \underbrace{\left\| \mathbf{W}^\text{ref} - \mathbf{W} \right\|^2}_{\text{Primitive data loss} } +
        \underbrace{\left\| S^\text{ref} - S \right\|^2}_{\text{Schlieren data loss}} 
        + \sum_i \underbrace{\left\| \mathcal{R}_i(\mathbf{W}, \boldsymbol{\theta}_s) \right\|^2}_{\text{Additional regularizations} }
    \end{aligned}
    \label{eq:loss_function}
\end{equation}
Here, $\boldsymbol{\mathcal{R}}_E$ is the operator of the steady-state Euler equation residual and
$\mathcal{R}_i$ are further regularization terms that improve stability and convergence
of the optimization process, as detailed in Section \ref{subsec:loss_function_computation}.
We solve this problem using gradient-descent optimization with the Adam optimizer \cite{Kingma2015}.

The focus of this work lies on the extension of the Optimizing a Discrete Loss (ODIL) \cite{Karnakov2023a, Karnakov2023b}
method for the solution of the previously described optimization problem (see Eqs. \eqref{eq:optimization_problem} and \eqref{eq:loss_function}) for compressible steady-state flows.
The extended ODIL framework is compared with PINNs \cite{Raissi2019} as a baseline method. We note that the implementation of PINNs for obstacle shape inference in compressible flows is a novel contribution, as PINNs so far have only been used for flow field inference for known obstacle shapes \cite{Jagtap2022,Mao2020}. The ODIL and PINN methods can be summarized as follows.
\begin{itemize}
    \item ODIL approximates the solution on a discrete mesh and uses traditional numerical methods to compute the PDE residual.
    \ref{appendix:numerical_methods} provides a detailed description of the numerical methods used in this work. The PDE residual is discretized by the CFD solver JAX-Fluids. Given its JAX \cite{jax2018a} backend, gradients through the CFD solver are easily computed via automatic differentiation (AD).
    During optimization, the discrete solution itself is treated as the set of parameters being optimized. In ODIL, the discrete solution is represented by a multigrid ansatz, as detailed in \ref{appendix:odil}.
    \item PINNs approximate the solution with a deep feedforward neural network.
    In contrast to ODIL, the neural network represents a continuous mapping from input variables to the solution, and the PDE residual is computed using AD.
    During optimization, the weights and biases of the neural network, denoted by $\boldsymbol{\theta}_\text{PINN}$, are the parameters that are being tuned.
    It is important to note that the shape parameters $\boldsymbol{\theta}_s$ are not encoded within the neural network itself,
    but are treated as additional parameters.
    These parameters are jointly optimized together with $\boldsymbol{\theta}_\text{PINN}$, allowing simultaneous inference of both the flow field and the obstacle geometry. \ref{appendix:pinn} provides details about the employed activation functions.
\end{itemize}

\subsection{Obstacle Shape Representation}
\label{subsection:shape_params}

We use the level-set method \cite{Osher1988} to represent the shape of the obstacle.
The fluid-solid interface $\Gamma$ is the zero level-set of a scalar signed distance function $\phi$, 
$\Gamma=\{\mathbf{x} \ | \ \phi(\mathbf{x})=0\}$.
Two distinct approaches are studied for the parameterization $\boldsymbol{\theta}_s$ of the level-set function.
\begin{itemize}
    \item \textit{Parametric shape representation}, where a small set of parameters defines the level-set function.
    Here, we investigated the flow around a cylinder, a sphere, and an ellipse.
    Taking into account fixed origins of the obstacles, cylinder and sphere are parameterized by their radius $\boldsymbol{\theta}_s = \left\{R\right\}$.
    The level-set function $\phi$ is analytically computed as
    $\phi(\mathbf{x},\boldsymbol{\theta}_s)=-R+\left\|\mathbf{x}\right\|$.
    For the ellipse, the parameters are the semi-major $A$,
    the ratio between semi-major $A$ and semi-minor $B$, $r_{AB} = B/A$,
    and the rotation angle $\lambda$, that is, $\boldsymbol{\theta}_s=\{A,r_{AB},\lambda\}$.
    Since there is no closed analytical expression for the signed distance functions of ellipses,
    we compute $\phi(\mathbf{x},\boldsymbol{\theta}_s)$ by discretizing the ellipse contour with a fixed set of 1000 points.
    The level-set function is then directly evaluated as the minimum signed distance from these points.

    Before evaluating the shape parameters are passed through activation functions.
    We denote the pre-activation shape parameters with $\hat{\boldsymbol{\theta}}_s$.
    The activation functions ensure (1) positivity of the radius and
    (2) a unique solution for the set of parameters of the ellipse,
    i.e., $R>0$, $A>0$, $0<r_{AB}<1$ and $-\pi/2<\hat{\lambda}<\pi/2$.
    \begin{equation}
    %   \begin{aligned}
        R = \text{softplus}(\hat{R}), \quad A = \text{softplus}(\hat{A}), \quad r_{AB} = \text{logistic}(\hat{r}_{AB}), \quad \lambda =\frac{\pi}{2} \tanh(\hat{\lambda}).
    %   \end{aligned}
      \label{eq:shape_params_activations}
    \end{equation}

    \item \textit{Free shape representation}, where the discrete level-set function $ \phi $ itself at each point of the computational mesh
    is treated as the set of optimization parameters. 
    This approach is only investigated for ODIL. 
    In contrast to the parametric shape representation, the signed distance property $\|\nabla \phi\| = 1$ is not automatically preserved during optimization.  
    Maintaining this property is crucial for correctly evaluating geometrical quantities such as interface normals $\mathbf{n}$ and apertures $A$ (see Fig.~\ref{fig:cut_cell}).  
    A common strategy in level-set methods is to restore the distance property by iteratively solving the reinitialization equation in pseudo-time \cite{Rouy1992}
    \begin{equation}
    \frac{\partial \phi}{\partial \tau} = - \text{sgn}(\phi^0) \left( \|\nabla \phi\| - 1 \right) = \mathcal{R}_\phi,
    \label{eq:levelset_reinit}
    \end{equation}
    as typically is done in fluid mechanics applications \cite{Sussman1994b, Hu2006} and shape optimization \cite{Allaire2004}.  
    In this work, we adopt a different strategy: rather than solving the reinitialization equation iteratively after each optimization step, we directly incorporate $\mathcal{R}_\phi$ as a regularization term in the loss function.    
\end{itemize}
  
\subsection{Loss Function}
\label{subsec:loss_function_computation}

The loss function has the following main contributions:
the residual loss of the PDE (steady-state compressible Euler equations) $\mathcal{L}_\text{E}$, the loss of primitive variable data
$ \mathcal{L}_{W}$, and the loss of schlieren data $\mathcal{L}_S$.
To improve the convergence of the optimization, we introduce additional regularization terms.  
(1) Since we only consider uniform inflow conditions, we regularize the total temperature using  
$ \mathcal{L}_{T_0}$, guiding the optimization toward solutions with constant total temperature, i.e.,  
$ \left\| \nabla T_0 \right\| = 0. $
(2) As previously described, $\phi$ must maintain its signed distance property for the free shape representation. To this end, we introduce the regularization $\mathcal{L}_\phi$ that minimizes $\left\| \mathcal{R}_\phi \right\|$.
(3) Finally, PINN requires an explicit regularization that enforces the boundary condition at the fluid-solid interface.  
Since we only consider inviscid flows governed by the Euler equations, we impose zero normal velocity at the interface.
This contribution is denoted by $ \mathcal{L}_{\Gamma}$.
Note that such explicit regularization of the interface is not required for ODIL, as
the interface condition is implicitly enforced by the residual loss of the PDE
through the employed immersed boundary method (see Section~\ref{subsection:level_set}).

The loss contributions are evaluated at specific spatial points within the domain $ \Omega \subset \mathbb{R}^d $ with $d\in\{2,3\}$.
First, we introduce the subdomains $\Omega_W\subset\Omega$,
$\Omega_S\subset\Omega$, and $\Omega_{T_0}\subset\Omega$. These subdomains represent the region where we enforce primitive variable measurements, schlieren measurements and the total temperature regularization, respectively.
We now define the following set of points:
\begin{itemize}
    \setlength{\itemsep}{0pt}
    \setlength{\parskip}{0pt}
    \setlength{\parsep}{0pt}
    \item $P_C = \{ \mathbf{x}_i \}_{i=1}^{N_C} \in \Omega$: Cell centers of the computational mesh used for ODIL. 
    % The domain $ \Omega $ is discretized using a uniform mesh identical to that of the reference solution.

    \item $P_E = \{ \mathbf{x}_i \}_{i=1}^{N_E} \in \Omega$: PDE residual points for PINN. 
    These are sampled randomly from a uniform distribution over $ \Omega $.

    \item $P_W = \{ \mathbf{x}_i \}_{i=1}^{N_W} \in \Omega_W \subset \Omega$: Primitive variable measurement points. 
    Although these are randomly sampled over $ \Omega_W $, we assume that they coincide with the cell centers of the reference solution.

    \item $P_S = \{ \mathbf{x}_i \}_{i=1}^{N_S} \in \Omega_S \subset \Omega$: Schlieren measurement points. 
    These are obtained by subsampling every $ k $-th cell center from $ P_C $ within the region $ \Omega_S $. As will be shown in the results, we use $k=2$ as default, however, we also investigate the effect of the variation of $k$ toward sparser point distributions.

    \item $P_{T_0} = \{ \mathbf{x}_i \}_{i=1}^{N_{T_0}} \in \Omega_{T_0} \subset \Omega$: Evaluation points for the total temperature regularization. 
    For ODIL, $ P_{T_0} $ coincides with $ P_C $; for PINN, these points are randomly sampled from $ \Omega_{T_0} $.

    \item $P_\Gamma = \{ \mathbf{x}_i \}_{i=1}^{N_{\Gamma}} \in \Gamma \subset \Omega$: Interface points used for PINN to impose solid boundary conditions. 
    These are uniformly distributed along the fluid-solid interface $ \Gamma $.
\end{itemize}
% \renewcommand{\arraystretch}{1.2}
% \begin{table}[!t]
%     \centering
%     \begin{tabular}{c| c c}
%          Method & ODIL & PINN \\
%          \hline
%          \hline
%          PDE residual & $P_C$ & $P_E$ \\
%          Fluid-solid interface condition & -- & $P_\Gamma$ \\
%          Level-set regularization & $P_C$ & -- \\
%          Primitive variable data & \multicolumn{2}{c}{$P_W$} \\
%          Schlieren data & \multicolumn{2}{c}{$P_S$} \\
%          Total temperature regularization & \multicolumn{2}{c}{$P_{T_0}$}\\
%     \end{tabular}
%     \caption{Notation of the training points and the associated loss contributions.}
%     \label{tab:training_points}
% \end{table}
Furthermore, we denote $ \mathbf{W}_i^\text{ODIL} $, $ \mathbf{W}_i^\text{PINN}$,
$ S_i^\text{ODIL}$,  and $S_i^\text{PINN}$ as ODIL and PINN approximations of the vector of primitive variables and
the schlieren at point $ \mathbf{x}_i $. 
The reference solution is indicated by $ \mathbf{W}_i^\text{ref} $ and $ S_i^\text{ref} $. 
The level-set function at point $ \mathbf{x}_i $ is written as $ \phi_i(\boldsymbol{\theta}_s) $.
The operator of the residual of the continuous Euler equation is given by $ \mathcal{R}_{E,j} $, with $ j = 1, \dots, N_{Eq} $ as equation index.
The corresponding discrete operator is given by $ \mathcal{R}_{E,j}^\Delta $ (see Eq. \eqref{eq:FVD} and \eqref{eq:FVD_levelset}).
Lastly, the operator of the discrete right-hand side of the reinitialization equation is denoted as $\mathcal{R}_\phi^\Delta$ (see Eq. \eqref{eq:levelset_reinitialization_discrete}).

\begin{figure}[H]
    \centering
    \begin{minipage}{0.48\textwidth}
    \begin{equation}
    \begin{aligned}
    \mathcal{L}_E^\text{ODIL} &= \frac{1}{N_{Eq} N_C} \sum_{j=1}^{N_{Eq}} \sum_{\mathbf{x}_i \in P_C} \left( \mathcal{R}_{E,j}^\Delta (\mathbf{W}_i^\text{ODIL},\phi_i(\boldsymbol{\theta}_s)) \right)^2 \\
    \mathcal{L}_W^\text{ODIL} &= \frac{1}{N_{Eq} N_W} \sum_{j=1}^{N_{Eq}} \sum_{\mathbf{x}_i \in P_W} \left( \mathbf{W}_{i,j}^\text{ODIL} - \mathbf{W}_{i,j}^\text{ref} \right)^2 \\
    \mathcal{L}_S^\text{ODIL} &= \frac{1}{N_S} \sum_{\mathbf{x}_i \in P_S} \left( S_i^\text{ODIL} - S_i^\text{ref} \right)^2 \\
    \mathcal{L}_{T_0}^\text{ODIL} &= \frac{1}{N_{T_0}} \sum_{\mathbf{x}_i \in P_{T_0}} \left( \left\| \nabla T_{0,i}^\text{ODIL} \right\| \right)^2 \\
    \mathcal{L}_\phi^\text{ODIL} &= \frac{1}{N_C} \sum_{\mathbf{x}_i \in P_C} \left( \mathcal{R}_\phi^\Delta(\phi_i) \right)^2 \\
    \mathcal{L}^\text{ODIL} &= \sum_k \omega_k^\text{ODIL} \mathcal{L}_k^\text{ODIL}, \quad k \in \{E,W,S,T_0,\phi\}
    \end{aligned}
    \label{eq:loss_odil}
    \end{equation}
    \end{minipage}
    \hfill
    \begin{minipage}{0.48\textwidth}
    \begin{equation}
    \begin{aligned}
    \mathcal{L}_E^\text{PINN} &= \frac{1}{N_{Eq} N_E} \sum_{j=1}^{N_{Eq}} \sum_{\mathbf{x}_i \in P_E} \left( \mathcal{R}_{E,j}(\mathbf{W}_i^\text{PINN}) \, \xi (\phi_i(\boldsymbol{\theta}_s)) \right)^2 \\
    \mathcal{L}_W^\text{PINN} &= \frac{1}{N_{Eq} N_W} \sum_{j=1}^{N_{Eq}} \sum_{\mathbf{x}_i \in P_W} \left( \mathbf{W}_{i,j}^\text{PINN} - \mathbf{W}_{i,j}^\text{ref} \right)^2 \\
    \mathcal{L}_S^\text{PINN} &= \frac{1}{N_S} \sum_{\mathbf{x}_i \in P_S} \left( S_i^\text{PINN} - S_i^\text{ref} \right)^2 \\
    \mathcal{L}_{T_0}^\text{PINN} &= \frac{1}{N_{T_0}} \sum_{\mathbf{x}_i \in P_{T_0}} \left( \left\| \nabla T_{0,i}^\text{PINN} \right\| \right)^2 \\
    \mathcal{L}_\Gamma^\text{PINN} &= \frac{1}{N_\Gamma} \sum_{\mathbf{x}_i \in P_\Gamma} \left( \mathbf{u}_i^\text{PINN} \cdot \mathbf{n}_i(\boldsymbol{\theta}_s) \right)^2 \\
    \mathcal{L}^\text{PINN} &= \sum_k \omega_k^\text{PINN} \mathcal{L}_k^\text{PINN}, \quad k \in \{E,W,\Gamma,S,T_0\}
    \end{aligned}
    \label{eq:loss_pinn}
    \end{equation}
    \end{minipage}
    % \caption{Side-by-side definition of ODIL and PINN loss function components.}
    \label{fig:loss_comparison}
\end{figure}

We make the following remarks about the loss formulations:
\begin{itemize}

    \item The total losses $\mathcal{L}^\text{ODIL}$ and $\mathcal{L}^\text{PINN}$ are weighted sums of the
    respective contributions. The loss weights $\omega_k$ are hyperparameters of the optimization process and depend
    on the specific case at hand.
    
    \item The function $ \xi(\phi(\boldsymbol{\theta}_s)) = 0.5 \left(1 + \tanh\left(\phi(\boldsymbol{\theta}_s) / h\right)\right) $ 
    in $\mathcal{L}_E^\text{PINN}$ acts as a smooth mask that transitions from $0$ inside the obstacle to $1$ outside. 
    The thickness of the transition region is controlled by the hyperparameter $ h $.
    Ideally, one would choose $ h \to 0 $, as the fluid-solid interface represents a discontinuity. 
    However, very small values of $ h $ hinder effective optimization of the shape parameters $ \boldsymbol{\theta}_s $, 
    since $ \lim_{h \to 0} \partial \xi / \partial \boldsymbol{\theta}_s(\mathbf{x}) = 0 $ for all $ \mathbf{x} \notin \Gamma $.
    In this work, we use $ h = 5 \cdot 10^{-3} $ to balance interface sharpness and gradient propagation during optimization.
    
    Minimizing the PDE residual loss $ \mathcal{L}^\text{PINN}_E $ alone tends to adjust the shape parameters toward larger obstacle sizes, i.e., increasing the radius for cylinders and spheres, and enlarging the semi-major and semi-minor axes for ellipses.
    This is because, in regions masked by $ \xi \approx 0 $, the PDE residual is effectively suppressed, thus reducing the loss. 
    In the extreme case where the obstacle spans the entire spatial domain $ \Omega $, $ \mathcal{L}^\text{PINN}_E $ approaches zero regardless of the actual flow field.  
    We combine $ \mathcal{L}^\text{PINN}_E $ with the interface regularization $ \mathcal{L}^\text{PINN}_\Gamma $, 
    which penalizes violations of the boundary condition at the fluid-solid interface. 
    This combination guides the optimization toward a meaningful local minimum, 
    located between the extremes of a vanishingly small and a domain-filling obstacle.  

    For ODIL, the immersed boundary method inherently incorporates both effects within the PDE residual loss $\mathcal{L}_E^\text{ODIL}$: 
    a masking effect, which tends to increase the obstacle size by suppressing the PDE residual inside the obstacle,
    and the enforcement of the interface condition at the fluid-solid boundary.

    % In our setup, optimization is initialized with a very small obstacle shape. 
    % This is essential: if the initial shape is larger than the reference, 
    % the optimization often fails to reduce it effectively. 
    % In contrast, starting from a small shape allows the optimizer to grow the obstacle as needed, 
    % driven by the interplay between the masking of the PDE residual physical constraint on the fluid-solid interface.
    % While this initialization strategy may appear restrictive, it aligns well with practical scenarios: 
    % in real applications, measurements are typically available only in the region surrounding the unknown object. 
    % Thus, it is reasonable to initialize the shape as a small obstacle placed within the measurement domain.

    \item The interface normal $\mathbf{n}(\boldsymbol{\theta}_s)$ required for the evaluation of $\mathcal{L}_\Gamma^\text{PINN}$
    is computed analytically.

    \item For both PINN and ODIL, the schlieren intensity $S=\left\|\nabla \rho\right\|$ is computed using
    second-order central finite-differences with the cell size of the underlying mesh of the reference solution.
    The absolute gradient of the total temperature $\left\|\nabla T_0 \right\|$ is computed
    with second-order central finite-differences for ODIL or AD for PINN.

    \item We reiterate that the level-set function is regularized via the loss term $ \mathcal{L}^\text{ODIL}_\phi $
    to preserve its signed distance property. 
    This regularization is only necessary for the free shape representation, 
    which is exclusively considered in the ODIL framework.

\end{itemize}

% \begin{table}[!b]
%     \centering
%         \begin{tabular}{r l}
%             Notation&Description\\
%             \hline
%             $P_C$&Points that represent the cell centers of the underyling mesh for ODIL\\
%             $P_E$&Points where the PDE residual for PINN is imposed\\
%             $P_W$&Points where measurements of primitive variables are imposed\\
%             $P_S$&Points where measurements of numerical schlieren are imposed\\
%             $P_{T_0}$&Points where the total temperature regularization is imposed\\
%             $P_{\Gamma}$&Points where the fluid-solid interface condition for PINN is imposed\\
%             \hline
%         \end{tabular}
%         \caption{Notation and description of subdomains, set of points, and loss contributions.}
%         \label{tab:point_description}
%     \end{table}
%     % 

%% file: sections/results.tex
\section{Results}
\label{sec:results}

\begin{figure}[!t]
  \centering
  \input{figures/schlieren.tex}
  \caption{Numerical schlieren of the reference solutions
  for the three flows under investigation.
  From left to right: Cylinder, parameterized by the radius $R$. Ellipse, parameterized by the semi-major $A$, semi-minor $B$ and rotation angle $\lambda$. Sphere, parameterized by $R$.}
  \label{fig:schlieren_canonical_cases}
\end{figure}
All cases under consideration have uniform inflow conditions at Mach number $M_\infty = 2$ with corresponding primitive variable
state $\mathbf{W}_\infty = [\rho, u, v, p]^T_\infty = [1, 2.366, 0, 1]^T$ in 2D and $\mathbf{W}_\infty=[\rho, u, v, w, p]^T_\infty = [1, 2.366, 0, 0, 1]^T$ in 3D.
Figure \ref{fig:schlieren_canonical_cases} depicts the numerical schlieren of the reference solutions.
These are generated with JAX-Fluids~\cite{Bezgin2022,Bezgin2024}
by integrating the unsteady compressible Euler equations to steady-state using traditional time-stepping methods.
They serve as measurements for the present study.
For all test cases, the computational domain $\Omega=\{\mathbf{x}\in[-0.3,0.3]^d\}$ is discretized with
a uniform mesh consisting of $128^d$ cells, where $d\in\{2,3\}$ is the dimension.
The same computational mesh is used for ODIL.
All computations are performed on a single NVIDIA A100 80GB GPU using
double-precision floating-point arithmetic (\textit{float64}).

\subsection{Parametric Shape Representation}
\label{subsec:results_parametric}

The shape parameters of the reference solutions
for the cylinder and sphere are $R_\text{ref} = 0.05$ and for the ellipse are $A_\text{ref} = 0.08$, $r_\text{AB,ref} = 0.375$ and $\lambda_\text{ref} = -30^\circ$. The corresponding initial guesses for the shape parameters are $R=0.01$, $A=0.01$, $r_{AB}=0.5$, and $\lambda = 0.0^\circ$.
Through a comprehensive hyperparameter sweep, we find the optimal optimization setup, including learning rate schedule, loss weights, and neural network sizes. Table \ref{tab:learning_rate_schedule} and \ref{tab:loss_weights} list the optimization setup.
The neural networks consist of 4 (cylinder), 5 (ellipse), and 6 (sphere)
hidden layers with 50 nodes each.

\input{./figures/parametric_shape_inference/error_table.tex} 
\subsubsection{Baseline Point Distribution}
\label{subsubsec:baseline_points_parametrc}
We first present the results for the baseline point distribution characterized
by primitive measurements provided in proximity to the obstacle and dense schlieren measurements. 
We define a narrow band region around the obstacle
$\Omega_W=\{\mathbf{x}\in\Omega \ | \  2 < \left\|\mathbf{x}\right\|/R_\text{ref} < 4\}$
where we randomly distribute the points $P_W$.
We use $N_W=25$, $N_W=30$, and $N_W=150$
points for the cylinder, ellipse and sphere, respectively.
Furthermore, we define $\Omega_{S}=\Omega_{T_0}=\{\mathbf{x}\in\Omega \ | \ 2 < \left\|\mathbf{x}\right\|/R_\text{ref} \}$
as the region where the points $P_S$ and $P_{T_0}$ are given.
For ODIL, $P_{T_0}=P_C\cap\Omega_{T_0}$ and $P_{S}=P_C|_2\cap\Omega_{S}$, where $P_C|_2$ is a subset of the cell centers of the underlying mesh consisting of every second cell center in each axis direction.
For PINN, particularly for the sphere case ($128^3\approx2\cdot 10^6$ cells)
using the same number of points results in significantly
higher wall clock times per optimization step compared to ODIL.
We find that reducing the number of points for PINN does not
significantly degrade the solution quality.
Through a thorough hyperparameter study,
we identify the optimal point distribution balancing accuracy
and wall clock time:
For the cylinder and ellipse,
we randomly distribute $N_E=15000$, $N_{T_0}=10000$,
and $N_{\Gamma}=1000$ points.
$P_S$ is chosen to be the same as for ODIL.
For the sphere, we randomly distribute $N_E=550000$,
$N_{T_0}=100000$, $N_{\Gamma}=20000$, and $N_{S}=200000$ points.
Note that $P_E\subset\Omega$, $P_{T_0}\subset\Omega_{T_0}$,
$P_{S}\subset \Omega_{S}$,
and $P_{\Gamma}\subset \Gamma$.

\begin{figure}[!t]
  \centering
  \input{figures/parametric_shape_inference/flowfield_baseline.tex}
  \caption{
    Density $\rho$ (left) and velocity $u$ (right) for the parametric shape representation
    using the baseline point distribution that is described in Section \ref{subsubsec:baseline_points_parametrc}.
    The solid black lines depict the fluid-solid interface.
    In the density plots, we depict the initial guess of the interface with red lines. The magenta colored points represent the measurement positions for the primitive variables $P_W$.
    The black dashed line represents the lower boundary
    of the domain regions $\Omega_{W}$ and $\Omega_{S}$,
    i.e., there are no flow measurements inside the dashed circle.
  }
  \label{fig:flowfield_parametric}
\end{figure}
\begin{figure}[!t]
  \centering
  \input{figures/parametric_shape_inference/shape_params_history.tex}
  \caption{
    Shape parameter history for the parametric shape representation.
    The dashed, horizontal lines indicate the reference values.
  }
  \label{fig:shape_params_history}
\end{figure}
\begin{figure}[!t]
  \centering
  \input{figures/parametric_shape_inference/loss_history.tex}
  \caption{
    Loss history for the parametric shape representation.
    The vertical dashed lines in the plots associated with ODIL
    indicate a switch from first-order upwind to second-order MUSCL discretization.
  }
  \label{fig:loss_history_parametric}
\end{figure}

Figure \ref{fig:flowfield_parametric}
shows the flow field for the cylinder, ellipse, and sphere.
Figure \ref{fig:shape_params_history} depicts the shape parameter history.
Figure \ref{fig:loss_history_parametric} shows the loss history.
Both ODIL and PINN accurately predict the shape parameters.
While PINN shows faster convergence for some cases, the wall clock time per optimization step is roughly 3 times larger compared to ODIL.
Qualitatively, the flow field reconstructions from both methods align
well with the reference solution.
It is important to note that, during optimization, PINN encounters a discrepancy between
the data and the PDE residual. Since the data is numerically generated,
it inherently contains discretization errors, whereas PINN aims to
satisfy the exact, continuous PDE residual.
Nonetheless, a closer examination of the inferred flow field of PINN
reveals errors in the wake and at the shock.
Especially for the sphere, PINN
does not capture the shock in the region close
to the stagnation point.
Note that flow measurements are only
distributed in $\Omega_W$ and $\Omega_{S}$,
i.e., there are no flow measurements at the shock
in the vicinity of the stagnation point of the sphere.
ODIL accurately captures the shock even in regions with no flow measurements.
The quantitative evaluation in Table \ref{tab:errors_parametric} shows
that ODIL achieves relative errors
an order of magnitude lower
than PINN across all flow variables.

\begin{figure}[!t]
  \centering
  \input{figures/parametric_shape_inference/flowfield_cylinder_variation.tex}
  \caption{
    Density $\rho$ for the parametric shape representation of the cylinder depending on sparsity
    of schlieren measurements and the region of flow measurements, as detailed in Section \ref{subsubsec:variation_points_parametrc}.
    The solid black lines depict the fluid-solid interface.
    The red line represents the initial guess of the interface.
    The magenta colored points
    represent the measurement position for the primitive
    variables $P_W$.
    The black points in the plots for $P_{S,02}$
    and $P_{S,12}$ depict the measurement position for the schlieren. For visual clarity, we omit them in the denser distributions.
    The black dashed line represents the lower boundary
    of the domain regions $\Omega_{W}$ and $\Omega_{S}$,
    i.e., there are no flow measurements inside the dashed circle.
  }
  \label{fig:flowfield_parametric_variation}
\end{figure}

\subsubsection{Variation of Point Distribution}
\label{subsubsec:variation_points_parametrc}
We use the cylinder case to investigate the influence of the point distribution on the inferred flow field and obstacle shape, specifically examining how increasing the distance of primitive measurements from the obstacle and the sparsity of schlieren measurements affect the results. To this end, we define the following domain regions.
\begin{equation}
  \begin{aligned}
    \Omega_{W,0}&=\{\mathbf{x}\in\Omega \ | \  2 < \left\|\mathbf{x}\right\|/R_\text{ref} < 4\} \quad &&\Omega_{S,0}=\{\mathbf{x}\in\Omega \ | \ 2 < \left\|\mathbf{x}\right\|/R_\text{ref} \} \\
    \Omega_{W,1}&=\{\mathbf{x}\in\Omega \ | \  3.6 < \left\|\mathbf{x}\right\|/R_\text{ref} < 4\} \quad &&\Omega_{S,1}=\{\mathbf{x}\in\Omega \ | \  3.6 < \left\|\mathbf{x}\right\|/R_\text{ref} \} 
  \end{aligned}
  \label{eq:parametric_domain_regions}
\end{equation}
We use $N_{W,0}=25$ and $N_{W,1}=70$. Note that the point distribution denoted with index 0 corresponds to the baseline point distribution that is presented in the previous subsection.
We adjust the sparsity of the schlieren measurements using three levels:
$P_{S,i0}=P_C|_2\cap\Omega_{S,i}$, $P_{S,i1}=P_C|_4\cap\Omega_{S,i}$, and $P_{S,i2}=P_C|_8\cap\Omega_{S,i}$, with $i\in\{0,1\}$,
where $P_C|_k$ is a subset of the cell centers of the underlying mesh consisting of every $k$-th cell center in each axis direction.
This corresponds to choosing the schlieren measurements $P_{S,i0}$ from a grid of $64^d$ points, $P_{S,i1}$ from a grid of $32^d$ points,
and $P_{S,i2}$ from a grid of $16^d$ points with $d\in\{2,3\}$.

Figure \ref{fig:flowfield_parametric_variation} presents the inferred density fields.
Both ODIL and PINN estimate reasonable cylinder radii across all point distributions.
As expected, the quality of the inferred flow field degrades for both methods as the distance of primitive measurements
from the obstacle increases and the schlieren points become sparser. PINN fails to capture the shock in regions lacking flow measurements.
For the point distributions associated with $\Omega_{W,1}$, $\Omega_{S,1}$ in particular, the solution exhibits noticeable smoothing at the shock discontinuity.
It is reported \cite{Fuks2020,Wassing2024,Wassing2024a} that PINNs struggle to find physical weak solutions to hyperbolic conservation laws
without additional regularization, which is consistent with the results we observe here.
In contrast, the numerical discretization in ODIL inherently provides regularization leading to sharp shock capturing even in regions without flow measurements.

\subsection{Free Shape Representation}
\label{subsec:results_free}

\begin{figure}[!t]
  \centering
  \input{figures/free_shape_inference/flowfield_rho_u.tex}
  \caption{
    Density $\rho$ (left) and velocity $u$ (right) of the ODIL and reference solution for the free shape representation.
    The solid black lines depict the interface.
    In the density plots, we depict the initial guess of the interface for ODIL with a red line, and the primitive variable measurement locations are shown in magenta colored points. The primitive measurement points are not shown for the sphere, since they are distributed in 3D (see Fig. \ref{fig:sphere_free_3D} for a 3D representation of the points). 
    The black dashed circles represent the lower boundary of the domain regions $\Omega_{W}$ and $\Omega_{S}$,
    i.e., there are no flow measurements inside the dashed circle.
    In the velocity plots for the reference solution, we also depict the inferred shape by ODIL with cyan colored dashed lines.
    }
  \label{fig:flowfield_free}
\end{figure}
\begin{figure}[!t]
  \centering
  \input{figures/free_shape_inference/measurement_noise/flowfield.tex}
  \caption{
    Density $\rho$ for the free shape representation of the cylinder for three point distributions
    $\left\{ P_{W,0}, P_{W,1}, P_{W,2} \right\}$ and four noise levels $\epsilon \in \left\{0.0, 0.1, 0.2, 0.3 \right\}$ on the primitive variable measurements.
    The solid black lines depict the fluid-solid interface.
    The red line represents the initial guess of the interface.
    The magenta colored points represent the measurement position for the primitive variables $P_W$.
    The black dashed line represents the boundary of the domain regions $\Omega_{W}$ and $\Omega_{S}$,
    i.e., there are no flow measurements inside the dashed circle.
  }
  \label{fig:free_noise}
\end{figure}
\begin{figure}[!t]
  \centering
  \input{figures/free_shape_inference/loss_history.tex}
  \caption{
    Loss history for the free shape representation.
    The vertical dashed lines in the plots associated with ODIL
    indicate a switch from first-order upwind to second-order MUSCL discretization.
  }
  \label{fig:loss_history_free}
\end{figure}
\begin{figure}[!t]
  \centering
  \begin{tikzpicture}
    \node[inner sep=0,line width=1pt] (A) at (0,0) {\includegraphics[scale=0.181,trim={400 400 400 400},clip]{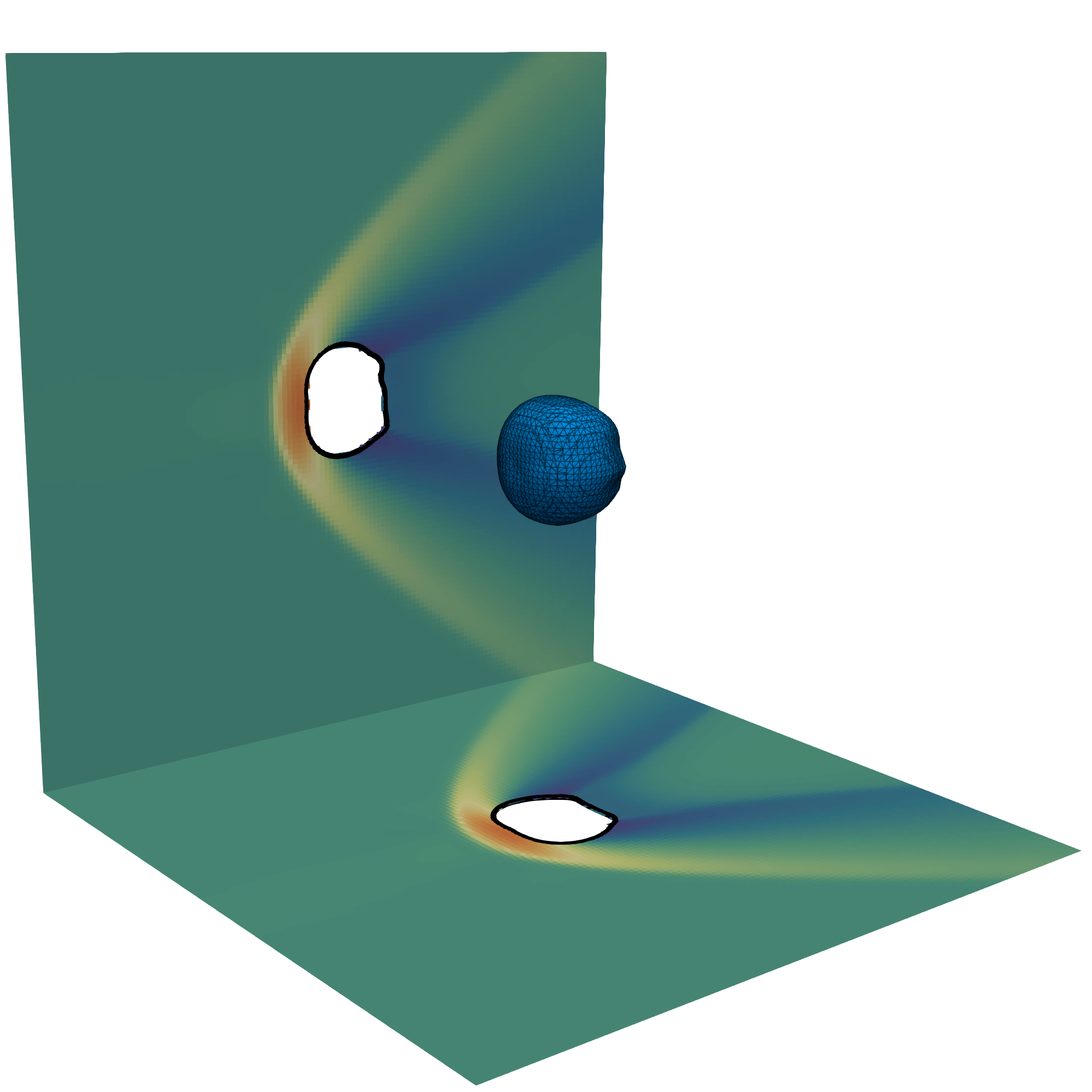}};
    \node[inner sep=0,line width=1pt, left = 0cm of A] (A1){\includegraphics[scale=0.18,trim={400 400 400 400},clip]{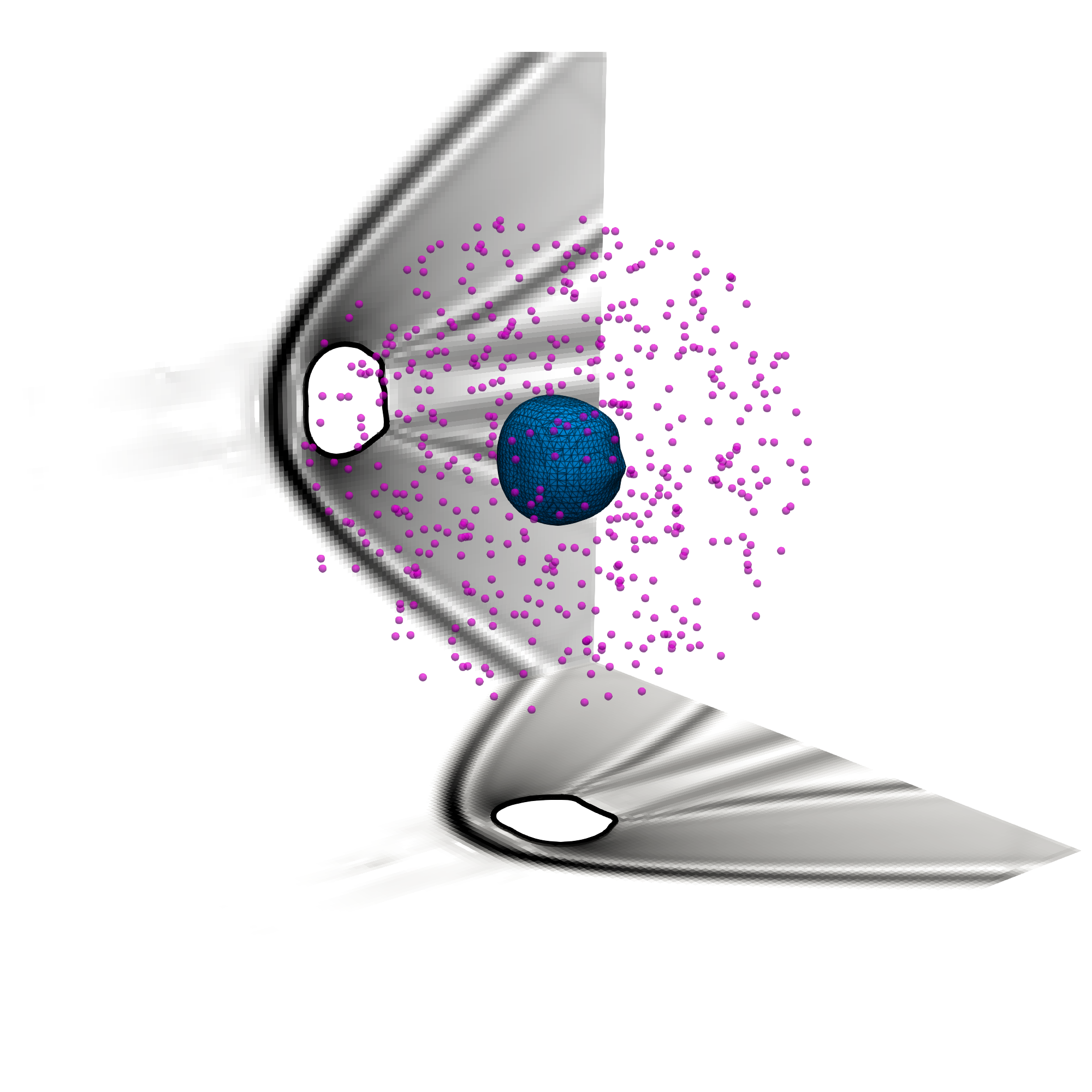}};
  \end{tikzpicture}
  \caption{
    Free shape representation for the sphere.
    The isosurface of the inferred shape is colored in blue.
    (Left) Projected contours of the numerical schlieren.
    The magenta colored points represent the primitive variable measurement locations.
    (Right) Projected contours of the pressure field.
    }
  \label{fig:sphere_free_3D}
\end{figure}

We now present results for the free shape representation, where the multigrid representation of the level-set function $\phi$ is directly optimized.
As previously, the initial guess for the obstacle shape is a small cylinder in 2D or sphere in 3D with radius $R=0.01$.
We define a narrow band region around the obstacle
$\Omega_W=\{\mathbf{x}\in\Omega \ | \  2 < \left\|\mathbf{x}\right\|/R_\text{ref} < 4\}$
in which we randomly distribute the points $P_W$.
We use $N_W=40$, $N_W=50$, and $N_W=500$
points for cylinder, ellipse and sphere, respectively.
Furthermore, we define $\Omega_{S}=\Omega_{T_0}=\{\mathbf{x}\in\Omega \ | \  \left\|\mathbf{x}\right\|/R_\text{ref} > 2\}$
as the region where we distribute the points $P_S$ and $P_{T_0}$.
We use $P_{T_0}=P_C\cap\Omega_{T_0}$ and $P_{S}=P_C|_2\cap\Omega_{S}$, where $P_C|_2$ is a subset
of the cell centers of the underlying mesh consisting of every second cell center in each axis direction.
Table \ref{tab:learning_rate_schedule} and \ref{tab:loss_weights} show the optimization setup, including learning rate schedule and loss weights, respectively.

Figure \ref{fig:flowfield_free}
shows density and velocity inferred by ODIL and the reference solution for all three shapes under consideration.
Figure \ref{fig:loss_history_free} shows the corresponding loss history.
In Figure \ref{fig:sphere_free_3D}, we depict a
3D visualization of the inferred solution for the spherical case.
The inferred flow fields align well with the reference,
with mean cell-wise relative errors of $\mathcal{O}(-3)$,
see Table \ref{tab:errors_free}.
While the shapes inferred by ODIL agree reasonably well with the reference shapes, noticeable discrepancies exist on the downstream side of the obstacles. For instance, the ODIL result for the cylinder shows a trailing edge that extends into the separation region of the reference flow field. Similarly, for the ellipse, ODIL infers a shape that is thickened in downstream direction. For the sphere case, the downstream side of the ODIL shape exhibits sharp edges and a flattened region.
In contrast, the upstream side of the inferred shapes closely matches the reference.
For the flows under investigation, the upstream geometry of the obstacle significantly affects the flow field at the upstream measurement points. In particular, the shock position is directly related to the location of the upstream stagnation point. In comparison, the flow field at the downstream measurement points is less sensitive to variations in the obstacle shape on its downstream side, allowing multiple candidate shapes to yield similar solutions.

\input{figures/free_shape_inference/error_table.tex}

We investigate the influence of the position of the
primitive measurement points, $P_W$,
by varying the random seed used for their distribution.
We also study the effect of
noise on the primitive measurements with
$\tilde{w}^\psi_\text{ref} = w^\psi_\text{ref}(1+\mathcal{U}(-\epsilon,\epsilon))$,
$\psi \in \{\rho,u,v,w,p\}$, where $\tilde{w}^\psi_\text{ref}$ are
the noisy primitive measurements and $\epsilon$
is the noise percentage.
Figure \ref{fig:free_noise} depicts the inferred density fields for three distinct random point distributions $P_{W,i}$, $i\in\{0,1,2\}$ and
noise percentages $\epsilon \in \{0.0,0.1,0.2,0.3\}$.
Even at high noise levels, ODIL is able to reconstruct plausible obstacle shapes. Although the noise introduces visible disturbances in the flow field, the overall solution remains reasonable.
The variation in random point distributions has a noticeable effect on both the inferred flow fields and obstacle shapes. Nevertheless, the key characteristics of the inferred cylinder shape, discussed earlier, persist across all distributions. Specifically, a trailing edge extending into the separation region downstream of the cylinder. The upstream side of the inferred shape continues to closely follow the reference.

%% file: figures/schlieren.tex
\begin{tikzpicture}
    % Spectral_r colormap from matplotlib
    \pgfplotsset{
        colormap={binary}{
        rgb255=(255, 255, 255)
        rgb255=(245, 245, 245)
        rgb255=(235, 235, 235)
        rgb255=(225, 225, 225)
        rgb255=(215, 215, 215)
        rgb255=(205, 205, 205)
        rgb255=(195, 195, 195)
        rgb255=(185, 185, 185)
        rgb255=(175, 175, 175)
        rgb255=(165, 165, 165)
        rgb255=(155, 155, 155)
        rgb255=(145, 145, 145)
        rgb255=(135, 135, 135)
        rgb255=(125, 125, 125)
        rgb255=(114, 114, 114)
        rgb255=(105, 105, 105)
        rgb255=( 95,  95,  95)
        rgb255=( 85,  85,  85)
        rgb255=( 75,  75,  75)
        rgb255=( 65,  65,  65)
        rgb255=( 55,  55,  55)
        rgb255=( 45,  45,  45)
        rgb255=( 34,  34,  34)
        rgb255=( 24,  24,  24)
        rgb255=( 15,  15,  15)
        rgb255=(  5,   5,   5)
        rgb255=(  0,   0,   0)
        }}
    \begin{groupplot}[
        group style={
            group size=3 by 1,
            horizontal sep=0.5cm,
            vertical sep=0.5cm
        }, 
        width=5.8cm, 
        height=5.8cm
        ]
        % \nextgroupplot[
        %     tick align=outside,
        %     tick pos=left,
        %     xtick style={draw=none},
        %     ytick style={draw=none},
        %     xticklabels = {},
        %     xtick = {},
        %     yticklabels = {},
        %     ytick = {},
        %     % title=(a),
        %     enlargelimits=false,
        %     colormap name=spectralr,
        %     axis equal image
        % ]
        % \addplot graphics [xmin=-0.3,xmax=0.3,ymin=-0.3,ymax=0.3] {figures/schlieren/expansionramp.pdf};
        % \nextgroupplot[
        %     tick align=outside,
        %     tick pos=left,
        %     xtick style={draw=none},
        %     ytick style={draw=none},
        %     xticklabels = {},
        %     xtick = {},
        %     yticklabels = {},
        %     ytick = {},
        %     % title=(b),
        %     enlargelimits=false,
        %     axis equal image
        % ]
        % \addplot graphics [xmin=-0.3,xmax=0.3,ymin=-0.3,ymax=0.3] {figures/schlieren/compressionramp.pdf};

        \nextgroupplot[
            tick align=outside,
            tick pos=left,
            xtick style={draw=none},
            ytick style={draw=none},
            xticklabels = {},
            xtick = {},
            yticklabels = {},
            ytick = {},
            % title=(c),
            enlargelimits=false,
            colormap name=binary,
            axis equal image
        ]
        \addplot graphics [xmin=-0.3,xmax=0.3,ymin=-0.3,ymax=0.3] {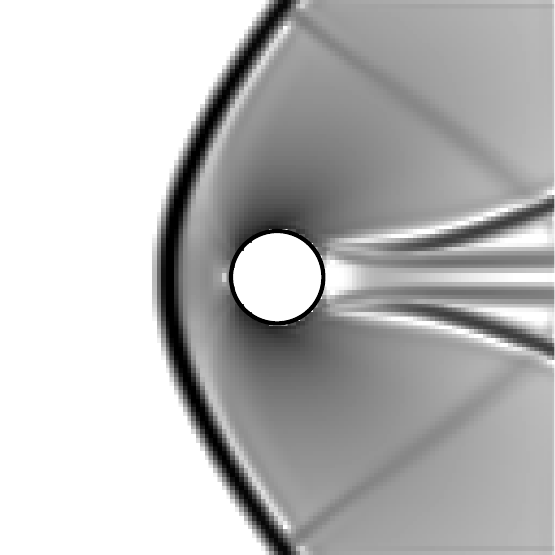};
        \addplot [
            black, 
            line width=1.0pt, 
            forget plot
        ] coordinates {
            (-0.3, -0.3)
            (-0.3, 0.3)
            (0.3, 0.3)
            (0.3, -0.3)
            (-0.3, -0.3)
        };    

        \nextgroupplot[
            tick align=outside,
            tick pos=left,
            xtick style={draw=none},
            ytick style={draw=none},
            xticklabels = {},
            xtick = {},
            yticklabels = {},
            ytick = {},
            % title=(d),
            enlargelimits=false,
            colormap name=binary,
            axis equal image,
            % colorbar,
            % colorbar style={
            %     at = {(1.05,0.0)},
            %     anchor=south west,
            %     ytick style={
            %         draw=none,
            %     },
            %     ytick={0,1,2},
            %     yticklabels = {$10^0$,$10^1$,$10^2$},
            %     width=0.3cm
            % },
            % point meta min=0,
            % point meta max=2
        ]
        \addplot graphics [xmin=-0.3,xmax=0.3,ymin=-0.3,ymax=0.3] {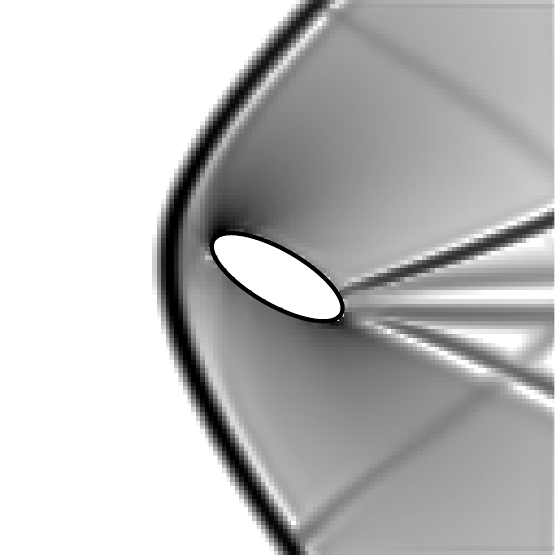};
        \addplot [
            black, 
            line width=1.0pt, 
            forget plot
        ] coordinates {
            (-0.3, -0.3)
            (-0.3, 0.3)
            (0.3, 0.3)
            (0.3, -0.3)
            (-0.3, -0.3)
        };

        % dashed “extensions” downwards from the top-right
        \draw[black, line width=0.3pt, dashed, TUMblack]
          (axis cs:0.06928203230275509,-0.04)
          -- (axis cs:0.01928203,-0.12660254);
        
        \draw[black, line width=0.3pt, dashed, TUMblack]
          (axis cs:0,0)
          -- (axis cs:-0.05,-0.08660254);
        
        % the A-arrow between those two new points, label below
        \draw[black, line width=0.5pt, <->, TUMblack]
          (axis cs:0.01928203,-0.12660254)
          -- (axis cs:-0.05,-0.08660254)
          node[below, midway] {$A$};

    % dashed guideline, reflected across the semimajor axis
    \draw[black, line width=0.3pt, dashed, TUMblack]
      (axis cs:-0.015,-0.02598076)
      -- (axis cs:0.11490381,-0.10098076);
    
    % main B‐segment (node on the left)
    \draw[black, line width=0.5pt, TUMblack]
      (axis cs:0.12990381,-0.075)
      -- (axis cs:0.11490381,-0.10098076)
      node[right, below, xshift=0.1cm] {$B$};
    
    % inward arrow at the reflected ellipse‐point
    \draw[black, line width=0.5pt, <-, TUMblack]
      (axis cs:0.11490381,-0.10098076)
      -- (axis cs:0.09240381,-0.13995190);

    % outward arrow at the reflected foot‐point
    \draw[black, line width=0.5pt, ->, TUMblack]
      (axis cs:0.15240381,-0.03602886)
      -- (axis cs:0.12990381,-0.075);
    
        \nextgroupplot[
           tick align=outside,
           tick pos=left,
           xtick style={draw=none},
           ytick style={draw=none},
           xticklabels = {},
           xtick = {},
           yticklabels = {},
           ytick = {},
        %    title=(e),
           enlargelimits=false,
           colormap name=binary,
           axis equal image,
           % colorbar,
           % colorbar style={
           %     at = {(1.05,0.0)},
           %     anchor=south west,
           %     ytick style={
           %         draw=none,
           %     },
           %     ytick={0,1,2},
           %     yticklabels = {$10^0$,$10^1$,$10^2$},
           %     width=0.3cm
           % },
           % point meta min=0,
           % point meta max=2
        ]
        % \addplot graphics [xmin=-0.3,xmax=0.3,ymin=-0.3,ymax=0.3,includegraphics={trim=0 0 0 0, clip},frame] {figures/schlieren/sphere.pdf};
        \addplot graphics [xmin=-0.3,xmax=0.3,ymin=-0.3,ymax=0.3] {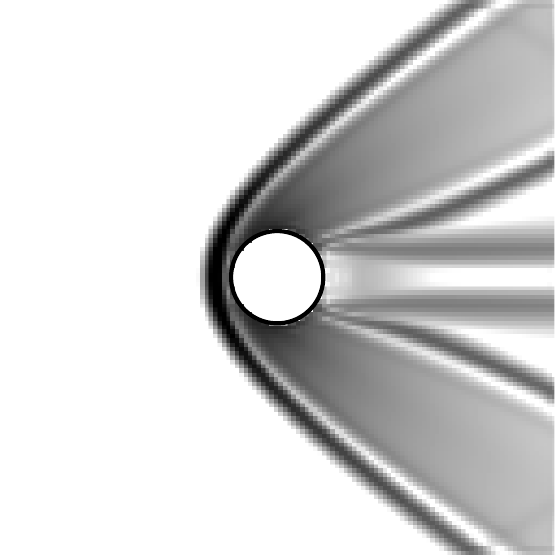};
        \addplot [
            black, 
            line width=1.0pt, 
            forget plot
        ] coordinates {
            (-0.3, -0.3)
            (-0.3, 0.3)
            (0.3, 0.3)
            (0.3, -0.3)
            (-0.3, -0.3)
        };

    \end{groupplot}

    \def\xshift{0.35cm}
    \def\yshift{0.8cm}

    \draw[black, line width=1pt, <->, TUMblack] ([yshift=-\yshift,xshift=-\xshift]$(group c1r1)$) -- ([yshift=-\yshift,xshift=\xshift]$(group c1r1)$) node[midway, below] {$R$};
    \draw[black, line width=0.3pt, dashed, TUMblack] ([xshift=-\xshift]$(group c1r1)$) --  ([yshift=-\yshift,xshift=-\xshift]$(group c1r1)$);
    \draw[black, line width=0.3pt, dashed, TUMblack] ([xshift=\xshift]$(group c1r1)$) --  ([yshift=-\yshift,xshift=\xshift]$(group c1r1)$);

    \draw[black, line width=0.5pt, <->] ([xshift=-0.2cm]$(group c1r1.south west)$) -- ([xshift=-0.2cm]$(group c1r1.north west)$) node[left, midway] {$0.6$};
    \draw[black, line width=0.5pt, <->] ([yshift=-0.2cm]$(group c1r1.south west)$) -- ([yshift=-0.2cm]$(group c1r1.south east)$) node[below, midway] {$0.6$};

    \draw[black, line width=1pt, ->, TUMblack] ($(group c1r1)$) -- ([xshift=0.6cm]$(group c1r1)$) node[above] {$x$};
    \draw[black, line width=1pt, ->, TUMblack] ($(group c1r1)$) -- ([yshift=0.6cm]$(group c1r1)$) node[right] {$y$};

    \draw[black, line width=1pt, ->, TUMblack] ($(group c2r1)$) -- ([xshift=0.6cm]$(group c2r1)$) node[above] {$x$};
    \draw[black, line width=1pt, ->, TUMblack] ($(group c2r1)$) -- ([yshift=0.6cm]$(group c2r1)$) node[right] {$y$};

    \draw[black, line width=1pt, ->, TUMblack] ($(group c3r1)$) -- ([xshift=0.6cm]$(group c3r1)$) node[above] {$x$};
    \draw[black, line width=1pt, ->, TUMblack] ($(group c3r1)$) -- ([yshift=0.6cm]$(group c3r1)$) node[right] {$y$};
    \node[inner sep=0pt, scale=0.6, line width=1pt, TUMblack] (A) at ($(group c3r1)$) {$\boldsymbol{\otimes}$};
    \node[yshift=-0.09cm, TUMblack] at (A.south) {$z$};

    \draw[black, line width=0.5pt, ->, TUMblack] ([yshift=-0.4cm,xshift=0.2cm]$(group c1r1.north west)$) -- ([yshift=-0.4cm,xshift=1.2cm]$(group c1r1.north west)$) node[below,midway] {$M_\infty=2$};

    \draw[black, dashed, TUMblack] ([yshift=0.9814954576223637cm,xshift=-1.7cm]$(group c2r1)$) -- ([yshift=-0.9814954576223637cm,xshift=1.7cm]$(group c2r1)$);
    \draw[black, dashed, TUMblack] ([xshift=-1.7cm]$(group c2r1)$) -- ([xshift=1.7cm]$(group c2r1)$);

    \draw[<-, line width=0.5pt, TUMblack] ([shift={(0,0)}]$(group c2r1)$) ++(180:1.3cm) arc[start angle=180, end angle=150, radius=1.3cm] node[midway,right] {$\lambda$};

    \def\xshift{0.35cm}
    \def\yshift{0.8cm}

    \draw[black, line width=1pt, <->, TUMblack] ([yshift=-\yshift,xshift=-\xshift]$(group c3r1)$) -- ([yshift=-\yshift,xshift=\xshift]$(group c3r1)$) node[midway, below] {$R$};
    \draw[black, line width=0.3pt, dashed, TUMblack] ([xshift=-\xshift]$(group c3r1)$) --  ([yshift=-\yshift,xshift=-\xshift]$(group c3r1)$);
    \draw[black, line width=0.3pt, dashed, TUMblack] ([xshift=\xshift]$(group c3r1)$) --  ([yshift=-\yshift,xshift=\xshift]$(group c3r1)$);

\end{tikzpicture}

%% file: figures/parametric_shape_inference/error_table.tex
\begingroup
\pgfkeys{/pgf/number format/.cd, sci, precision=2, zerofill}

% cylinder odil
\pgfplotstableread{./figures/parametric_shape_inference/cylinder/odil/error_history_50000.txt}\datatable
\pgfplotstablegetrowsof{\datatable}
\pgfmathtruncatemacro{\lastrow}{\pgfplotsretval-1}
\pgfplotstablegetelem{\lastrow}{[index]0}\of\datatable
\pgfmathsetmacro{\rhoodilcylinder}{\pgfplotsretval}
\pgfplotstablegetelem{\lastrow}{[index]1}\of\datatable
\pgfmathsetmacro{\podilcylinder}{\pgfplotsretval}
\pgfplotstablegetelem{\lastrow}{[index]3}\of\datatable
\pgfmathsetmacro{\uodilcylinder}{\pgfplotsretval}
\pgfplotstablegetelem{\lastrow}{[index]4}\of\datatable
\pgfmathsetmacro{\vodilcylinder}{\pgfplotsretval}
% ellipse odil
\pgfplotstableread{./figures/parametric_shape_inference/ellipse/odil/error_history_150000.txt}\datatable
\pgfplotstablegetrowsof{\datatable}
\pgfmathtruncatemacro{\lastrow}{\pgfplotsretval-1}
\pgfplotstablegetelem{\lastrow}{[index]0}\of\datatable
\pgfmathsetmacro{\rhoodilellipse}{\pgfplotsretval}
\pgfplotstablegetelem{\lastrow}{[index]1}\of\datatable
\pgfmathsetmacro{\podilellipse}{\pgfplotsretval}
\pgfplotstablegetelem{\lastrow}{[index]3}\of\datatable
\pgfmathsetmacro{\uodilellipse}{\pgfplotsretval}
\pgfplotstablegetelem{\lastrow}{[index]4}\of\datatable
\pgfmathsetmacro{\vodilellipse}{\pgfplotsretval}
% sphere odil
\pgfplotstableread{./figures/parametric_shape_inference/sphere/odil/error_history_30000.txt}\datatable
\pgfplotstablegetrowsof{\datatable}
\pgfmathtruncatemacro{\lastrow}{\pgfplotsretval-1}
\pgfplotstablegetelem{\lastrow}{[index]0}\of\datatable
\pgfmathsetmacro{\rhoodilsphere}{\pgfplotsretval}
\pgfplotstablegetelem{\lastrow}{[index]1}\of\datatable
\pgfmathsetmacro{\podilsphere}{\pgfplotsretval}
\pgfplotstablegetelem{\lastrow}{[index]3}\of\datatable
\pgfmathsetmacro{\uodilsphere}{\pgfplotsretval}
\pgfplotstablegetelem{\lastrow}{[index]4}\of\datatable
\pgfmathsetmacro{\vodilsphere}{\pgfplotsretval}
\pgfplotstablegetelem{\lastrow}{[index]5}\of\datatable
\pgfmathsetmacro{\wodilsphere}{\pgfplotsretval}
% cylinder pinn
\pgfplotstableread{./figures/parametric_shape_inference/cylinder/pinn/error_history_50000.txt}\datatable
\pgfplotstablegetrowsof{\datatable}
\pgfmathtruncatemacro{\lastrow}{\pgfplotsretval-1}
\pgfplotstablegetelem{\lastrow}{[index]0}\of\datatable
\pgfmathsetmacro{\rhopinncylinder}{\pgfplotsretval}
\pgfplotstablegetelem{\lastrow}{[index]1}\of\datatable
\pgfmathsetmacro{\ppinncylinder}{\pgfplotsretval}
\pgfplotstablegetelem{\lastrow}{[index]4}\of\datatable
\pgfmathsetmacro{\upinncylinder}{\pgfplotsretval}
\pgfplotstablegetelem{\lastrow}{[index]5}\of\datatable
\pgfmathsetmacro{\vpinncylinder}{\pgfplotsretval}
% ellipse odil
\pgfplotstableread{./figures/parametric_shape_inference/ellipse/pinn/error_history_150000.txt}\datatable
\pgfplotstablegetrowsof{\datatable}
\pgfmathtruncatemacro{\lastrow}{\pgfplotsretval-1}
\pgfplotstablegetelem{\lastrow}{[index]0}\of\datatable
\pgfmathsetmacro{\rhopinnellipse}{\pgfplotsretval}
\pgfplotstablegetelem{\lastrow}{[index]1}\of\datatable
\pgfmathsetmacro{\ppinnellipse}{\pgfplotsretval}
\pgfplotstablegetelem{\lastrow}{[index]4}\of\datatable
\pgfmathsetmacro{\upinnellipse}{\pgfplotsretval}
\pgfplotstablegetelem{\lastrow}{[index]5}\of\datatable
\pgfmathsetmacro{\vpinnellipse}{\pgfplotsretval}
% sphere odil
\pgfplotstableread{./figures/parametric_shape_inference/sphere/pinn/error_history_30000.txt}\datatable
\pgfplotstablegetrowsof{\datatable}
\pgfmathtruncatemacro{\lastrow}{\pgfplotsretval-1}
\pgfplotstablegetelem{\lastrow}{[index]0}\of\datatable
\pgfmathsetmacro{\rhopinnsphere}{\pgfplotsretval}
\pgfplotstablegetelem{\lastrow}{[index]1}\of\datatable
\pgfmathsetmacro{\ppinnsphere}{\pgfplotsretval}
\pgfplotstablegetelem{\lastrow}{[index]4}\of\datatable
\pgfmathsetmacro{\upinnsphere}{\pgfplotsretval}
\pgfplotstablegetelem{\lastrow}{[index]5}\of\datatable
\pgfmathsetmacro{\vpinnsphere}{\pgfplotsretval}
\pgfplotstablegetelem{\lastrow}{[index]6}\of\datatable
\pgfmathsetmacro{\wpinnsphere}{\pgfplotsretval}

\begin{table}[!b]
  \centering
  \begin{tabular}{p{4em} c c c c c c}
    \hline
    Method&Shape&$\rho$ & $u$ & $v$ & $w$ &$p$ \\
    \hline
    \multirow{3}{4em}{ODIL}&Cylinder&\print{\rhoodilcylinder}&\print{\uodilcylinder}&\print{\vodilcylinder}&-&\print{\podilcylinder} \\
    &Ellipse&\print{\rhoodilellipse}&\print{\uodilellipse}&\print{\vodilellipse}&-&\print{\podilellipse} \\
    &Sphere&\print{\rhoodilsphere}&\print{\uodilsphere}&\print{\vodilsphere}&\print{\wodilsphere}&\print{\podilsphere} \\
    \hline
    \multirow{3}{4em}{PINN}&Cylinder&\print{\rhopinncylinder}&\print{\upinncylinder}&\print{\vpinncylinder}&-&\print{\ppinncylinder} \\
    &Ellipse&\print{\rhopinnellipse}&\print{\upinnellipse}&\print{\vpinnellipse}&-&\print{\ppinnellipse} \\
    &Sphere&\print{\rhopinnsphere}&\print{\upinnsphere}&\print{\vpinnsphere}&\print{\wpinnsphere}&\print{\ppinnsphere} \\
    \hline
  \end{tabular}
  \caption{
    Mean cell-wise relative error between ODIL/PINN and reference for low-dimensional shape parameterization
    for the baseline point distribution, see Section \ref{subsubsec:baseline_points_parametrc}.
    The corresponding flow fields are shown in Figure \ref{fig:flowfield_parametric}.
    }
  \label{tab:errors_parametric}
\end{table}

\endgroup

%% file: figures/parametric_shape_inference/flowfield_baseline.tex
\begin{tikzpicture}
    % \pgfmathdeclarefunction{lg10}{1}{%
    %     \pgfmathparse{ln(#1)/ln(10)}%
    % }
    
    % Spectral_r colormap from matplotlib
    \pgfplotsset{
        colormap={spectralr}{rgb255=(94,79,162) rgb255=(68,112,177) rgb255=(59,146,184) rgb255=(89,180,170)
            rgb255=(126,203,164) rgb255=(166,219,164) rgb255=(202,233,157) rgb255=(232,246,156) 
            rgb255=(247,252,179) rgb255=(254,245,175) rgb255=(254,227,145) rgb255=(253,200,119) 
            rgb255=(252,170,95) rgb255=(247,131,77) rgb255=(236,97,69) rgb255=(218,70,76)
            rgb255=(190,36,73) rgb255=(158,1,66)}}
    
    \begin{groupplot}[
        group style={
            group size=6 by 3,
            horizontal sep=0.05cm,
            vertical sep=0.35cm
        }, 
        width=3.9cm, 
        height=3.9cm
        ]

        \nextgroupplot[
            tick align=outside,
            tick pos=left,
            xtick style={draw=none},
            ytick style={draw=none},
            xticklabels = {},
            xtick = {},
            yticklabels = {},
            ytick = {},
            title=Reference,
            % ylabel style={yshift=2pt}
            % ylabel=$\rho$,
            enlargelimits=false,
            colormap name=spectralr,
            axis equal image,
        ]
        \addplot graphics [xmin=-0.3,xmax=0.3,ymin=-0.3,ymax=0.3] {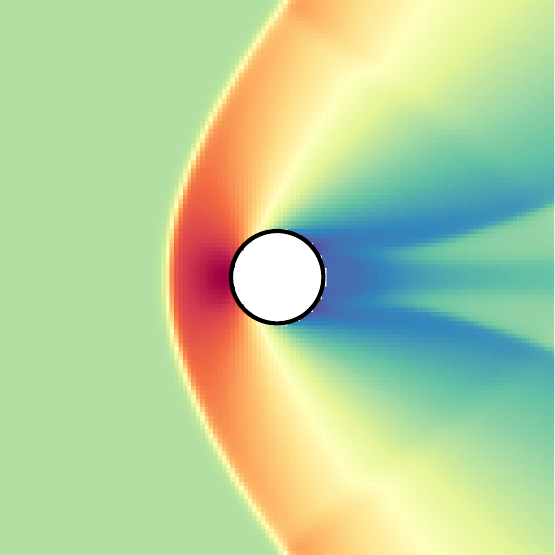};

        \nextgroupplot[
            tick align=outside,
            tick pos=left,
            xtick style={draw=none},
            ytick style={draw=none},
            xticklabels = {},
            xtick = {},
            yticklabels = {},
            ytick = {},
            title=ODIL,
            % title=$u$,
            enlargelimits=false,
            colormap name=spectralr,
            axis equal image,
        ]
        \addplot graphics [xmin=-0.3,xmax=0.3,ymin=-0.3,ymax=0.3] {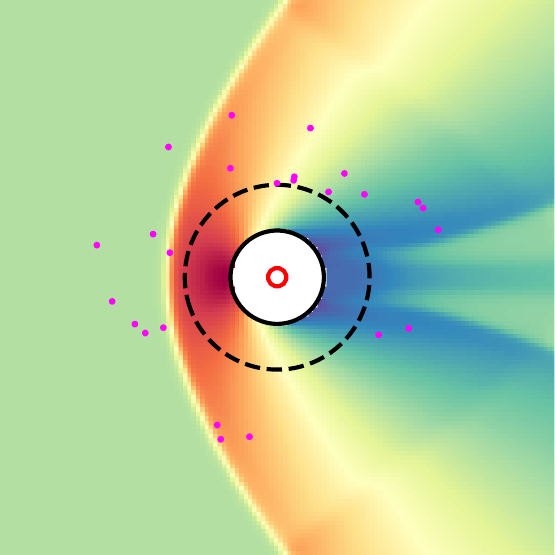};

        \nextgroupplot[
            tick align=outside,
            tick pos=left,
            xtick style={draw=none},
            ytick style={draw=none},
            xticklabels = {},
            xtick = {},
            yticklabels = {},
            ytick = {},
            title = PINN,
            enlargelimits=false,
            colormap name=spectralr,
            axis equal image,
            colorbar,
            colorbar style={
                at={(1.02,0.0)},
                anchor=south west,
                ytick style={
                    draw=none,
                },
                ytick={0.07,1.56,3.05},
                yticklabels = {0.07,1.56,3.05},
                width=0.2cm
            },
            point meta min=0.07,
            point meta max=3.05
        ]
        \addplot graphics [xmin=-0.3,xmax=0.3,ymin=-0.3,ymax=0.3] {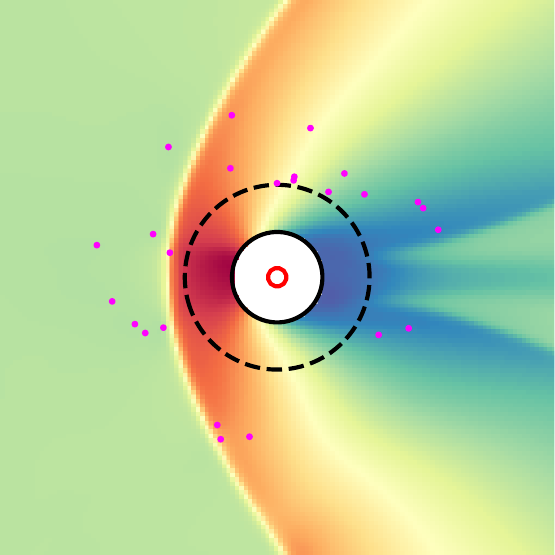};

        \nextgroupplot[
            xshift=1.2cm,
            tick align=outside,
            tick pos=left,
            xtick style={draw=none},
            ytick style={draw=none},
            xticklabels = {},
            xtick = {},
            yticklabels = {},
            ytick = {},
            title=Reference,
            enlargelimits=false,
            colormap name=spectralr,
            axis equal image,
        ]
        \addplot graphics [xmin=-0.3,xmax=0.3,ymin=-0.3,ymax=0.3] {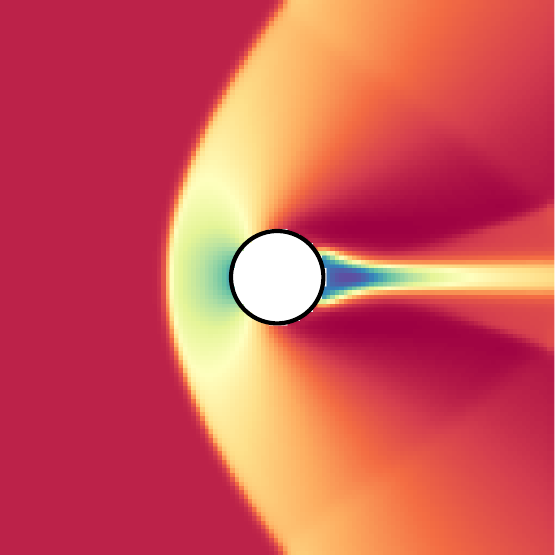};

        \nextgroupplot[
            tick align=outside,
            tick pos=left,
            xtick style={draw=none},
            ytick style={draw=none},
            xticklabels = {},
            xtick = {},
            yticklabels = {},
            ytick = {},
            title=ODIL,
            enlargelimits=false,
            colormap name=spectralr,
            axis equal image,
        ]
        \addplot graphics [xmin=-0.3,xmax=0.3,ymin=-0.3,ymax=0.3] {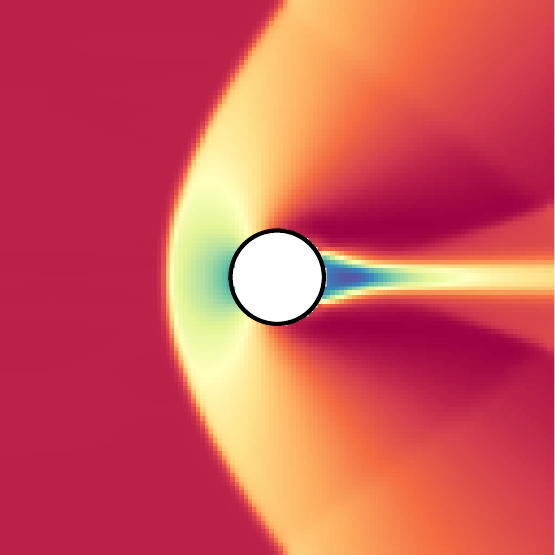};

        \nextgroupplot[
            tick align=outside,
            tick pos=left,
            xtick style={draw=none},
            ytick style={draw=none},
            xticklabels = {},
            xtick = {},
            yticklabels = {},
            ytick = {},
            title = PINN,
            % title=$\log_{10}\left(\vert \text{rel. error} \vert \right)$,
            enlargelimits=false,
            colormap name=spectralr,
            colorbar,
            colorbar style={
                at={(1.02,0.0)},
                anchor=south west,
                ytick style={
                    draw=none,
                },
                ytick={-0.32,1.1,2.52},
                yticklabels = {-0.32,1.1,2.52},
                width=0.2cm
            },
            point meta min=-0.32,
            point meta max=2.52
        ]
        \addplot graphics [xmin=-0.3,xmax=0.3,ymin=-0.3,ymax=0.3] {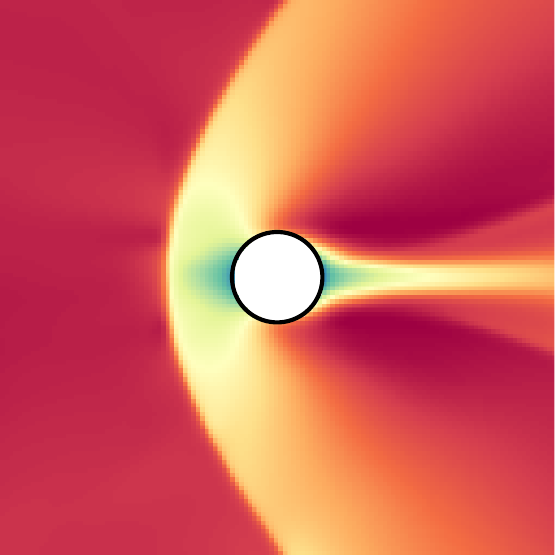};

        \nextgroupplot[
            tick align=outside,
            tick pos=left,
            xtick style={draw=none},
            ytick style={draw=none},
            xticklabels = {},
            xtick = {},
            yticklabels = {},
            ytick = {},
            % title=reference,
            % ylabel style={yshift=2pt}
            % ylabel=$\rho$,
            enlargelimits=false,
            colormap name=spectralr,
            axis equal image,
        ]
        \addplot graphics [xmin=-0.3,xmax=0.3,ymin=-0.3,ymax=0.3] {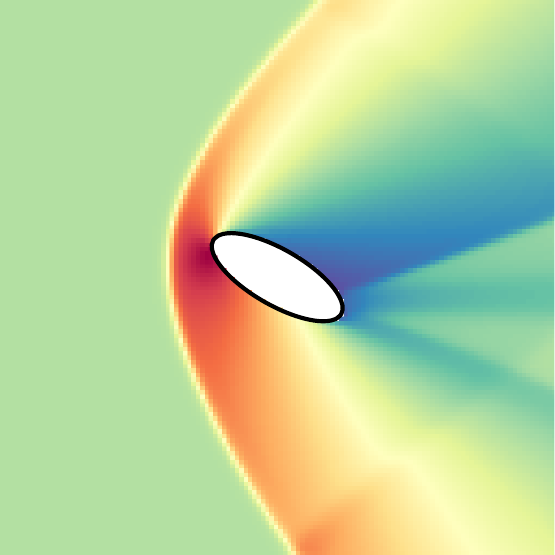};

        \nextgroupplot[
            tick align=outside,
            tick pos=left,
            xtick style={draw=none},
            ytick style={draw=none},
            xticklabels = {},
            xtick = {},
            yticklabels = {},
            ytick = {},
            % title=ODIL,
            % title=$u$,
            enlargelimits=false,
            colormap name=spectralr,
            axis equal image,
        ]
        \addplot graphics [xmin=-0.3,xmax=0.3,ymin=-0.3,ymax=0.3] {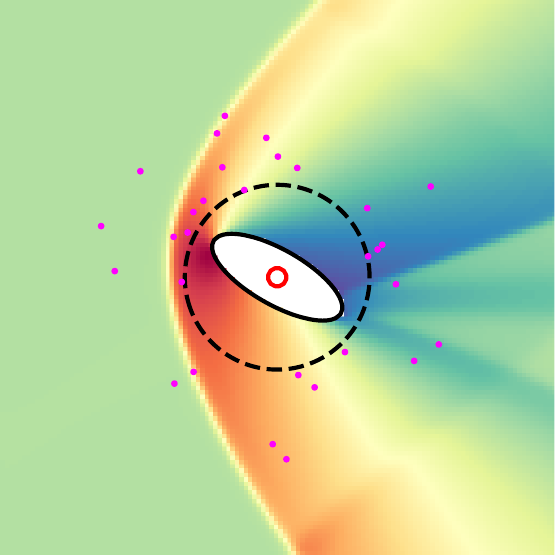};

        \nextgroupplot[
            tick align=outside,
            tick pos=left,
            xtick style={draw=none},
            ytick style={draw=none},
            xticklabels = {},
            xtick = {},
            yticklabels = {},
            ytick = {},
            % title = PINN,
            enlargelimits=false,
            colormap name=spectralr,
            axis equal image,
            colorbar,
            colorbar style={
                at={(1.02,0.0)},
                anchor=south west,
                ytick style={
                    draw=none,
                },
                ytick={0.07,1.55,3.03},
                yticklabels = {0.07,1.55,3.03},
                width=0.2cm
            },
            point meta min=0.07,
            point meta max=3.03
        ]
        \addplot graphics [xmin=-0.3,xmax=0.3,ymin=-0.3,ymax=0.3] {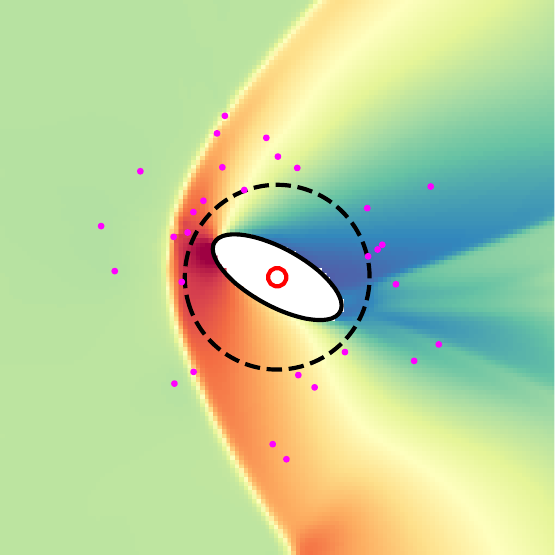};

        \nextgroupplot[
            tick align=outside,
            tick pos=left,
            xtick style={draw=none},
            ytick style={draw=none},
            xticklabels = {},
            xtick = {},
            yticklabels = {},
            ytick = {},
            enlargelimits=false,
            colormap name=spectralr,
            axis equal image,
        ]
        \addplot graphics [xmin=-0.3,xmax=0.3,ymin=-0.3,ymax=0.3] {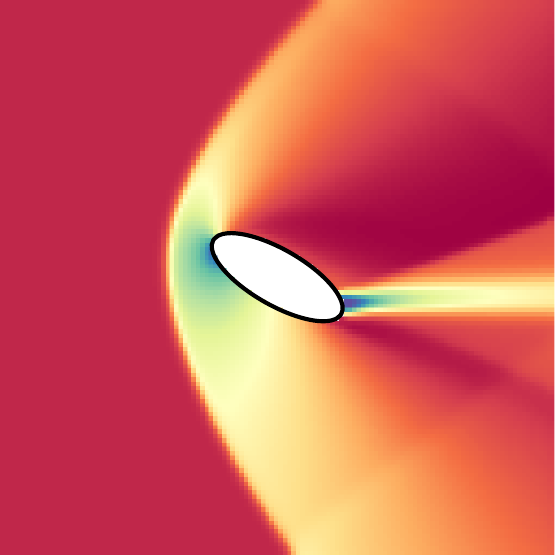};

        \nextgroupplot[
            tick align=outside,
            tick pos=left,
            xtick style={draw=none},
            ytick style={draw=none},
            xticklabels = {},
            xtick = {},
            yticklabels = {},
            ytick = {},
            % title=ODIL,
            enlargelimits=false,
            colormap name=spectralr,
            axis equal image,
        ]
        \addplot graphics [xmin=-0.3,xmax=0.3,ymin=-0.3,ymax=0.3] {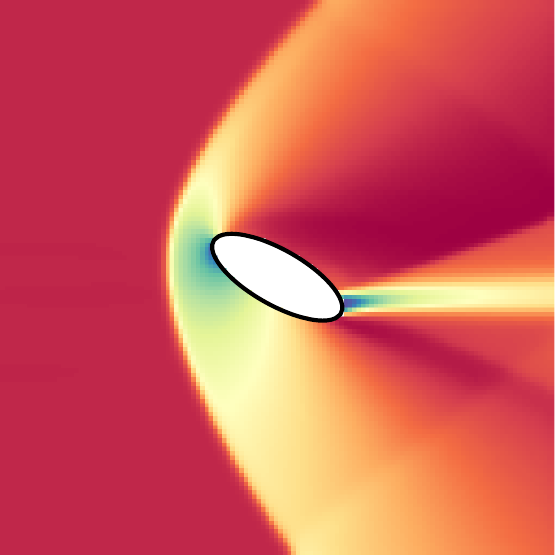};

        \nextgroupplot[
            tick align=outside,
            tick pos=left,
            xtick style={draw=none},
            ytick style={draw=none},
            xticklabels = {},
            xtick = {},
            yticklabels = {},
            ytick = {},
            % title = PINN,
            % title=$\log_{10}\left(\vert \text{rel. error} \vert \right)$,
            enlargelimits=false,
            colormap name=spectralr,
            colorbar,
            colorbar style={
                at={(1.02,0.0)},
                anchor=south west,
                ytick style={
                    draw=none,
                },
                ytick={0.01,1.27,2.53},
                yticklabels = {0.01,1.27,2.53},
                width=0.2cm
            },
            point meta min=0.01,
            point meta max=2.53
        ]
        \addplot graphics [xmin=-0.3,xmax=0.3,ymin=-0.3,ymax=0.3] {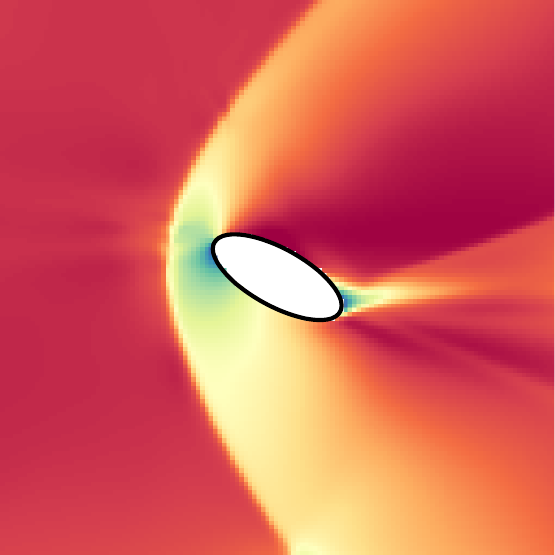};

        \nextgroupplot[
            tick align=outside,
            tick pos=left,
            xtick style={draw=none},
            ytick style={draw=none},
            xticklabels = {},
            xtick = {},
            yticklabels = {},
            ytick = {},
            % title=reference,
            % ylabel style={yshift=2pt}
            % ylabel=$\rho$,
            enlargelimits=false,
            colormap name=spectralr,
            axis equal image,
        ]
        \addplot graphics [xmin=-0.3,xmax=0.3,ymin=-0.3,ymax=0.3] {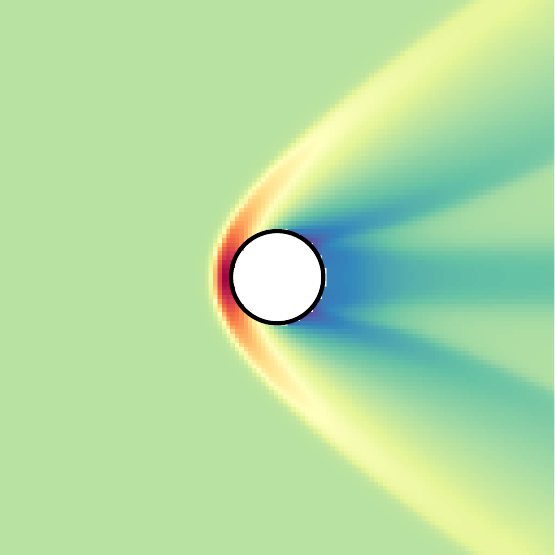};

        \nextgroupplot[
            tick align=outside,
            tick pos=left,
            xtick style={draw=none},
            ytick style={draw=none},
            xticklabels = {},
            xtick = {},
            yticklabels = {},
            ytick = {},
            % title=ODIL,
            % title=$u$,
            enlargelimits=false,
            colormap name=spectralr,
            axis equal image,
        ]
        \addplot graphics [xmin=-0.3,xmax=0.3,ymin=-0.3,ymax=0.3] {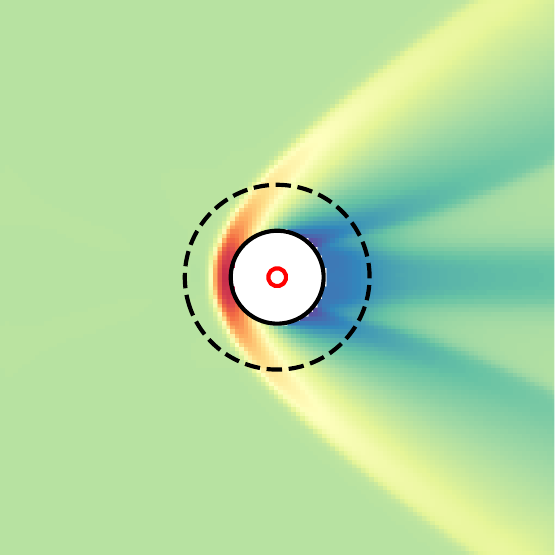};

        \nextgroupplot[
            tick align=outside,
            tick pos=left,
            xtick style={draw=none},
            ytick style={draw=none},
            xticklabels = {},
            xtick = {},
            yticklabels = {},
            ytick = {},
            % title = PINN,
            enlargelimits=false,
            colormap name=spectralr,
            axis equal image,
            colorbar,
            colorbar style={
                at={(1.02,0.0)},
                anchor=south west,
                ytick style={
                    draw=none,
                },
                ytick={0.12,1.47,2.82},
                yticklabels = {0.12,1.47,2.82},
                width=0.2cm
            },
            point meta min=0.12,
            point meta max=2.82
        ]
        \addplot graphics [xmin=-0.3,xmax=0.3,ymin=-0.3,ymax=0.3] {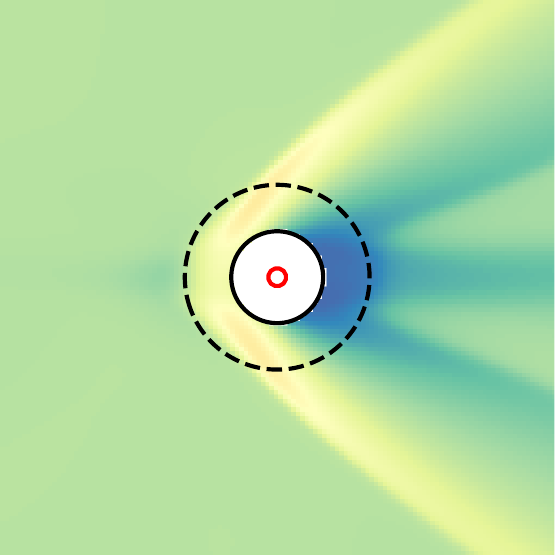};

        \nextgroupplot[
            tick align=outside,
            tick pos=left,
            xtick style={draw=none},
            ytick style={draw=none},
            xticklabels = {},
            xtick = {},
            yticklabels = {},
            ytick = {},
            enlargelimits=false,
            colormap name=spectralr,
            axis equal image,
        ]
        \addplot graphics [xmin=-0.3,xmax=0.3,ymin=-0.3,ymax=0.3] {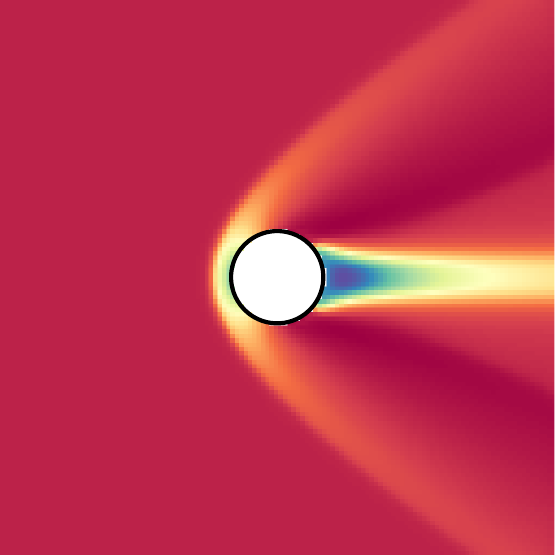};

        \nextgroupplot[
            tick align=outside,
            tick pos=left,
            xtick style={draw=none},
            ytick style={draw=none},
            xticklabels = {},
            xtick = {},
            yticklabels = {},
            ytick = {},
            % title=ODIL,
            enlargelimits=false,
            colormap name=spectralr,
            axis equal image,
        ]
        \addplot graphics [xmin=-0.3,xmax=0.3,ymin=-0.3,ymax=0.3] {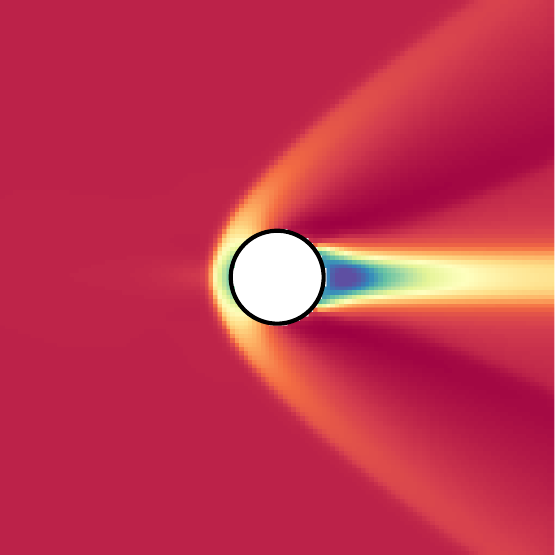};

        \nextgroupplot[
            tick align=outside,
            tick pos=left,
            xtick style={draw=none},
            ytick style={draw=none},
            xticklabels = {},
            xtick = {},
            yticklabels = {},
            ytick = {},
            % title = PINN,
            % title=$\log_{10}\left(\vert \text{rel. error} \vert \right)$,
            enlargelimits=false,
            colormap name=spectralr,
            colorbar,
            colorbar style={
                at={(1.02,0.0)},
                anchor=south west,
                ytick style={
                    draw=none,
                },
                ytick={-0.52,1.01,2.54},
                yticklabels = {-0.52,1.01,2.54},
                width=0.2cm
            },
            point meta min=-0.52,
            point meta max=2.54
        ]
        \addplot graphics [xmin=-0.3,xmax=0.3,ymin=-0.3,ymax=0.3] {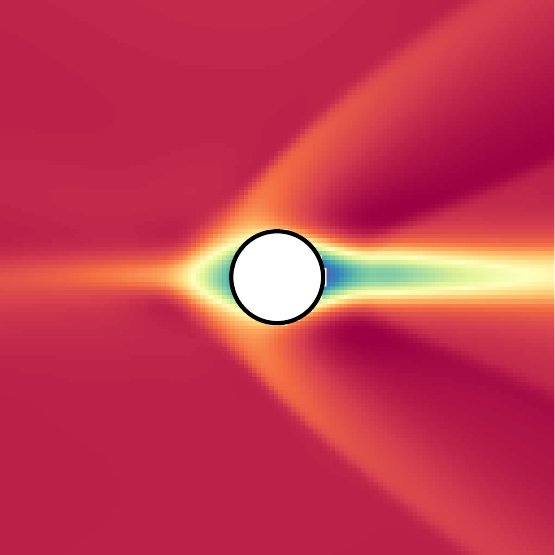};

    \end{groupplot}

    % \node[above, yshift=2cm] at ($(group c1r1.north east)!0.5!(group c4r1.north west)$) {Sphere};
    % \node[above, yshift=1cm] at ($(group c1r1.north east)!0.5!(group c3r1.north west)$) {Flow field};
    % \node[above, yshift=1cm] at ($(group c4r1.north east)!0.5!(group c5r1.north west)$) {$\log_{10}(|\text{rel. error}|)$};
    % \node[above, xshift=0.0cm, rotate=90] at ($(group c1r1.south west)!0.5!(group c1r1.north west)$) {$\rho$};
    % \node[above, xshift=0.0cm, rotate=90] at ($(group c1r2.south west)!0.5!(group c1r2.north west)$) {$u$};

\end{tikzpicture}

%% file: figures/parametric_shape_inference/shape_params_history.tex
\begin{tikzpicture}
    
    % \definecolor{color0}{RGB}{0,101,189}
    % \definecolor{color1}{RGB}{227,114,34}
    % \definecolor{color2}{RGB}{162,173,0}

    \begin{groupplot}[
        group style={group size=4 by 1, horizontal sep=1.3cm},
        width=4.2cm, height=4.2cm]
      
    \nextgroupplot[
    tick align=outside,
    tick pos=left,
    legend style={at={(1.0,0.0)},draw=black,anchor=south east,font=\footnotesize},
    xtick style={color=black},
    xlabel=step,
    ylabel=$R$,
    % title=(a)
    ]

    \addplot [line width=1.0pt, TUMblue] table[x index = {0}, y index = {1}]
    {figures/parametric_shape_inference/cylinder/odil/radius_history_50000.txt};
    \addlegendentry{ODIL}
    \addplot [line width=1.0pt, TUMorange] table[x index = {0}, y index = {1}]
    {figures/parametric_shape_inference/cylinder/pinn/radius_history_50000.txt};
    \addlegendentry{PINN}
    \addplot [line width=1.0pt, black, dashed] coordinates {(0,0.05) (50000,0.05)};

    \nextgroupplot[
        xshift=-0.8cm,
        tick align=outside,
        tick pos=left,
        ytick={0.02,0.04},
        yticklabels={},
        xtick style={color=black},
        xlabel=step,
        % title=(d)
        ]
    \addplot [line width=1.0pt, TUMblue] table[x index = {0}, y index = {1}]
    {figures/parametric_shape_inference/sphere/odil/radius_history_30000.txt};
    \addplot [line width=1.0pt, TUMorange] table[x index = {0}, y index = {1}]
    {figures/parametric_shape_inference/sphere/pinn/radius_history_30000.txt};
    \addplot [line width=1.0pt, black, dashed] coordinates {(0,0.05) (30000,0.05)};

    \nextgroupplot[
    tick align=outside,
    tick pos=left,
    xtick style={color=black},
    xlabel=step,
    ylabel=$A\text{,} \ B$,
    % title=(b)
    ]

    \addplot [line width=1.0pt, TUMblue] table[x index = {0}, y index = {1}]
    {figures/parametric_shape_inference/ellipse/odil/radius_history_150000.txt};
    \addplot [line width=1.0pt, TUMblue] table[x index = {0}, y index = {2}]
    {figures/parametric_shape_inference/ellipse/odil/ratio_history_150000.txt};
    \addplot [line width=1.0pt, TUMorange] table[x index = {0}, y index = {1}]
    {figures/parametric_shape_inference/ellipse/pinn/radius_history_150000.txt};
    \addplot [line width=1.0pt, TUMorange] table[x index = {0}, y index = {2}]
    {figures/parametric_shape_inference/ellipse/pinn/ratio_history_150000.txt};
    \addplot [line width=1.0pt, black, dashed] coordinates {(0,0.03) (150000,0.03)};
    \addplot [line width=1.0pt, black, dashed] coordinates {(0,0.08) (150000,0.08)};

    \nextgroupplot[
    tick align=outside,
    tick pos=left,
    xshift=0.3cm,
    xtick style={color=black},
    xlabel=step,
    ylabel=$\lambda$ in degree,
    % title=(c)
    ]
    \addplot [line width=1.0pt, TUMblue] table[x index = {0}, y index = {1}]
    {figures/parametric_shape_inference/ellipse/odil/angle_history_150000.txt};
    \addplot [line width=1.0pt, TUMorange] table[x index = {0}, y index = {1}]
    {figures/parametric_shape_inference/ellipse/pinn/angle_history_150000.txt};
    \addplot [line width=1.0pt, black, dashed] coordinates {(0,-30) (150000,-30)};
    
    \end{groupplot}

\node[above, yshift=0.5cm] at ($(group c1r1.north)$) {Cylinder};
\node[above, yshift=0.5cm] at ($(group c3r1.north east)!0.5!(group c4r1.north west)$) {Ellipse};
\node[above, yshift=0.5cm] at ($(group c2r1.north)$) {Sphere};
\end{tikzpicture}

%% file: figures/parametric_shape_inference/loss_history.tex
\begin{tikzpicture}

\begin{groupplot}[
    group style={group size=3 by 2, horizontal sep=1.3cm, vertical sep=1.7cm},
    width=4.2cm, height=4.2cm]
    
\nextgroupplot[
tick align=outside,
tick pos=left,
legend style={at={(1.2,1.1)},draw=black,anchor=north east,font=\footnotesize,legend columns=2},
xtick style={color=black},
ymin=1e-6, ymax=1e4,
% xmin=0.0, xmax=1e5,
log basis x={10},
log basis y={10},
% xtick={0,40000,30000},
% xmode=log,
ymode=log,
ylabel=$\omega \mathcal{L}$,
% title=(a)
]

\addplot [line width=1.0pt, TUMblue] table[skip first n=1,
    x index = {0}, y index = {2}
    ] {figures/parametric_shape_inference/cylinder/odil/loss_history_50000.txt};
\addlegendentry{$E$};
\addplot [line width=1.0pt, TUMorange] table[skip first n=1,
    x index = {0}, y index = {3}
    ] {figures/parametric_shape_inference/cylinder/odil/loss_history_50000.txt};
\addlegendentry{$W$};
\addplot [line width=1.0pt, TUMgreen] table[skip first n=1,
    x index = {0}, y index = {4}
    ] {figures/parametric_shape_inference/cylinder/odil/loss_history_50000.txt};
\addlegendentry{$S$};
\addplot [line width=1.0pt, TUMgray] table[skip first n=1,
    x index = {0}, y index = {6}
    ] {figures/parametric_shape_inference/cylinder/odil/loss_history_50000.txt};
\addlegendentry{$T_0$};
\addplot [line width=1.0pt, black, dashed] coordinates {(20000,1e-6) (20000,1e4)};

\nextgroupplot[
tick align=outside,
tick pos=left,
xtick style={color=black},
ymin=1e-6, ymax=1e4,
% xmin=0.0, xmax=1e5,
log basis x={10},
log basis y={10},
% xtick={0,40000,30000},
% xmode=log,
ymode=log,
yticklabels={},
% ylabel=$\omega \mathcal{L}$,
% title=(b)
]

\addplot [line width=1.0pt, TUMblue] table[skip first n=1,
    x index = {0}, y index = {2}
    ] {figures/parametric_shape_inference/ellipse/odil/loss_history_150000.txt};
% \addlegendentry{$E$};
\addplot [line width=1.0pt, TUMorange] table[skip first n=1,
    x index = {0}, y index = {3}
    ] {figures/parametric_shape_inference/ellipse/odil/loss_history_150000.txt};
% \addlegendentry{$W$};
\addplot [line width=1.0pt, TUMgreen] table[skip first n=1,
    x index = {0}, y index = {4}
    ] {figures/parametric_shape_inference/ellipse/odil/loss_history_150000.txt};
% \addlegendentry{$S$};
\addplot [line width=1.0pt, TUMgray] table[skip first n=1,
    x index = {0}, y index = {6}
    ] {figures/parametric_shape_inference/ellipse/odil/loss_history_150000.txt};
% \addlegendentry{$T_0$};
\addplot [line width=1.0pt, black, dashed] coordinates {(20000,1e-6) (20000,1e4)};

\nextgroupplot[
tick align=outside,
tick pos=left,
xtick style={color=black},
ymin=1e-6, ymax=1e4,
% xmin=0.0, xmax=1e5,
log basis x={10},
log basis y={10},
% xtick={0,40000,30000},
% xmode=log,
ymode=log,
yticklabels={},
% ylabel=$\omega \mathcal{L}$,
% title=(b)
]

\addplot [line width=1.0pt, TUMblue] table[skip first n=1,
    x index = {0}, y index = {2}
    ] {figures/parametric_shape_inference/sphere/odil/loss_history_30000.txt};
% \addlegendentry{$E$};
\addplot [line width=1.0pt, TUMorange] table[skip first n=1,
    x index = {0}, y index = {3}
    ] {figures/parametric_shape_inference/sphere/odil/loss_history_30000.txt};
% \addlegendentry{$W$};
\addplot [line width=1.0pt, TUMgreen] table[skip first n=1,
    x index = {0}, y index = {4}
    ] {figures/parametric_shape_inference/sphere/odil/loss_history_30000.txt};
% \addlegendentry{$S$};
\addplot [line width=1.0pt, TUMgray] table[skip first n=1,
    x index = {0}, y index = {6}
    ] {figures/parametric_shape_inference/sphere/odil/loss_history_30000.txt};
% \addlegendentry{$T_0$};
\addplot [line width=1.0pt, black, dashed] coordinates {(5000,1e-6) (5000,1e4)};

\nextgroupplot[
    tick align=outside,
    tick pos=left,
    legend style={at={(1.2,1.1)},draw=black,anchor=north east,font=\footnotesize,legend columns=2},
    xtick style={color=black},
    ymin=1e-4, ymax=1e4,
    % xmin=0.0, xmax=1e5,
    log basis x={10},
    log basis y={10},
    % xmode=log,
    ymode=log,
    xlabel=step,
    % title=(c),
    ylabel=$\omega \mathcal{L}$,
    % ylabel=$\mathcal{L}$
    ]

\addplot [line width=1.0pt, TUMblue] table[skip first n=1,
x index = {11}, y index = {0}
] {figures/parametric_shape_inference/cylinder/pinn/loss_history_50000.txt};
\addlegendentry{$E$};
\addplot [line width=1.0pt, TUMorange] table[skip first n=1,
    x index = {11}, y index = {6}
    ] {figures/parametric_shape_inference/cylinder/pinn/loss_history_50000.txt};
\addlegendentry{$W$};
\addplot [line width=1.0pt, TUMgreen] table[skip first n=1,
    x index = {11}, y index = {7}
    ] {figures/parametric_shape_inference/cylinder/pinn/loss_history_50000.txt};
\addlegendentry{$S$};
\addplot [line width=1.0pt, TUMgray] table[skip first n=1,
    x index = {11}, y index = {12}
    ] {figures/parametric_shape_inference/cylinder/pinn/loss_history_50000.txt};
\addlegendentry{$T_0$};
\addplot [line width=1.0pt, TUMpink] table[skip first n=1,
x index = {11}, y index = {10}
] {figures/parametric_shape_inference/cylinder/pinn/loss_history_50000.txt};
\addlegendentry{$\mathbf{u}_\perp$};

\nextgroupplot[
    tick align=outside,
    tick pos=left,
    xtick style={color=black},
    ymin=1e-4, ymax=1e4,
    % xmin=0.0, xmax=1e5,
    log basis x={10},
    log basis y={10},
    % xmode=log,
    ymode=log,
    xlabel=step,
    % title=(d),
    yticklabels={},
    % ylabel=$\mathcal{L}$
    ]

\addplot [line width=1.0pt, TUMblue] table[skip first n=1,
x index = {11}, y index = {0}
] {figures/parametric_shape_inference/ellipse/pinn/loss_history_150000.txt};
\addplot [line width=1.0pt, TUMorange] table[skip first n=1,
    x index = {11}, y index = {6}
    ] {figures/parametric_shape_inference/ellipse/pinn/loss_history_150000.txt};
\addplot [line width=1.0pt, TUMgreen] table[skip first n=1,
    x index = {11}, y index = {7}
    ] {figures/parametric_shape_inference/ellipse/pinn/loss_history_150000.txt};
\addplot [line width=1.0pt, TUMgray] table[skip first n=1,
    x index = {11}, y index = {12}
    ] {figures/parametric_shape_inference/ellipse/pinn/loss_history_150000.txt};
\addplot [line width=1.0pt, TUMpink] table[skip first n=1,
x index = {11}, y index = {10}
] {figures/parametric_shape_inference/ellipse/pinn/loss_history_150000.txt};

\nextgroupplot[
tick align=outside,
tick pos=left,
xtick style={color=black},
ymin=1e-4, ymax=1e4,
% xmin=0.0, xmax=1e5,
log basis x={10},
log basis y={10},
% xtick={0,40000,30000},
% xmode=log,
ymode=log,
xlabel=step,
yticklabels={},
% ylabel=$\omega \mathcal{L}$,
% title=(b)
]

\addplot [line width=1.0pt, TUMblue] table[skip first n=1,
x index = {10}, y index = {0}
] {figures/parametric_shape_inference/sphere/pinn/loss_history_30000.txt};
\addplot [line width=1.0pt, TUMorange] table[skip first n=1,
    x index = {10}, y index = {6}
    ] {figures/parametric_shape_inference/sphere/pinn/loss_history_30000.txt};
\addplot [line width=1.0pt, TUMgreen] table[skip first n=1,
    x index = {10}, y index = {7}
    ] {figures/parametric_shape_inference/sphere/pinn/loss_history_30000.txt};
\addplot [line width=1.0pt, TUMgray] table[skip first n=1,
    x index = {10}, y index = {11}
    ] {figures/parametric_shape_inference/sphere/pinn/loss_history_30000.txt};
\addplot [line width=1.0pt, TUMpink] table[skip first n=1,
x index = {10}, y index = {9}
] {figures/parametric_shape_inference/sphere/pinn/loss_history_30000.txt};

\end{groupplot}
\node[above, yshift=0.5cm] at ($(group c1r1.north)$) {Cylinder};
\node[above, yshift=0.5cm] at ($(group c2r1.north)$) {Ellipse};
\node[above, yshift=0.5cm] at ($(group c3r1.north)$) {Sphere};

\node[above, xshift=-2cm, rotate=90] at ($(group c1r1.north west)!0.5!(group c1r1.south west)$) {ODIL};
\node[above, xshift=-2cm, rotate=90] at ($(group c1r2.north west)!0.5!(group c1r2.south west)$) {PINN};

\end{tikzpicture}

%% file: figures/parametric_shape_inference/flowfield_cylinder_variation.tex
\begin{tikzpicture}
    % \pgfmathdeclarefunction{lg10}{1}{%
    %     \pgfmathparse{ln(#1)/ln(10)}%
    % }
    
    % Spectral_r colormap from matplotlib
    \pgfplotsset{
        colormap={spectralr}{rgb255=(94,79,162) rgb255=(68,112,177) rgb255=(59,146,184) rgb255=(89,180,170)
            rgb255=(126,203,164) rgb255=(166,219,164) rgb255=(202,233,157) rgb255=(232,246,156) 
            rgb255=(247,252,179) rgb255=(254,245,175) rgb255=(254,227,145) rgb255=(253,200,119) 
            rgb255=(252,170,95) rgb255=(247,131,77) rgb255=(236,97,69) rgb255=(218,70,76)
            rgb255=(190,36,73) rgb255=(158,1,66)}}
    
    \begin{groupplot}[
        group style={
            group size=6 by 2,
            horizontal sep=0.05cm,
            vertical sep=0.35cm
        }, 
        width=3.9cm, 
        height=3.9cm
        ]

        \nextgroupplot[
            tick align=outside,
            tick pos=left,
            xtick style={draw=none},
            ytick style={draw=none},
            xticklabels = {},
            xtick = {},
            yticklabels = {},
            ytick = {},
            % title=reference,
            % ylabel style={yshift=2pt}
            % ylabel=ODIL,
            enlargelimits=false,
            colormap name=spectralr,
            axis equal image,
        ]
        \addplot graphics [xmin=-0.3,xmax=0.3,ymin=-0.3,ymax=0.3] {figures/parametric_shape_inference/cylinder/odil/density_50000_xy_odil.pdf};

        \nextgroupplot[
            tick align=outside,
            tick pos=left,
            xtick style={draw=none},
            ytick style={draw=none},
            xticklabels = {},
            xtick = {},
            yticklabels = {},
            ytick = {},
            % title=ODIL,
            % title=$u$,
            enlargelimits=false,
            colormap name=spectralr,
            axis equal image,
        ]
        \addplot graphics [xmin=-0.3,xmax=0.3,ymin=-0.3,ymax=0.3] {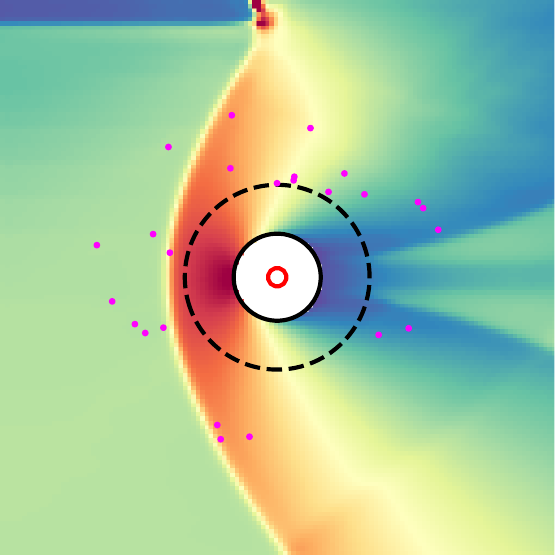};

        \nextgroupplot[
            tick align=outside,
            tick pos=left,
            xtick style={draw=none},
            ytick style={draw=none},
            xticklabels = {},
            xtick = {},
            yticklabels = {},
            ytick = {},
            % title = PINN,
            enlargelimits=false,
            colormap name=spectralr,
            axis equal image
        ]
        \addplot graphics [xmin=-0.3,xmax=0.3,ymin=-0.3,ymax=0.3] {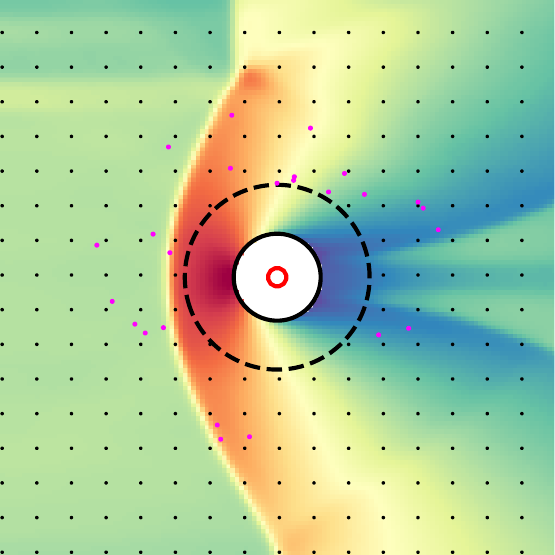};

        \nextgroupplot[
            xshift=0.4cm,
            tick align=outside,
            tick pos=left,
            xtick style={draw=none},
            ytick style={draw=none},
            xticklabels = {},
            xtick = {},
            yticklabels = {},
            ytick = {},
            enlargelimits=false,
            colormap name=spectralr,
            axis equal image,
        ]
        \addplot graphics [xmin=-0.3,xmax=0.3,ymin=-0.3,ymax=0.3] {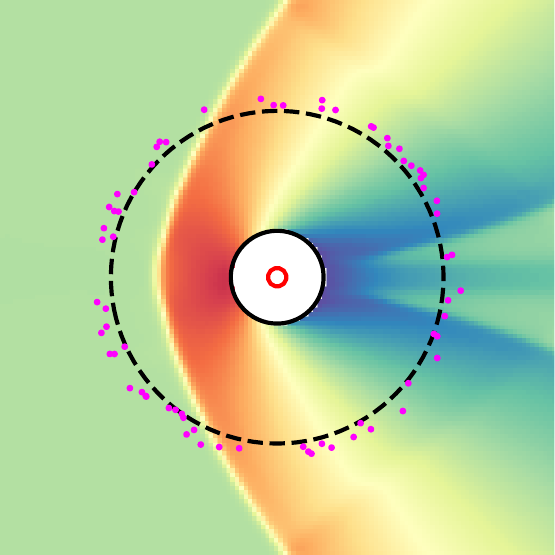};

        \nextgroupplot[
            tick align=outside,
            tick pos=left,
            xtick style={draw=none},
            ytick style={draw=none},
            xticklabels = {},
            xtick = {},
            yticklabels = {},
            ytick = {},
            % title=ODIL,
            enlargelimits=false,
            colormap name=spectralr,
            axis equal image,
        ]
        \addplot graphics [xmin=-0.3,xmax=0.3,ymin=-0.3,ymax=0.3] {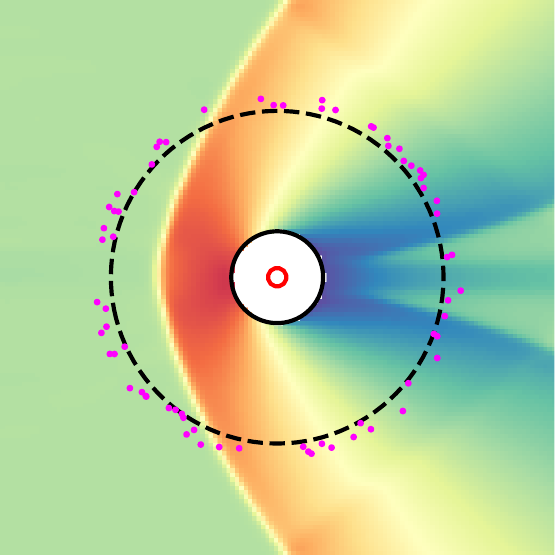};

        \nextgroupplot[
            tick align=outside,
            tick pos=left,
            xtick style={draw=none},
            ytick style={draw=none},
            xticklabels = {},
            xtick = {},
            yticklabels = {},
            ytick = {},
            % title = PINN,
            % title=$\log_{10}\left(\vert \text{rel. error} \vert \right)$,
            enlargelimits=false,
            colormap name=spectralr,
            colorbar,
            colorbar style={
                at={(1.02,0.0)},
                anchor=south west,
                ytick style={
                    draw=none,
                },
                ytick={0.07,1.56,3.05},
                yticklabels = {0.07,1.56,3.05},
                width=0.2cm
            },
            point meta min=0.07,
            point meta max=3.05
        ]
        \addplot graphics [xmin=-0.3,xmax=0.3,ymin=-0.3,ymax=0.3] {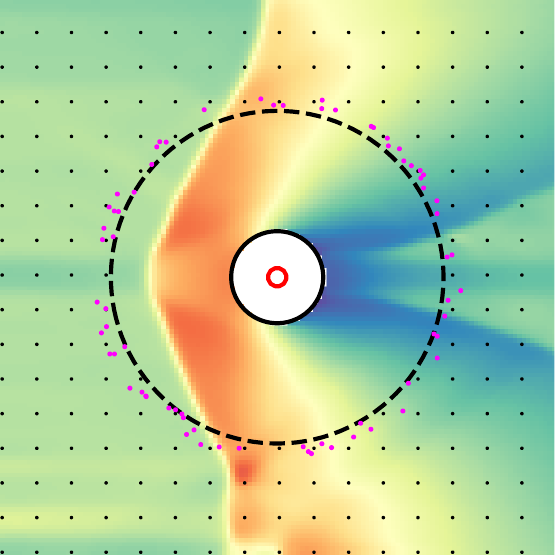};

        \nextgroupplot[
            tick align=outside,
            tick pos=left,
            xtick style={draw=none},
            ytick style={draw=none},
            xticklabels = {},
            xtick = {},
            yticklabels = {},
            ytick = {},
            % title=reference,
            % ylabel style={yshift=2pt}
            % ylabel=PINN,
            enlargelimits=false,
            colormap name=spectralr,
            axis equal image,
        ]
        \addplot graphics [xmin=-0.3,xmax=0.3,ymin=-0.3,ymax=0.3] {figures/parametric_shape_inference/cylinder/pinn/density_50000_xy_pinn.pdf};

        \nextgroupplot[
            tick align=outside,
            tick pos=left,
            xtick style={draw=none},
            ytick style={draw=none},
            xticklabels = {},
            xtick = {},
            yticklabels = {},
            ytick = {},
            % title=ODIL,
            % title=$u$,
            enlargelimits=false,
            colormap name=spectralr,
            axis equal image,
        ]
        \addplot graphics [xmin=-0.3,xmax=0.3,ymin=-0.3,ymax=0.3] {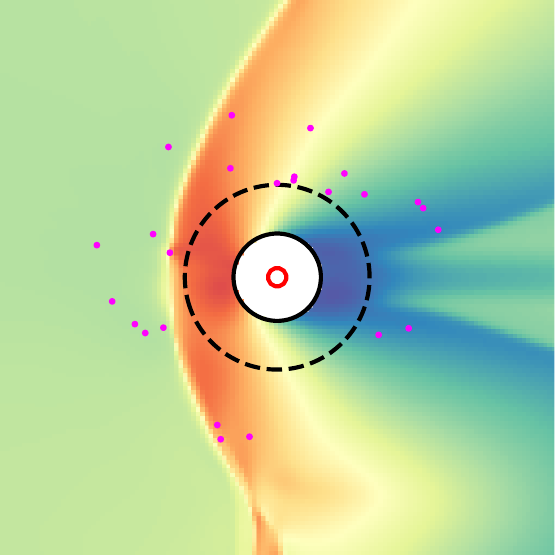};

        \nextgroupplot[
            tick align=outside,
            tick pos=left,
            xtick style={draw=none},
            ytick style={draw=none},
            xticklabels = {},
            xtick = {},
            yticklabels = {},
            ytick = {},
            % title = PINN,
            enlargelimits=false,
            colormap name=spectralr,
            axis equal image
        ]
        \addplot graphics [xmin=-0.3,xmax=0.3,ymin=-0.3,ymax=0.3] {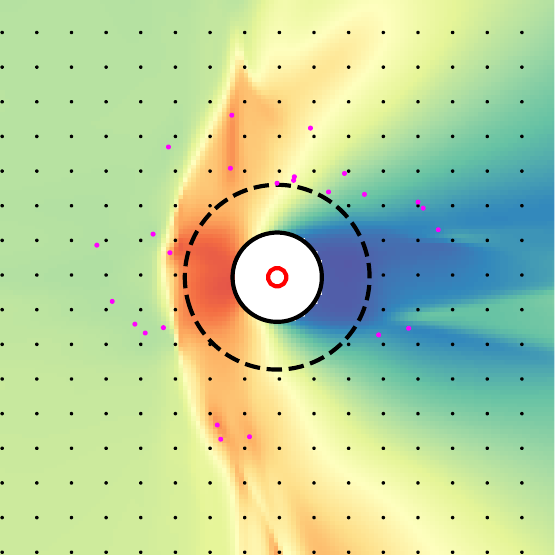};

        \nextgroupplot[
            tick align=outside,
            tick pos=left,
            xtick style={draw=none},
            ytick style={draw=none},
            xticklabels = {},
            xtick = {},
            yticklabels = {},
            ytick = {},
            enlargelimits=false,
            colormap name=spectralr,
            axis equal image,
        ]
        \addplot graphics [xmin=-0.3,xmax=0.3,ymin=-0.3,ymax=0.3] {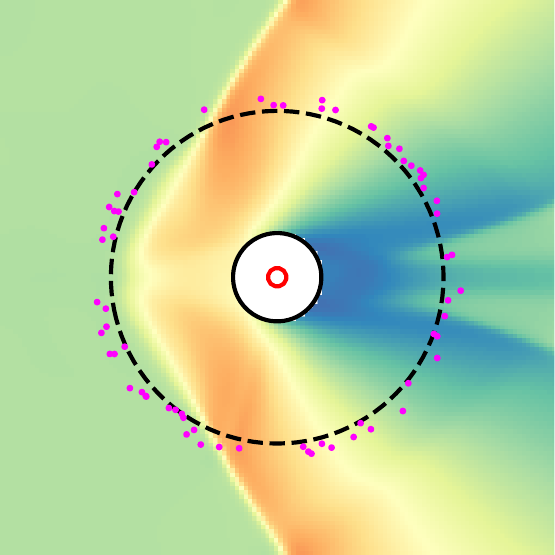};

        \nextgroupplot[
            tick align=outside,
            tick pos=left,
            xtick style={draw=none},
            ytick style={draw=none},
            xticklabels = {},
            xtick = {},
            yticklabels = {},
            ytick = {},
            % title=ODIL,
            enlargelimits=false,
            colormap name=spectralr,
            axis equal image,
        ]
        \addplot graphics [xmin=-0.3,xmax=0.3,ymin=-0.3,ymax=0.3] {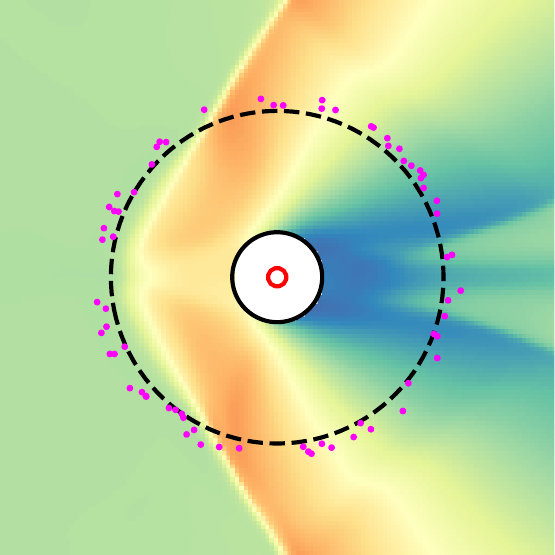};

        \nextgroupplot[
            tick align=outside,
            tick pos=left,
            xtick style={draw=none},
            ytick style={draw=none},
            xticklabels = {},
            xtick = {},
            yticklabels = {},
            ytick = {},
            % title = PINN,
            % title=$\log_{10}\left(\vert \text{rel. error} \vert \right)$,
            enlargelimits=false,
            colormap name=spectralr,
            % colorbar,
            % colorbar style={
            %     at={(1.02,0.0)},
            %     % yshift=0.3cm,
            %     anchor=south west,
            %     ytick style={
            %         draw=none,
            %     },
            %     ytick={0.12,1.47,2.82},
            %     yticklabels = {0.12,1.47,2.82},
            %     width=0.2cm
            % },
            % point meta min=0.12,
            % point meta max=2.82
        ]
        \addplot graphics [xmin=-0.3,xmax=0.3,ymin=-0.3,ymax=0.3] {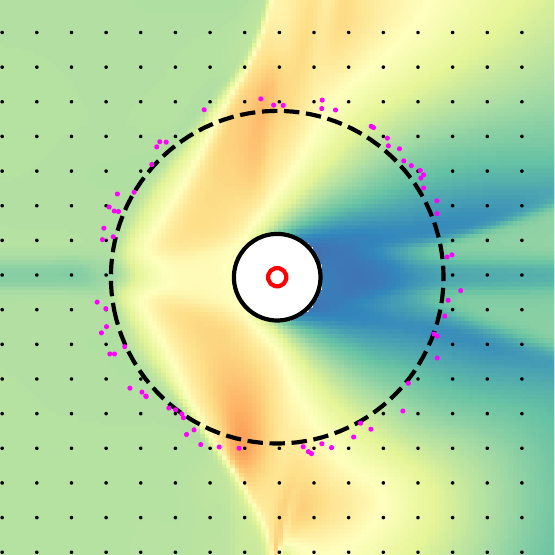};

    \end{groupplot}

    \node[above, yshift=1.5cm] at ($(group c1r1.north east)!0.5!(group c3r1.north west)$) {$\Omega_{W,0}$, $\Omega_{S,0}$};
    \node[above, yshift=1.5cm] at ($(group c4r1.north east)!0.5!(group c6r1.north west)$) {$\Omega_{W,1}$, $\Omega_{S,1}$};
    \draw[->, line width=0.5pt] ($(group c1r1.north)+(0,0.9cm)$) -- ($(group c3r1.north)+(0,0.9cm)$) node[above, midway] {increasing sparsity of schlieren};
    \draw[->, line width=0.5pt] ($(group c4r1.north)+(0,0.9cm)$) -- ($(group c6r1.north)+(0,0.9cm)$) node[above, midway] {increasing sparsity of schlieren};

    \node[above, yshift=0.05cm] at ($(group c1r1.north)$) {$P_{S,00}$};
    \node[above, yshift=0.05cm] at ($(group c2r1.north)$) {$P_{S,01}$};
    \node[above, yshift=0.05cm] at ($(group c3r1.north)$) {$P_{S,02}$};
    \node[above, yshift=0.05cm] at ($(group c4r1.north)$) {$P_{S,10}$};
    \node[above, yshift=0.05cm] at ($(group c5r1.north)$) {$P_{S,11}$};
    \node[above, yshift=0.05cm] at ($(group c6r1.north)$) {$P_{S,12}$};

    \node[above, xshift=-0.1cm, rotate=90] at ($(group c1r1.south west)!0.5!(group c1r1.north west)$) {ODIL};
    \node[above, xshift=-0.1cm, rotate=90] at ($(group c1r2.south west)!0.5!(group c1r2.north west)$) {PINN};

\end{tikzpicture}

%% file: figures/free_shape_inference/flowfield_rho_u.tex
\begin{tikzpicture}
    % \pgfmathdeclarefunction{lg10}{1}{%
    %     \pgfmathparse{ln(#1)/ln(10)}%
    % }
    
    % Spectral_r colormap from matplotlib
    \pgfplotsset{
        colormap={spectralr}{rgb255=(94,79,162) rgb255=(68,112,177) rgb255=(59,146,184) rgb255=(89,180,170)
            rgb255=(126,203,164) rgb255=(166,219,164) rgb255=(202,233,157) rgb255=(232,246,156) 
            rgb255=(247,252,179) rgb255=(254,245,175) rgb255=(254,227,145) rgb255=(253,200,119) 
            rgb255=(252,170,95) rgb255=(247,131,77) rgb255=(236,97,69) rgb255=(218,70,76)
            rgb255=(190,36,73) rgb255=(158,1,66)}}
    
    \begin{groupplot}[
        group style={
            group size=4 by 6,
            horizontal sep=0.2cm,
            vertical sep=0.5cm
        }, 
        width=4.2cm, 
        height=4.2cm
        ]
        \nextgroupplot[
            tick align=outside,
            tick pos=left,
            xtick style={draw=none},
            ytick style={draw=none},
            xticklabels = {},
            xtick = {},
            yticklabels = {},
            ytick = {},
            title=ODIL,
            ylabel=Cylinder,
            enlargelimits=false,
            colormap name=spectralr,
            axis equal image,
        ]
        \addplot graphics [xmin=-0.3,xmax=0.3,ymin=-0.3,ymax=0.3] {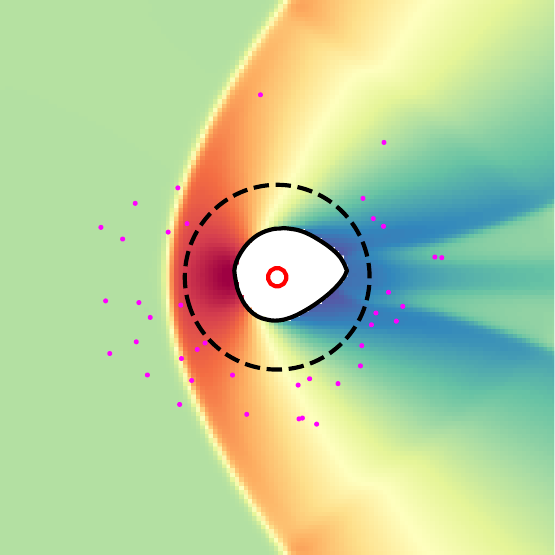};

        \nextgroupplot[
            tick align=outside,
            tick pos=left,
            xtick style={draw=none},
            ytick style={draw=none},
            xticklabels = {},
            xtick = {},
            yticklabels = {},
            ytick = {},
            title=Reference,
            enlargelimits=false,
            colormap name=spectralr,
            axis equal image,
            colorbar,
            colorbar,
            colorbar style={
                at={(1.05,0.0)},
                anchor=south west,
                ytick style={
                    draw=none,
                },
                ytick={0.07,1.56,3.05},
                yticklabels = {0.07,1.56,3.05},
                width=0.2cm
            },
            point meta min=0.07,
            point meta max=3.05
        ]
        \addplot graphics [xmin=-0.3,xmax=0.3,ymin=-0.3,ymax=0.3] {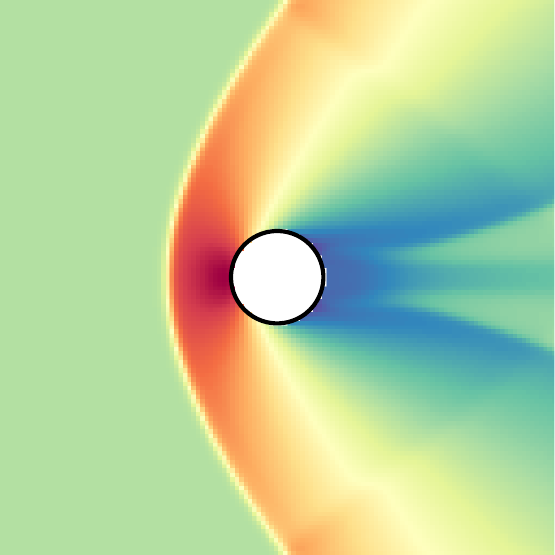};

        \nextgroupplot[
            xshift=1.5cm,
            tick align=outside,
            tick pos=left,
            xtick style={draw=none},
            ytick style={draw=none},
            xticklabels = {},
            xtick = {},
            yticklabels = {},
            ytick = {},
            title=ODIL,
            enlargelimits=false,
            colormap name=spectralr,
        ]
        \addplot graphics [xmin=-0.3,xmax=0.3,ymin=-0.3,ymax=0.3] {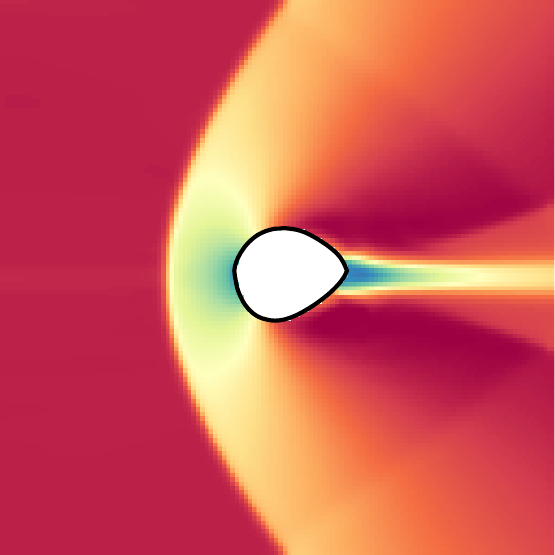};

        \nextgroupplot[
            tick align=outside,
            tick pos=left,
            xtick style={draw=none},
            ytick style={draw=none},
            xticklabels = {},
            xtick = {},
            yticklabels = {},
            ytick = {},
            title=Reference,
            enlargelimits=false,
            colormap name=spectralr,
            axis equal image,
            colorbar,
            colorbar style={
                at={(1.05,0.0)},
                anchor=south west,
                ytick style={
                    draw=none,
                },
                ytick={-0.32,1.1,2.52},
                yticklabels = {-0.32,1.1,2.52},
                width=0.2cm
            },
            point meta min=-0.32,
            point meta max=2.52
        ]
        \addplot graphics [xmin=-0.3,xmax=0.3,ymin=-0.3,ymax=0.3] {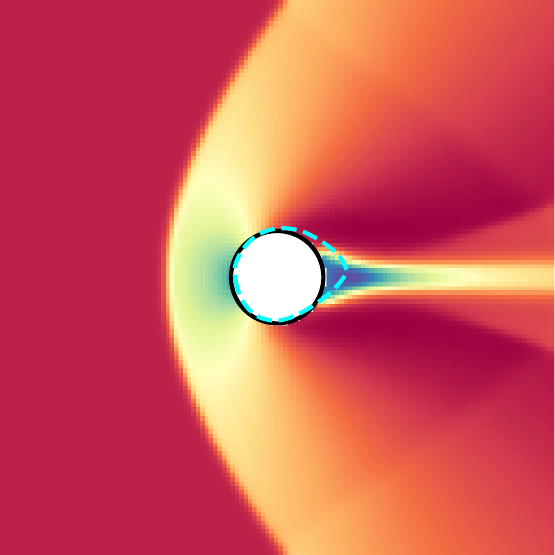};

        \nextgroupplot[
            tick align=outside,
            tick pos=left,
            xtick style={draw=none},
            ytick style={draw=none},
            xticklabels = {},
            xtick = {},
            yticklabels = {},
            ytick = {},
            ylabel=Ellipse,
            enlargelimits=false,
            colormap name=spectralr,
            axis equal image,
        ]
        \addplot graphics [xmin=-0.3,xmax=0.3,ymin=-0.3,ymax=0.3] {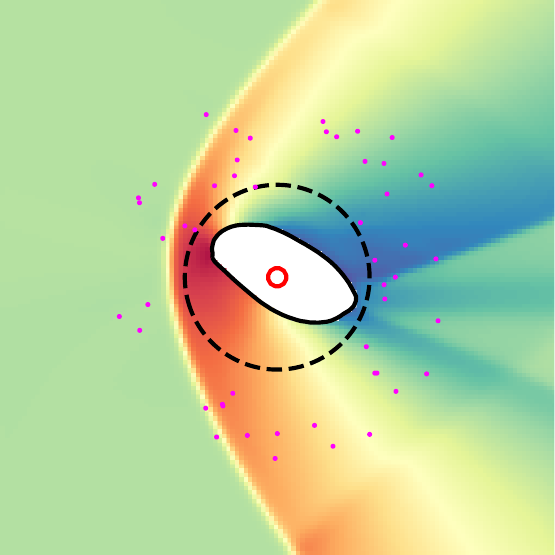};

        \nextgroupplot[
            tick align=outside,
            tick pos=left,
            xtick style={draw=none},
            ytick style={draw=none},
            xticklabels = {},
            xtick = {},
            yticklabels = {},
            ytick = {},
            enlargelimits=false,
            colormap name=spectralr,
            axis equal image,
            colorbar,
            colorbar style={
                at={(1.05,0.0)},
                anchor=south west,
                ytick style={
                    draw=none,
                },
                ytick={0.07,1.55,3.03},
                yticklabels = {0.07,1.55,3.03},
                width=0.2cm
            },
            point meta min=0.07,
            point meta max=3.03
        ]
        \addplot graphics [xmin=-0.3,xmax=0.3,ymin=-0.3,ymax=0.3] {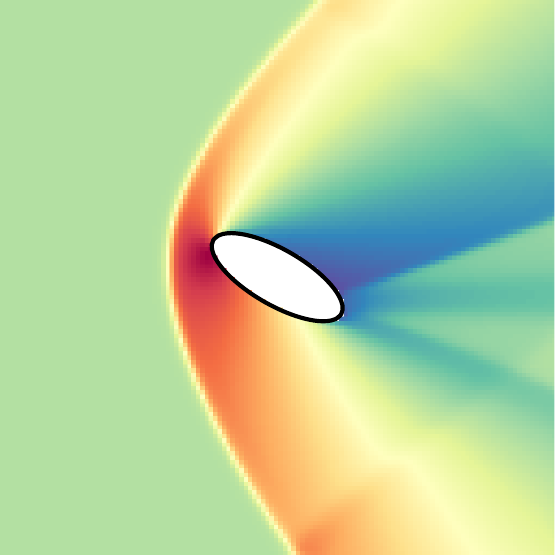};

        \nextgroupplot[
            tick align=outside,
            tick pos=left,
            xtick style={draw=none},
            ytick style={draw=none},
            xticklabels = {},
            xtick = {},
            yticklabels = {},
            ytick = {},
            enlargelimits=false,
            colormap name=spectralr,
        ]
        \addplot graphics [xmin=-0.3,xmax=0.3,ymin=-0.3,ymax=0.3] {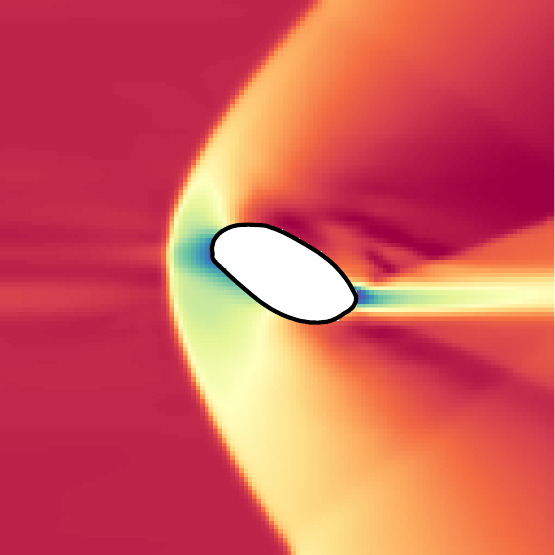};

        \nextgroupplot[
            tick align=outside,
            tick pos=left,
            xtick style={draw=none},
            ytick style={draw=none},
            xticklabels = {},
            xtick = {},
            yticklabels = {},
            ytick = {},
            enlargelimits=false,
            colormap name=spectralr,
            axis equal image,
            colorbar,
            colorbar style={
                at={(1.05,0.0)},
                anchor=south west,
                ytick style={
                    draw=none,
                },
                ytick={0.01,1.27,2.53},
                yticklabels = {0.01,1.27,2.53},
                width=0.2cm
            },
            point meta min=0.01,
            point meta max=2.53
        ]
        \addplot graphics [xmin=-0.3,xmax=0.3,ymin=-0.3,ymax=0.3] {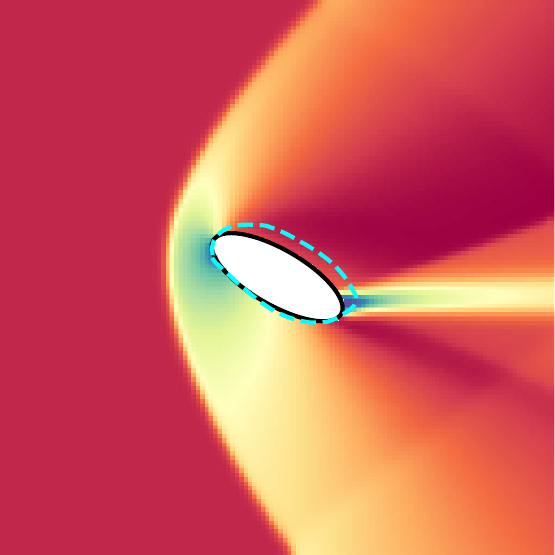};

        \nextgroupplot[
            tick align=outside,
            tick pos=left,
            xtick style={draw=none},
            ytick style={draw=none},
            xticklabels = {},
            xtick = {},
            yticklabels = {},
            ytick = {},
            ylabel=Sphere,
            enlargelimits=false,
            colormap name=spectralr,
            axis equal image,
        ]
        \addplot graphics [xmin=-0.3,xmax=0.3,ymin=-0.3,ymax=0.3] {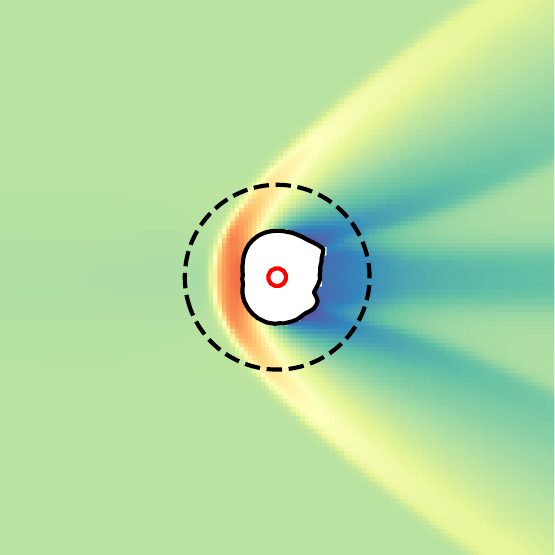};

        \nextgroupplot[
            tick align=outside,
            tick pos=left,
            xtick style={draw=none},
            ytick style={draw=none},
            xticklabels = {},
            xtick = {},
            yticklabels = {},
            ytick = {},
            enlargelimits=false,
            colormap name=spectralr,
            axis equal image,
            colorbar,
            colorbar style={
                at={(1.05,0.0)},
                anchor=south west,
                ytick style={
                    draw=none,
                },
                ytick={0.12,1.47,2.82},
                yticklabels = {0.12,1.47,2.82},
                width=0.2cm
            },
            point meta min=0.12,
            point meta max=2.82
        ]
        \addplot graphics [xmin=-0.3,xmax=0.3,ymin=-0.3,ymax=0.3] {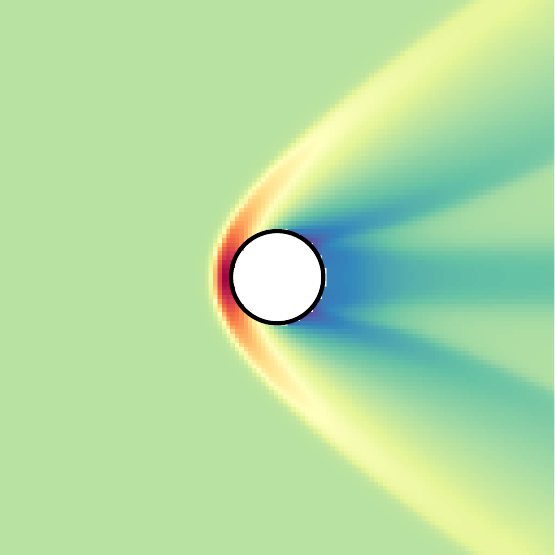};

        \nextgroupplot[
            tick align=outside,
            tick pos=left,
            xtick style={draw=none},
            ytick style={draw=none},
            xticklabels = {},
            xtick = {},
            yticklabels = {},
            ytick = {},
            enlargelimits=false,
            colormap name=spectralr,
        ]
        \addplot graphics [xmin=-0.3,xmax=0.3,ymin=-0.3,ymax=0.3] {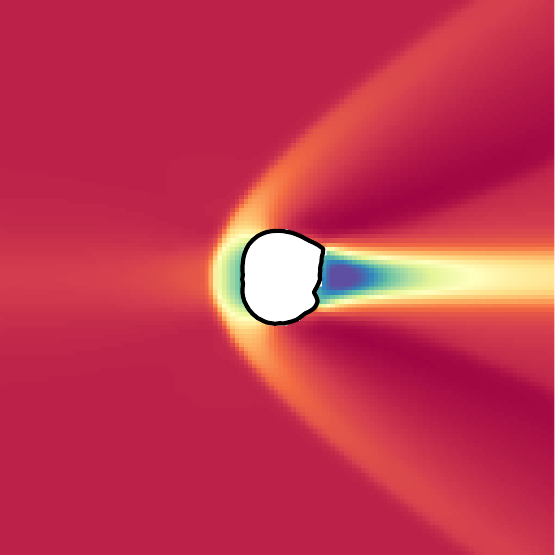};

        \nextgroupplot[
            tick align=outside,
            tick pos=left,
            xtick style={draw=none},
            ytick style={draw=none},
            xticklabels = {},
            xtick = {},
            yticklabels = {},
            ytick = {},
            enlargelimits=false,
            colormap name=spectralr,
            axis equal image,
            colorbar,
            colorbar style={
                at={(1.05,0.0)},
                anchor=south west,
                ytick style={
                    draw=none,
                },
                ytick={-0.52,1.01,2.54},
                yticklabels = {-0.52,1.01,2.54},
                width=0.2cm
            },
            point meta min=-0.52,
            point meta max=2.54
        ]
        \addplot graphics [xmin=-0.3,xmax=0.3,ymin=-0.3,ymax=0.3] {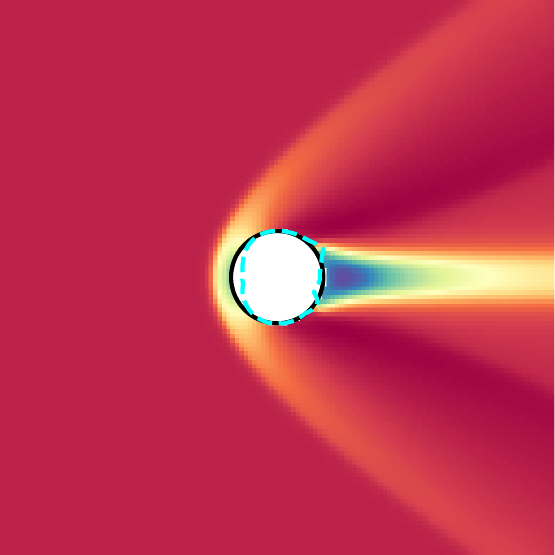};

    \end{groupplot}

    % \node[above, yshift=1cm] at ($(group c1r1.north east)!0.5!(group c2r1.north west)$) {$\rho$};
    % \node[above, yshift=1cm] at ($(group c3r1.north east)!0.5!(group c4r1.north west)$) {$u$};
\end{tikzpicture}

%% file: figures/free_shape_inference/measurement_noise/flowfield.tex
\begin{tikzpicture}
    % \pgfmathdeclarefunction{lg10}{1}{%
    %     \pgfmathparse{ln(#1)/ln(10)}%
    % }
    
    % Spectral_r colormap from matplotlib
    \pgfplotsset{
        colormap={spectralr}{rgb255=(94,79,162) rgb255=(68,112,177) rgb255=(59,146,184) rgb255=(89,180,170)
            rgb255=(126,203,164) rgb255=(166,219,164) rgb255=(202,233,157) rgb255=(232,246,156) 
            rgb255=(247,252,179) rgb255=(254,245,175) rgb255=(254,227,145) rgb255=(253,200,119) 
            rgb255=(252,170,95) rgb255=(247,131,77) rgb255=(236,97,69) rgb255=(218,70,76)
            rgb255=(190,36,73) rgb255=(158,1,66)}}
    
    \begin{groupplot}[
        group style={
            group size=4 by 3,
            horizontal sep=0.2cm,
            vertical sep=0.2cm
        }, 
        width=4.2cm, 
        height=4.21cm
        ]

        \nextgroupplot[
            tick align=outside,
            tick pos=left,
            xtick style={draw=none},
            ytick style={draw=none},
            xticklabels = {},
            xtick = {},
            yticklabels = {},
            ytick = {},
            title={$\epsilon=0.0$},
            ylabel=$P_{W,0}$,
            enlargelimits=false,
            colormap name=spectralr,
            axis equal image,
        ]
        \addplot graphics [xmin=-0.3,xmax=0.3,ymin=-0.3,ymax=0.3] {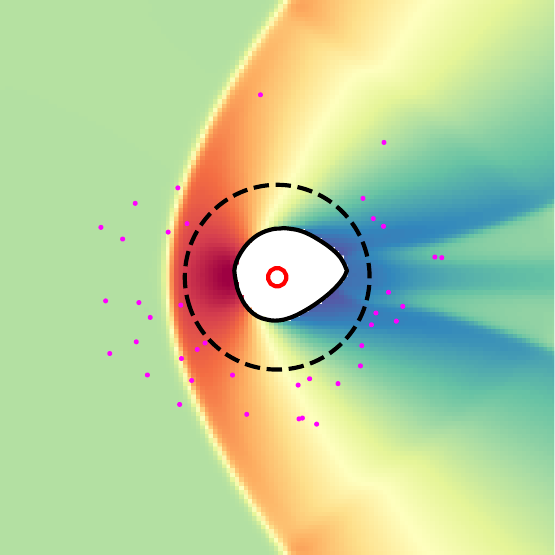};

        \nextgroupplot[
            tick align=outside,
            tick pos=left,
            xtick style={draw=none},
            ytick style={draw=none},
            xticklabels = {},
            xtick = {},
            yticklabels = {},
            ytick = {},
            title={$\epsilon=0.1$},
            enlargelimits=false,
            colormap name=spectralr,
            axis equal image
        ]
        \addplot graphics [xmin=-0.3,xmax=0.3,ymin=-0.3,ymax=0.3] {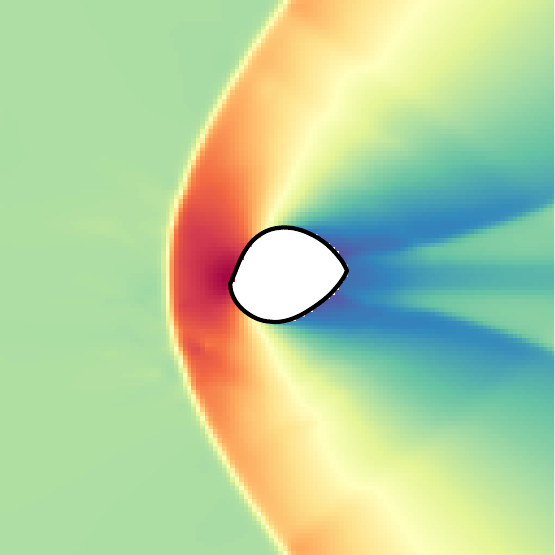};

        \nextgroupplot[
            tick align=outside,
            tick pos=left,
            xtick style={draw=none},
            ytick style={draw=none},
            xticklabels = {},
            xtick = {},
            yticklabels = {},
            ytick = {},
            title={$\epsilon=0.2$},
            enlargelimits=false,
            colormap name=spectralr,
        ]
        \addplot graphics [xmin=-0.3,xmax=0.3,ymin=-0.3,ymax=0.3] {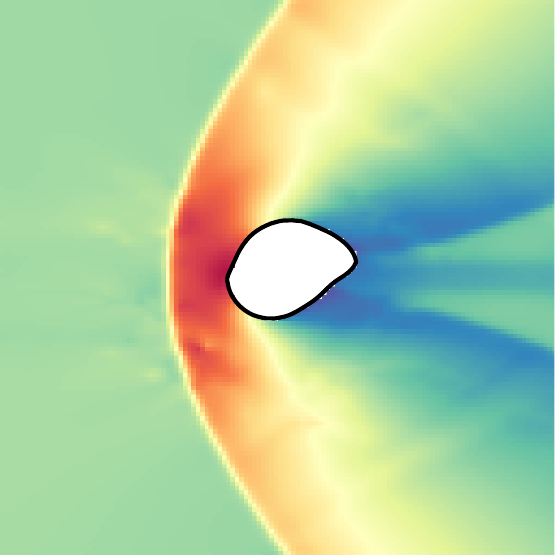};

        \nextgroupplot[
            tick align=outside,
            tick pos=left,
            xtick style={draw=none},
            ytick style={draw=none},
            xticklabels = {},
            xtick = {},
            yticklabels = {},
            ytick = {},
            title={$\epsilon=0.3$},
            enlargelimits=false,
            colormap name=spectralr,
            axis equal image,
            colorbar,
            colorbar style={
                at={(1.05,0.0)},
                anchor=south west,
                ytick style={
                    draw=none,
                },
                ytick={0.07,1.56,3.05},
                yticklabels = {0.07,1.56,3.05},
                width=0.2cm
            },
            point meta min=0.07,
            point meta max=3.05
        ]
        \addplot graphics [xmin=-0.3,xmax=0.3,ymin=-0.3,ymax=0.3] {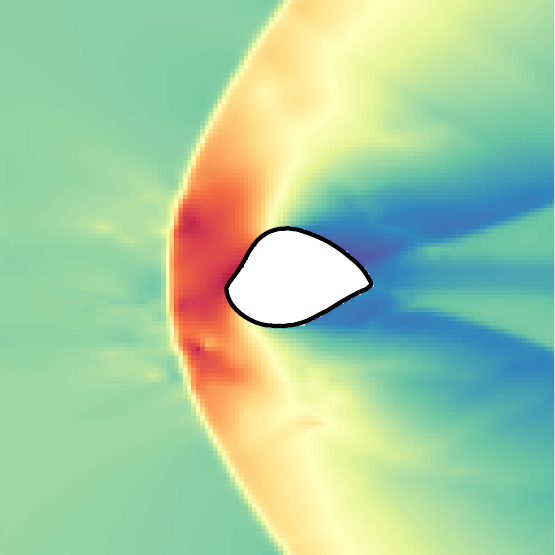};

        \nextgroupplot[
            tick align=outside,
            tick pos=left,
            xtick style={draw=none},
            ytick style={draw=none},
            xticklabels = {},
            xtick = {},
            yticklabels = {},
            ytick = {},
            % title=$0\%$,
            ylabel=$P_{W,1}$,
            enlargelimits=false,
            colormap name=spectralr,
            axis equal image,
        ]
        \addplot graphics [xmin=-0.3,xmax=0.3,ymin=-0.3,ymax=0.3] {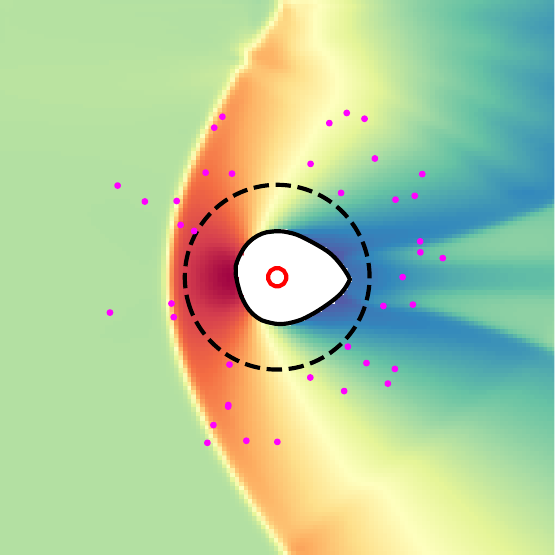};

        \nextgroupplot[
            tick align=outside,
            tick pos=left,
            xtick style={draw=none},
            ytick style={draw=none},
            xticklabels = {},
            xtick = {},
            yticklabels = {},
            ytick = {},
            % title=$10\%$,
            enlargelimits=false,
            colormap name=spectralr,
            axis equal image
        ]
        \addplot graphics [xmin=-0.3,xmax=0.3,ymin=-0.3,ymax=0.3] {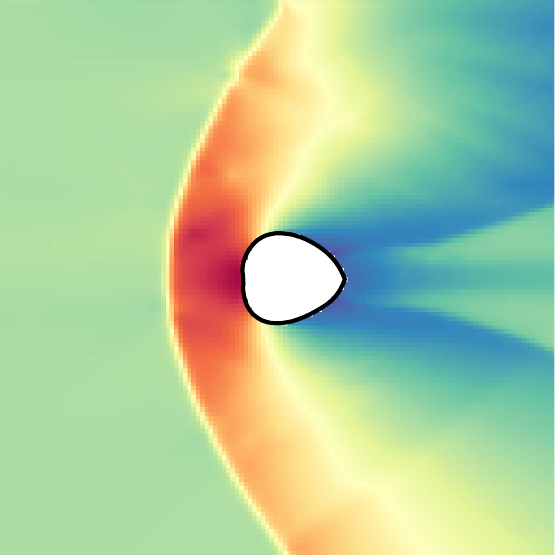};

        \nextgroupplot[
            tick align=outside,
            tick pos=left,
            xtick style={draw=none},
            ytick style={draw=none},
            xticklabels = {},
            xtick = {},
            yticklabels = {},
            ytick = {},
            % title=$20\%$,
            enlargelimits=false,
            colormap name=spectralr,
        ]
        \addplot graphics [xmin=-0.3,xmax=0.3,ymin=-0.3,ymax=0.3] {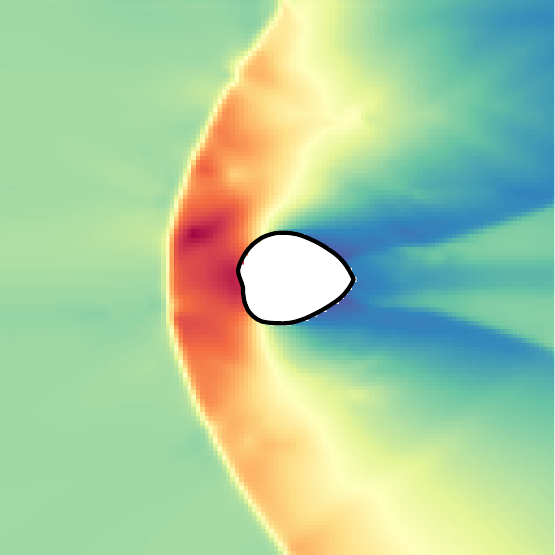};

        \nextgroupplot[
            tick align=outside,
            tick pos=left,
            xtick style={draw=none},
            ytick style={draw=none},
            xticklabels = {},
            xtick = {},
            yticklabels = {},
            ytick = {},
            % title=$30\%$,
            enlargelimits=false,
            colormap name=spectralr,
            axis equal image,
            % colorbar,
            % colorbar style={
            %     at={(1.05,0.0)},
            %     anchor=south west,
            %     ytick style={
            %         draw=none,
            %     },
            %     ytick={0.07,1.56,3.05},
            %     yticklabels = {0.07,1.56,3.05},
            %     width=0.2cm
            % },
            % point meta min=0.07,
            % point meta max=3.05
        ]
        \addplot graphics [xmin=-0.3,xmax=0.3,ymin=-0.3,ymax=0.3] {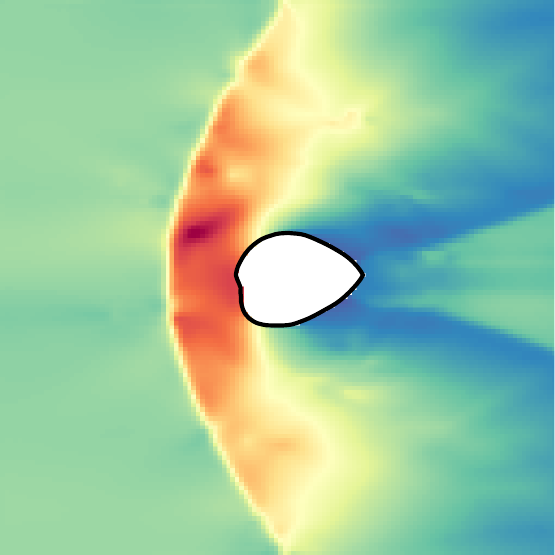};

        \nextgroupplot[
            tick align=outside,
            tick pos=left,
            xtick style={draw=none},
            ytick style={draw=none},
            xticklabels = {},
            xtick = {},
            yticklabels = {},
            ytick = {},
            % title=$0\%$,
            ylabel=$P_{W,2}$,
            enlargelimits=false,
            colormap name=spectralr,
            axis equal image,
        ]
        \addplot graphics [xmin=-0.3,xmax=0.3,ymin=-0.3,ymax=0.3] {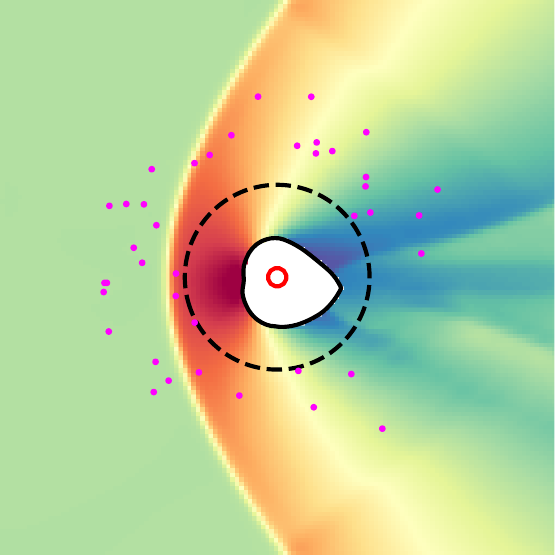};

        \nextgroupplot[
            tick align=outside,
            tick pos=left,
            xtick style={draw=none},
            ytick style={draw=none},
            xticklabels = {},
            xtick = {},
            yticklabels = {},
            ytick = {},
            % title=$10\%$,
            enlargelimits=false,
            colormap name=spectralr,
            axis equal image
        ]
        \addplot graphics [xmin=-0.3,xmax=0.3,ymin=-0.3,ymax=0.3] {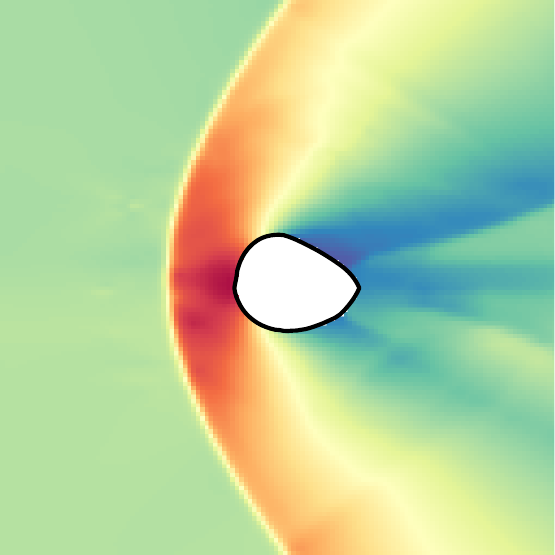};

        \nextgroupplot[
            tick align=outside,
            tick pos=left,
            xtick style={draw=none},
            ytick style={draw=none},
            xticklabels = {},
            xtick = {},
            yticklabels = {},
            ytick = {},
            % title=$20\%$,
            enlargelimits=false,
            colormap name=spectralr,
        ]
        \addplot graphics [xmin=-0.3,xmax=0.3,ymin=-0.3,ymax=0.3] {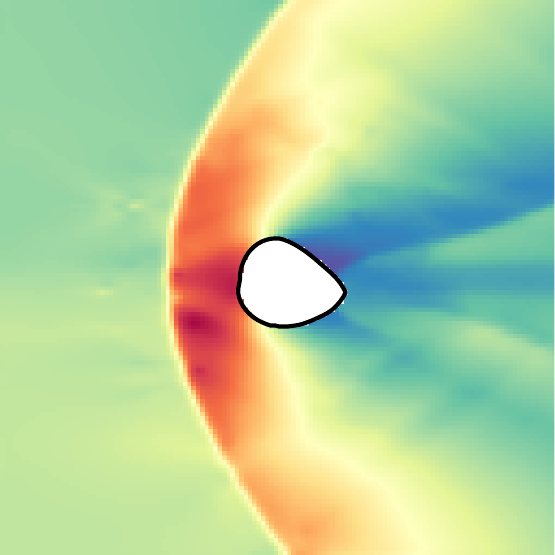};

        \nextgroupplot[
            tick align=outside,
            tick pos=left,
            xtick style={draw=none},
            ytick style={draw=none},
            xticklabels = {},
            xtick = {},
            yticklabels = {},
            ytick = {},
            % title=$30\%$,
            enlargelimits=false,
            colormap name=spectralr,
            axis equal image,
            % colorbar,
            % colorbar style={
            %     at={(1.05,0.0)},
            %     anchor=south west,
            %     ytick style={
            %         draw=none,
            %     },
            %     ytick={0.07,1.56,3.05},
            %     yticklabels = {0.07,1.56,3.05},
            %     width=0.2cm
            % },
            % point meta min=0.07,
            % point meta max=3.05
        ]
        \addplot graphics [xmin=-0.3,xmax=0.3,ymin=-0.3,ymax=0.3] {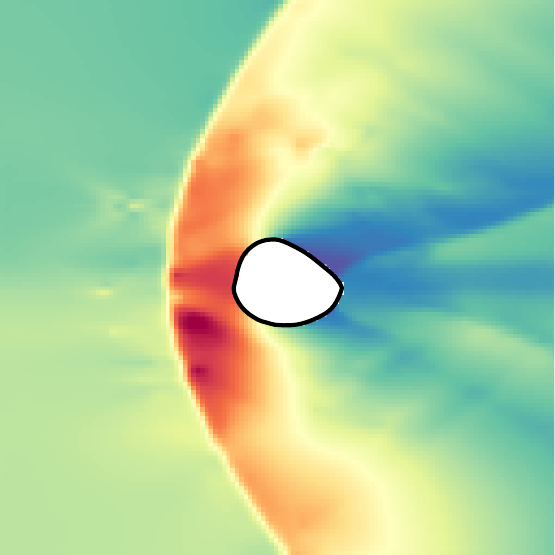};

    \end{groupplot}

    \draw[->, line width=0.5pt] ($(group c1r1.north east)+(0,0.9cm)$) -- ($(group c4r1.north west)+(0,0.9cm)$) node[above, midway] {increasing noise};
    
\end{tikzpicture}

%% file: figures/free_shape_inference/loss_history.tex
\begin{tikzpicture}

    \begin{groupplot}[
        group style={group size=3 by 1, horizontal sep=1.5cm},
        width=4.2cm, height=4.2cm]
      
    \nextgroupplot[
        tick align=outside,
        tick pos=left,
        legend style={at={(1.2,1.1)},draw=black,anchor=north east,font=\footnotesize,legend columns=2},
        xtick style={color=black},
        ymin=1e-4, ymax=1e4,
        xticklabel style={/pgf/number format/fixed},
        log basis y={10},
        ymode=log,
        xlabel=step,
        ylabel=$\omega \mathcal{L}$,
        title=Cylinder
        ]

    \addplot [line width=1.0pt, TUMblue] table[skip first n=1,
        x index = {0}, y index = {2}
        ] {figures/free_shape_inference/cylinder/odil/loss_history_50000.txt};
    \addlegendentry{$E$};
    \addplot [line width=1.0pt, TUMorange] table[skip first n=1,
        x index = {0}, y index = {3}
        ] {figures/free_shape_inference/cylinder/odil/loss_history_50000.txt};
    \addlegendentry{$W$};
    \addplot [line width=1.0pt, TUMgreen] table[skip first n=1,
        x index = {0}, y index = {4}
        ] {figures/free_shape_inference/cylinder/odil/loss_history_50000.txt};
    \addlegendentry{$S$};
    \addplot [line width=1.0pt, TUMgray] table[skip first n=1,
        x index = {0}, y index = {6}
        ] {figures/free_shape_inference/cylinder/odil/loss_history_50000.txt};
    \addlegendentry{$T_0$};
    \addplot [line width=1.0pt, TUMpink] table[skip first n=1,
        x index = {0}, y index = {5}
        ] {figures/free_shape_inference/cylinder/odil/loss_history_50000.txt};
    \addlegendentry{$\phi$};
    \addplot [line width=1.0pt, black, dashed] coordinates {(15000,1e-7) (15000,1e4)};
    % \node[left, anchor=north east, black] at (axis cs: 40000,1e4) {LO};
    % \node[right, anchor=north west, black] at (axis cs: 40000,1e4) {HO};

    \nextgroupplot[
        xticklabel style={/pgf/number format/fixed},
        tick align=outside,
        tick pos=left,
        xtick style={color=black},
        ymin=1e-4, ymax=1e4,
        log basis y={10},
        % xtick={0,40000,80000},
        % xmode=log,
        title=Ellipse,
        ymode=log,
        xlabel=step,
        % title=(b)
        ]

    \addplot [line width=1.0pt, TUMblue] table[skip first n=1,
        x index = {0}, y index = {2}
        ] {figures/free_shape_inference/ellipse/odil/loss_history_50000.txt};
    \addplot [line width=1.0pt, TUMorange] table[skip first n=1,
        x index = {0}, y index = {3}
        ] {figures/free_shape_inference/ellipse/odil/loss_history_50000.txt};
    \addplot [line width=1.0pt, TUMgreen] table[skip first n=1,
        x index = {0}, y index = {4}
        ] {figures/free_shape_inference/ellipse/odil/loss_history_50000.txt};
    \addplot [line width=1.0pt, TUMgray] table[skip first n=1,
        x index = {0}, y index = {6}
        ] {figures/free_shape_inference/ellipse/odil/loss_history_50000.txt};
    \addplot [line width=1.0pt, TUMpink] table[skip first n=1,
        x index = {0}, y index = {5}
        ] {figures/free_shape_inference/ellipse/odil/loss_history_50000.txt};
    \addplot [line width=1.0pt, black, dashed] coordinates {(15000,1e-7) (15000,1e4)};
    % \node[left, anchor=north east, black] at (axis cs: 40000,1e4) {LO};
    % \node[right, anchor=north west, black] at (axis cs: 40000,1e4) {HO};

    \nextgroupplot[
    xticklabel style={/pgf/number format/fixed},
    tick align=outside,
    tick pos=left,
    xtick style={color=black},
    ymin=1e-5, ymax=1e2,
    log basis y={10},
    % xtick={0,40000,80000},
    % xmode=log,
    title=Sphere,
    ymode=log,
    xlabel=step,
    % title=(c)
    ]

    \addplot [line width=1.0pt, TUMblue] table[skip first n=1,
        x index = {0}, y index = {2}
        ] {figures/free_shape_inference/sphere/odil/loss_history_30000.txt};
    \addplot [line width=1.0pt, TUMorange] table[skip first n=1,
        x index = {0}, y index = {3}
        ] {figures/free_shape_inference/sphere/odil/loss_history_30000.txt};
    \addplot [line width=1.0pt, TUMgreen] table[skip first n=1,
        x index = {0}, y index = {4}
        ] {figures/free_shape_inference/sphere/odil/loss_history_30000.txt};
    \addplot [line width=1.0pt, TUMgray] table[skip first n=1,
        x index = {0}, y index = {6}
        ] {figures/free_shape_inference/sphere/odil/loss_history_30000.txt};
    \addplot [line width=1.0pt, TUMpink] table[skip first n=1,
        x index = {0}, y index = {5}
        ] {figures/free_shape_inference/sphere/odil/loss_history_30000.txt};
    \addplot [line width=1.0pt, black, dashed] coordinates {(10000,1e-7) (10000,1e4)};
    % \node[left, anchor=north east, black] at (axis cs: 30000,1e3) {LO};
    % \node[right, anchor=north west, black] at (axis cs: 30000,1e3) {HO};

    \end{groupplot}

\end{tikzpicture}

%% file: figures/free_shape_inference/error_table.tex
\begingroup
\pgfkeys{/pgf/number format/.cd, sci, precision=2, zerofill}

% cylinder 
\pgfplotstableread{./figures/free_shape_inference/cylinder/odil/error_history_50000.txt}\datatable
\pgfplotstablegetrowsof{\datatable}
\pgfmathtruncatemacro{\lastrow}{\pgfplotsretval-1}
\pgfplotstablegetelem{\lastrow}{[index]0}\of\datatable
\pgfmathsetmacro{\rhocylinder}{\pgfplotsretval}
\pgfplotstablegetelem{\lastrow}{[index]1}\of\datatable
\pgfmathsetmacro{\pcylinder}{\pgfplotsretval}
\pgfplotstablegetelem{\lastrow}{[index]3}\of\datatable
\pgfmathsetmacro{\ucylinder}{\pgfplotsretval}
\pgfplotstablegetelem{\lastrow}{[index]4}\of\datatable
\pgfmathsetmacro{\vcylinder}{\pgfplotsretval}
% ellipse 
\pgfplotstableread{./figures/free_shape_inference/ellipse/odil/error_history_50000.txt}\datatable
\pgfplotstablegetrowsof{\datatable}
\pgfmathtruncatemacro{\lastrow}{\pgfplotsretval-1}
\pgfplotstablegetelem{\lastrow}{[index]0}\of\datatable
\pgfmathsetmacro{\rhoellipse}{\pgfplotsretval}
\pgfplotstablegetelem{\lastrow}{[index]1}\of\datatable
\pgfmathsetmacro{\pellipse}{\pgfplotsretval}
\pgfplotstablegetelem{\lastrow}{[index]3}\of\datatable
\pgfmathsetmacro{\uellipse}{\pgfplotsretval}
\pgfplotstablegetelem{\lastrow}{[index]4}\of\datatable
\pgfmathsetmacro{\vellipse}{\pgfplotsretval}
% sphere 
\pgfplotstableread{./figures/free_shape_inference/sphere/odil/error_history_30000.txt}\datatable
\pgfplotstablegetrowsof{\datatable}
\pgfmathtruncatemacro{\lastrow}{\pgfplotsretval-1}
\pgfplotstablegetelem{\lastrow}{[index]0}\of\datatable
\pgfmathsetmacro{\rhosphere}{\pgfplotsretval}
\pgfplotstablegetelem{\lastrow}{[index]1}\of\datatable
\pgfmathsetmacro{\psphere}{\pgfplotsretval}
\pgfplotstablegetelem{\lastrow}{[index]3}\of\datatable
\pgfmathsetmacro{\usphere}{\pgfplotsretval}
\pgfplotstablegetelem{\lastrow}{[index]4}\of\datatable
\pgfmathsetmacro{\vsphere}{\pgfplotsretval}
\pgfplotstablegetelem{\lastrow}{[index]5}\of\datatable
\pgfmathsetmacro{\wsphere}{\pgfplotsretval}

\begin{table}[!t]
  \centering
  \begin{tabular}{r c c c c c}
    \hline
    Shape&$\rho$ & $u$ & $v$ & $w$ &$p$ \\
    \hline
    Cylinder&\print{\rhocylinder}&\print{\ucylinder}&\print{\vcylinder}&-&\print{\pcylinder} \\
    Ellipse&\print{\rhoellipse}&\print{\uellipse}&\print{\vellipse}&-&\print{\pellipse} \\
    Sphere&\print{\rhosphere}&\print{\usphere}&\print{\vsphere}&\print{\wsphere}&\print{\psphere} \\
    \hline
  \end{tabular}
  \caption{
    Mean cell-wise relative error between ODIL and reference for
    free shape parameterization.
    The corresponding flow fields are shown in Figs \ref{fig:flowfield_free} and \ref{fig:sphere_free_3D}.
    }
  \label{tab:errors_free}
\end{table}
\endgroup

%% file: sections/conclusion.tex
\section{Conclusion}
\label{sec:conclusion}

We have extended the Optimizing a Discrete Loss
(ODIL) method ~\cite{Karnakov2023b, Karnakov2023a} to infer flow fields and
obstacle shapes in systems governed by the steady-state compressible Euler equations.
ODIL is designed to minimize a loss function incorporating the discrete approximation of
the underlying partial differential equations (PDE) using gradient-descent-based optimization.
We use JAX-Fluids ~\cite{Bezgin2022, Bezgin2024} to discretize the PDE residual. JAX-Fluids is a differentiable CFD solver for compressible two-phase flows that implements shock-capturing discretizations and a level-set-based sharp-interface immersed boundary method.
The gradients of the PDE residual that are required by the optimization are computed via automatic differentiation (AD) through JAX-Fluids. 
By combining sparse, scattered measurements of primitive
variables and numerical schlieren data with the discrete
PDE residual, we address complex inverse supersonic
flow problems, such as joint inference of flow fields and obstacle shapes.

ODIL inherently incorporates the numerical schemes used to discretize the PDE residual in the optimization process.
In particular, for the inference of flow fields featuring shock discontinuities, we highlight three key
factors that make the present approach powerful: (1) The conservative finite-volume
Godunov-type flux guides the optimization toward physical
weak solutions. (2) Employing high-order shock-capturing
spatial reconstructions ensures accuracy and stability around shocks.
(3) For the joint inference of flow fields and obstacle shapes,
the level-set-based sharp-interface representation of the solid body is
crucial for accurate inference of the fluid-solid interface. 

We explored two optimization scenarios:
(1) \textit{Parametric shape representation}, where a
small set of parameters defining the obstacle shape is learned alongside the flow field.
(2) \textit{Free shape representation}, involving direct
optimization of the level-set field in each cell of the underlying mesh.
For the former, we compared ODIL with Physics-informed Neural Networks (PINN).
Both approaches accurately reconstruct obstacle shapes, but ODIL performs better in inferring sharp shock fronts, particularly in regions where no flow measurements are present.
In these areas, ODIL provides more precise shock inference.
In the more complex case of free shape representation,
we showed that accurate shape inference is still possible, albeit with a
slightly increased number of measurement points for the primitive variables.

Our results indicate that the combination of ODIL and JAX-Fluids is a potent method for studying inverse problems in supersonic flows.
JAX-Fluids is capable of representing material interfaces
via a sharp-interface level-set model
or a five-equation diffuse-interface model.
We envision future directions in which this framework will extend to two-phase
and hypersonic flows, incorporate heterogeneous experimental data, and deploy in massively parallel computing architectures.

%% file: sections/acknowledgements.tex
\section*{Acknowledgments}
NAA acknowledges support by the European Research Council (ERC) Advanced Investigator Grant GENUFASD (project number: 101094463).

%% file: sections/appendix.tex
\appendix

\section{Optimization Setups}

Tables \ref{tab:learning_rate_schedule} and \ref{tab:loss_weights}
depict the optimization setups, including learning rate schedule and the loss weights, respectively.
We use the Adam \cite{Kingma2015} optimizer for all cases.
For the learning rate schedule, we either use a constant value 
or a schedule that consists of a constant value followed
by an exponential decay.
For the latter, the learning rate $\eta$ is a function of the optimization step $s$.
\begin{equation}
\eta(s) \;=\;
\begin{cases}
\eta_0, & 0 \le s < s_{0}\\
\text{max}(\eta_0 \cdot a_\text{decay}^{\displaystyle(s - s_0)/s_\text{decay}}, \eta_\text{end}), & s \ge s_{0}
\end{cases}
\label{eq:lr_schedule_threephase}
\end{equation}
Here, $\eta_0$ is the initial learning rate, $a_\text{decay}$ is the decay rate, $s_0$ is the step indicating the start of the exponential decay, and $s_\text{decay}$ is the amount of decay transition steps.
The learning rate is lower-bounded by $\eta_\text{end}$, which serves as a clipping threshold during the decay phase.

\renewcommand{\arraystretch}{1.2} % Increase row height locally

\begin{table}[!t]
    \centering
    \small
    \begin{tabular}{c c c c c c c c c c}
        \hline
        Shape representation & Method & Shape & Steps & \multicolumn{5}{c}{Learning rate schedule} & $s_{FO\rightarrow SO}$ \\
        &&& & $\eta_0$ & $a_\text{decay}$ & $s_0$ & $s_\text{decay}$ & $\eta_\text{end}$ & \\
        \hline\hline
        \multirow{3}{*}{Parametric}
          & \multirow{3}{*}{ODIL}
            & Cylinder & $5\cdot10^{4}$ & $5\cdot10^{-4}$ & $10^{-2}$  & $4\cdot10^{4}$  & $10^{4}$      & $5\cdot10^{-7}$ & $2\cdot10^4$ \\
          &                    & Ellipse  & $15\cdot10^{4}$& $5\cdot10^{-4}$ & $10^{-2}$  & $13\cdot10^{4}$ & $10^{4}$      & $5\cdot10^{-7}$ & $2\cdot10^4$    \\
          &                    & Sphere   & $3\cdot10^{4}$ & $10^{-3}$       & $10^{-2}$  & $2\cdot10^{4}$  & $10^{4}$      & $10^{-5}$ & $5\cdot10^3$    \\
        \hline
        \multirow{3}{*}{Parametric}
          & \multirow{3}{*}{PINN}
            & Cylinder & $5\cdot10^{4}$ & $10^{-4}$       & $-$        & $-$            & $-$           & $-$ & $-$               \\
          &                    & Ellipse  & $15\cdot10^{4}$& $2\cdot10^{-4}$ & $10^{-2}$  & $5\cdot10^{4}$  & $10^{4}$      & $2\cdot 10^{-5}$ & $-$              \\
          &                    & Sphere   & $3\cdot10^{4}$ & $5\cdot10^{-4}$ & $10^{-2}$  & $2\cdot10^{4}$  & $10^{4}$      & $5\cdot 10^{-4}$  & $-$             \\
        \hline
        \multirow{3}{*}{Free}
          & \multirow{3}{*}{ODIL}
          & Cylinder & $5\cdot10^4$ & $2\cdot10^{-4}$ & $10^{-2}$ & $2\cdot10^4$ & $2\cdot10^{4}$ & $2\cdot10^{-7}$ & $1.5\cdot10^4$ \\
          &  & Ellipse  & $5\cdot10^4$ & $2\cdot10^{-4}$ & $10^{-2}$ & $2\cdot10^4$ & $2\cdot10^{4}$ & $2\cdot10^{-7}$& $1.5\cdot10^4$ \\
          &  & Sphere   & $3\cdot10^4$ & $2\cdot10^{-4}$ & $10^{-2}$ & $2\cdot10^4$ & $10^{4}$ & $2\cdot10^{-6}$ & $10^4$         \\
        \hline
    \end{tabular}
    \caption{Learning rate schedule and switch between first-order and second-order MUSCL discretization $s_{FO\rightarrow SO}$. The corresponding results for the parametric and free shape representation are presented in Section \ref{subsec:results_parametric} and \ref{subsec:results_free}.}
    \label{tab:learning_rate_schedule}
\end{table}

\begin{table}[!t]
    \centering
    \small
    \begin{tabular}{c c c c c c c c c}
        \hline
        Shape representation & Method & Shape & \multicolumn{6}{c}{Loss weights} \\
        &&& $\omega_E$ & $\omega_W$ & $\omega_S$ & $\omega_{T_0}$ & $\omega_\Gamma$ & $\omega_{\phi}$ \\
        \hline\hline
        \multirow{3}{*}{Parametric}
          & \multirow{3}{*}{ODIL}
            & Cylinder & $1$               & $10^3$            & $10^{-1}$         & $10^{-3}$       & $-$              & $-$            \\
          &                    & Ellipse  & $1$               & $10^3$            & $10^{-1}$         & $10^{-3}$       & $-$              & $-$            \\
          &                    & Sphere   & $1$               & $10^3$            & $10^{-1}$         & $5\cdot10^{-2}$ & $-$              & $-$            \\
        \hline
        \multirow{3}{*}{Parametric}
          & \multirow{3}{*}{PINN}
            & Cylinder & $10^{-1}$   & $10^3$            & $5\cdot10^{-1}$   & $5\cdot10^{-1}$ & $10$             & $-$            \\
          &                    & Ellipse  & $10^{-1}$   & $10^3$            & $5\cdot10^{-1}$   & $5\cdot10^{-1}$ & $10$             & $-$            \\
          &                    & Sphere   & $10^{-1}$   & $10^3$            & $5\cdot10^{-1}$   & $5\cdot10^{-1}$ & $1$              & $-$            \\
        \hline
        \multirow{3}{*}{Free}
          & \multirow{3}{*}{ODIL}
            & Cylinder & $1$               & $10^3$            & $10^{-1}$         & $10^{-3}$       & $-$              & $10^3$         \\
          &                    & Ellipse  & $1$               & $10^3$            & $10^{-1}$         & $10^{-3}$       & $-$              & $10^{2}$ \\
          &                    & Sphere   & $1$               & $10^{1}$          & $10^{-1}$         & $5\cdot10^{-2}$ & $-$              & $10^{1}$       \\
        \hline
    \end{tabular}
    \caption{Loss weights $\omega$ for the results of the parametric and free shape representation presented in Section \ref{subsec:results_parametric} and \ref{subsec:results_free}.}
    \label{tab:loss_weights}
\end{table}

% \section{Forward Time-stepping Solution of the Inferred ODIL Shapes}
% \label{appendix:forward_time_stepping_solution_of_odil}

% As discussed in Section~\ref{subsec:results_free}, the downstream side of the inferred obstacle shapes shows noticeable mismatches compared to the reference. This is due to the inherently ill-posed nature of the optimization problem, which lacks sufficient regularization. In particular, multiple obstacle shapes can produce similar downstream flow fields at the primitive measurement points.
% To verify this, we integrate the inferred shapes to steady-state using conventional time-stepping methods implemented in JAX-Fluids. As shown in Figure~\ref{fig:comparison_time_stepping}, the flow fields at the downstream primitive measurement points closely match the reference, despite the differences in the obstacle shapes. This confirms that the optimization procedure identifies valid solutions in terms of matching the measured flow field.
% % 
% \begin{figure}[!t]
%   \centering
%   \input{figures/free_shape_inference/comparison.tex}
%   \caption{Density $\rho$ (left) and velocity $u$ (right) ofthe inferred ODIL shape, integrated to steady-state using conventional time-stepping, and the reference. The solid black lines depict the fluid-solid interface. The mangenta colored points in the density plots represent the primitive variable measurement points. The dashed cyan colored lines in the velocity plots for the reference depict the inferred ODIL interface.}
%   \label{fig:comparison_time_stepping}
% \end{figure}
% % 

\section{Numerical Methods}
\label{appendix:numerical_methods}

\subsection{Godunov-type Finite-Volume Formulation}
\label{subsection:high_order_godunov}

The steady-state compressible Euler equations \eqref{eq:DiffConsLaw1} are discretized on a Cartesian grid using a high-order Godunov-type finite-volume formulation~\cite{Toro2009a}.
\begin{align}
    \boldsymbol{\mathcal{R}}_E^\Delta = \frac{1}{\Delta x} \left( \mathbf{F}_{i-\frac{1}{2},j,k} - \mathbf{F}_{i+\frac{1}{2},j,k} \right)
  + \frac{1}{\Delta y} \left( \mathbf{G}_{i,j-\frac{1}{2},k}  - \mathbf{G}_{i,j+\frac{1}{2},k} \right) 
  + \frac{1}{\Delta z} \left( \mathbf{H}_{i,j,k-\frac{1}{2}}  - \mathbf{H}_{i,j,k+\frac{1}{2}} \right)
\label{eq:FVD} 
\end{align}
The cell centers of the finite-volumes are indexed by $(i,j,k)$.
In this work, we restrict ourselves to squared domains in 2D and cubed domains in 3D, respectively.
In addition, we use uniform meshes throughout this work, i.e., $\Delta x = \Delta y = \Delta z$.

To evaluate the convective fluxes at cell faces, 
we use first-order (FO) upwind reconstruction  or second-order (SO) MUSCL reconstruction
in combination with the local Lax-Friedrichs (LLF) flux function (also referred to as Rusanov flux function).
% A detailed description of the discretization of the convective fluxes is given in \ref{appendix:cell_face_flux}.
We summarize the flux calculation step-by-step for the
flux $\mathbf{F}_{i+\frac{1}{2},j,k}$,
i.e., the convective flux in $x$-direction at the cell face $x_{i+\frac{1}{2},j,k}$,
% see Sec. \ref{subsubsec:high_order_godunov} and Eq. \eqref{eq:FVD}.
The procedure is applied dimension-by-dimension,
and the calculation of $\mathbf{G}$ and $\mathbf{H}$
in $y$- and $z$-directions is analogous.
To improve legibility, we drop the indices $j$ and $k$ for $y$- and $z$-directions, respectively,
for the remainder of this section.
The spatial reconstruction is applied independently to each component of the primitive variables.
Let us denote by $\psi$ any flow variable to be reconstructed, i.e., $\psi \in \left\{\rho, u, v, w, p\right\}$.
First-order (FO) and second-order (SO) reconstructions read
\begin{equation}
\begin{aligned}
    \psi_{i + \frac{1}{2}}^{-,\text{FO}} &= \psi_i, &&\psi_{i + \frac{1}{2}}^{+,\text{FO}} = \psi_{i+1}, \\
    \psi_{i + \frac{1}{2}}^{-,\text{SO}} &= \psi_i + 0.5 \phi(r_i^-) \left( \psi_i - \psi_{i-1} \right), &&\psi_{i + \frac{1}{2}}^{+,\text{SO}} = \psi_{i+1} - 0.5 \phi(r_{i+1}^+) \left( \psi_{i+2} - \psi_{i+1} \right).
\end{aligned}
\end{equation}
Here, $r$ is the ratio of adjacent differences defined by 
\begin{align}
    r_i^- = \frac{\psi_{i+1} - \psi_i}{\psi_i - \psi_{i-1}}, \quad r_{i+1}^+ = \frac{\psi_{i+1} - \psi_i}{\psi_{i+2} - \psi_{i+1}},
\end{align}
and $\phi(r) = \text{max} \left[0, \text{min} \left(1, r\right) \right]$
is the minmod limiter function.
The spatial reconstruction yields the reconstructed vector of primitive variables $\mathbf{W}_{i+\frac{1}{2}}^{\mp}$.
The corresponding vector of conservative variables $\mathbf{U}_{i+\frac{1}{2}}^{\mp}$ can
be computed from $\mathbf{W}_{i+\frac{1}{2}}^{\mp}$ in a straightforward manner.
Then, the flux at the cell face $x_{i+\frac{1}{2}}$ is given by the chosen (numerical) flux function,
$\mathbf{F}_{i+\frac{1}{2}} = \mathbf{F}^{\text{num}} \left( \mathbf{U}_{i+\frac{1}{2}}^{-}, 
\mathbf{U}_{i+\frac{1}{2}}^{+} \right)$.
We use the local Lax-Friedrichs flux function,
\begin{align}
    \mathbf{F}^{\text{num}}\left( \mathbf{U}_L, \mathbf{U}_R \right) = \mathbf{F}^{\text{LLF}} \left( \mathbf{U}_L, \mathbf{U}_R \right) = 
    \frac{1}{2} \left( \mathbf{F}_L + \mathbf{F}_R \right)
    - \frac{1}{2} \alpha \left( \mathbf{U}_R - \mathbf{U}_L \right),
\end{align}
where $\alpha = \text{max}( \vert u_L \vert + c_L, \vert u_R \vert + c_R)$ 
is an estimate of the maximum local wave speed.

To evaluate cell-face fluxes close to domain boundaries,
we append a fixed amount of halo cells $N_h$ to the grid
in each axis direction. We use the halo cells
to impose user-specified boundary conditions.
In this work, we use Dirichlet and zero-gradient (constant extrapolation) boundary conditions.
For Dirichlet boundary conditions, we assign fixed values to the halo cells
directly.
Zero-gradient boundary conditions are imposed
by populating halo cells with values
from adjacent cells inside the domain.
% We denote the halo update procedure with the operator $\mathcal{H}$.

\subsection{Immersed Solid Boundary Method}
\label{subsection:level_set}

We use the conservative sharp-interface approach \cite{Hu2006} to model immersed solid boundaries.
This approach uses the level-set method \cite{Osher1988} to model the fluid-solid interface.
The interface is implicitly given by a scalar function  $\phi(\mathbf{x})$ that satisfies the signed distance property
$\left\|\nabla \phi\right\| = 1$.
Its zero level-set defines the interface location $\Gamma(\mathbf{x})=\{\mathbf{x} \ | \ \phi(\mathbf{x})=0\}$.
% The distinction between the two phases is given by the sign of $\phi$.
% We choose fluid phase and solid body to correspond to the positive $\phi > 0$
% and negative $\phi < 0$ level-set, respectively.
Figure \ref{fig:cut_cell} shows a  finite-volume cell $(i,j,k)$ that contains an interface segment $\Gamma$. 
We refer to such cells as \textit{cut-cells}. 
Cells that do not contain an interface segment are referred to as \textit{full cells}.
% The gray region in the figure
% indicates the volumetric portion $\alpha_{i,j,k}$ of the cell that is occupied by the fluid phase. 
We define apertures $A_{i\pm\frac{1}{2},j,k}$, $A_{i,j\pm\frac{1}{2},k}$,
and $A_{i,j,k\pm\frac{1}{2}}$ as the cell face fractions that are covered by the fluid phase.
\begin{figure}[t]
    \centering
    \input{figures/cutcell.tex}
    \caption{
        Schematic finite-volume discretization for cut-cell $(i,j,k)$ on a Cartesian grid.
        The black dots represent the cell centers. 
        The blue line indicates the interface,
        and the orange line depicts the linear
        approximation of the interface. 
        The figure illustrates a two-dimensional
        slice in the $(x,y)$-plane.
    }
    \label{fig:cut_cell}
\end{figure}
For full cells, the steady-state compressible Euler equations are discretized as described by Eq. \eqref{eq:FVD}.
For cut-cells, the following modification to the
equation is made.
\begin{align}
    \begin{aligned}
        \boldsymbol{\mathcal{R}}_\text{E}^\Delta = &\frac{1}{\Delta x} \left( A_{i-\frac{1}{2},j,k} \mathbf{F}^{c}_{i-\frac{1}{2},j,k} - A_{i+\frac{1}{2},j,k} \mathbf{F}_{i+\frac{1}{2},j,k}  \right)
        + \frac{1}{\Delta y} \left( A_{i,j-\frac{1}{2},k} \mathbf{G}^{c}_{i,j-\frac{1}{2},k} - A_{i,j+\frac{1}{2},k} \mathbf{G}_{i,j+\frac{1}{2},k}  \right) \\
        + &\frac{1}{\Delta z} \left( A_{i,j,k-\frac{1}{2}} \mathbf{H}^{c}_{i,j,k-\frac{1}{2}} - A_{i,j,k+\frac{1}{2}} \mathbf{H}_{i,j,k+\frac{1}{2}}  \right) 
        + \frac{1}{\Delta x \Delta y \Delta z} \mathbf{X}_{i,j,k}(\Delta \mathbf{\Gamma}_{i,j,k}) \\
    \end{aligned}
    \label{eq:FVD_levelset}
\end{align}
Here, we weigh the cell face fluxes $\mathbf{F}_{\dots}$, 
$\mathbf{G}_{\dots}$, $\mathbf{H}_{\dots}$ with the corresponding
apertures $A_{\dots}$. The term $\mathbf{X}$ denotes the convective interface flux.
The interface flux $\mathbf{X}_{i,j,k}$ reads
% To solve the discrete steady Euler equation \eqref{eq:FVD_levelset},
% we compute the interface flux as follows.
\begin{equation}
    \mathbf{X} = 
    \begin{bmatrix}
        0 \\
        p_\Gamma \Delta \mathbf{\Gamma} \\
        p_\Gamma \Delta \mathbf{\Gamma} \cdot \mathbf{v}_\Gamma
    \end{bmatrix}, \qquad
    \Delta \mathbf{\Gamma}_{i,j,k} =
    \begin{bmatrix}
        \left(A_{i+\frac{1}{2},j,k} - A_{i-\frac{1}{2},j,k} \right)\Delta y \Delta z \\
        \left(A_{i,j+\frac{1}{2},k} - A_{i,j-\frac{1}{2},k} \right)\Delta x \Delta z \\
        \left(A_{i,j,k+\frac{1}{2}} - A_{i,j,k-\frac{1}{2}} \right)\Delta x \Delta y \\
    \end{bmatrix}.
    \label{eq:interface_flux}
\end{equation}
Here, $p_\Gamma$, $\mathbf{v}_\Gamma$, and $\Delta \mathbf{\Gamma}$
denote the interface pressure, interface velocity,
and the projection of the interface segment length, respectively.
The interface pressure is approximated with the cell center pressure of the present cut-cell.
In this work, we only consider static solid bodies, hence $\mathbf{v}_\Gamma = 0$. 
The computation of the apertures $A_{i\pm\frac{1}{2},j,k},A_{i,j\pm\frac{1}{2},k},A_{i,j,k\pm\frac{1}{2}}$
is based on the marching squares approach \cite{Lorensen1987a}.
More details about the immersed boundary method can be found in \cite{Bezgin2022,Bezgin2024}.

The discretization of the level-set reinitialization equation \eqref{eq:levelset_reinit} reads \cite{Rouy1992}
\begin{equation}
    \mathcal{R}_\phi^\Delta = \begin{cases}
        \sqrt{\text{max}((a^+)^2,(b^-)^2) + \text{max}((c^+)^2,(d^-)^2) + \text{max}((e^+)^2,(f^-)^2)} - 1, \quad \phi_{i,j,k} \geq 0 \\      
        \sqrt{\text{max}((a^-)^2,(b^+)^2) + \text{max}((c^-)^2,(d^+)^2) + \text{max}((e^-)^2,(f^+)^2)} - 1, \quad \phi_{i,j,k}  < 0
    \end{cases},
\label{eq:levelset_reinitialization_discrete}
\end{equation}
with $g^+=\text{max}(g,0)$ and $g^-=\text{min}(g,0)$.
The one-sided derivatives are defined as
\begin{align}
    a &=(\phi_{i,j,k} - \phi_{i-1,j,k})/\Delta x, \quad b =(\phi_{i+1,j,k} - \phi_{i,j,k})/\Delta x, \notag \\
    c &=(\phi_{i,j,k} - \phi_{i,j-1,k})/\Delta y, \quad d =(\phi_{i,j+1,k} - \phi_{i,j,k})/\Delta y, \notag \\
    e &=(\phi_{i,j,k} - \phi_{i,j,k-1})/\Delta z, \quad f =(\phi_{i,j,k+1} - \phi_{i,j,k})/\Delta z. \notag
\end{align}

\section{Optimizing a Discrete Loss}
\label{appendix:odil}

The Optimizing a Discrete Loss (ODIL) method solves the optimization problem defined in Eq.~\eqref{eq:optimization_problem} and Eq.~\eqref{eq:loss_function} by approximating the solution on a discrete mesh. The discrete partial differential equation (PDE) residual is computed with the numerical methods outlined in \ref{appendix:numerical_methods}. During optimization,  the discrete solution itself is treated as a set of tunable parameters.
% The residual of the partial differential equation (PDE) is computed using conventional numerical methods.
To accelerate convergence, we employ a multigrid representation of the discrete solution, following the methodology of~\cite{Karnakov2023a,Karnakov2023b}. For a uniform grid with $N$ cells in each spatial direction, we define a hierarchy of coarser grids with resolutions $N_{(i)} = N / 2^{i-1}$, where $i \in \{1, 2, \dots, L\}$ and $L$ is the total number of levels.
The multigrid decomposition operator $\mathcal{M}$ applied to the solution $u$ is defined as
\begin{equation}
    \label{eq:multigrid_decomposition}
    \mathcal{M}(u_{(1)}, \dots, u_{(L)}) = u_{(1)} + T_{(1)} u_{(2)} + \dots + T_{(1)} T_{(2)} \dots T_{(L-1)} u_{(L)},
\end{equation}
where $u_{(i)}$ denotes the parameters on the grid with resolution $N_{(i)}$, and $T_{(i)}$ is a linear interpolation operator mapping from grid $N_{(i+1)}$ to $N_{(i)}$. The solution $u$ is reconstructed from the multigrid components as $u = \mathcal{M}(u_{(1)}, \dots, u_{(L)})$.
% \begin{equation}
%     \label{eq:multigrid_decomposition1}
%     u = \mathcal{M}(u_{(1)}, \dots, u_{(L)}).
% \end{equation}

In this work, the multigrid parameters include the primitive variable vector $\mathbf{W}_{(i)}^\text{ODIL}$ and the level-set field $\phi_{(i)}^\text{ODIL}$ (the latter is used only in the free shape representation, see Section~\ref{subsection:shape_params}). We initialize $\mathbf{W}_{(i)}^\text{ODIL}$ and $\phi_{(i)}^\text{ODIL}$ to zero for all $i \in \{1,2,\dots, L-1\}$. At the finest level $L$, $\mathbf{W}_{(L)}^\text{ODIL}$ is initialized to represent a uniform flow field with zero velocity, unit pressure, and unit density. Similarly, $\phi_{(L)}^\text{ODIL}$ is initialized to represent a circle (in 2D) or a sphere (in 3D) with a specified radius.
To ensure positivity of the density and pressure fields, we apply a \textit{softplus} activation to these parameters before evaluating the discrete PDE residual. Finally, the loss function (see Eq.~\eqref{eq:loss_odil}) is evaluated on the reconstructed fields.

We start the optimization with first-order (FO)
spatial reconstruction and subsequently switch to second-order (SO) spatial
reconstruction.
This approach adds further regularization,
leading to better convergence.
To this end, the PDE residual in Eq. \eqref{eq:loss_odil} is computed as
\begin{equation}
    \mathcal{L}^\text{ODIL}_E = \beta \mathcal{L}^\text{ODIL,FO}_E + (1-\beta) \mathcal{L}^\text{ODIL,SO}_E,
\end{equation}
where $\beta$ is a function of the optimization step $s$. We use
\begin{align}
    \beta \left(s\right) = \frac{1}{1+\exp(-0.01 \cdot (s - s_{FO\rightarrow SO}))},
    \label{eq:switch_reconstructions}
\end{align}
such that the switch from first- to second-order is centered around optimization step $s_{FO\rightarrow SO}$.

\section{Physics-informed Neural Networks}
\label{appendix:pinn}
Physics-informed Neural Networks \cite{Raissi2019} are deep feedforward networks that map input variables to the solution vector of
a given system of partial differential equations.
For the present application of steady-state compressible flows governed by the Euler equations,
the network maps the spatial coordinates $\mathbf{x}$ to the vector of primitive variables $\mathbf{W}^\text{PINN}$.
For the hidden layers of the neural network, we use layer-wise locally adaptive \cite{Jagtap2019} hyperbolic tangent activation functions following the setup in \cite{Jagtap2022}, i.e., with a scaling factor of 10 and initial values for the adaptive activation parameters of 0.1.
For the output layer, we employ a \textit{softplus} activation function for the density and the pressure, ensuring positivity.
We use a linear activation function for velocity components.

%% file: figures/cutcell.tex
\begin{tikzpicture}

    \definecolor{color0}{RGB}{0,101,189}
    \definecolor{color1}{RGB}{227,114,34}
    \definecolor{color2}{RGB}{162,173,0}

    \coordinate (NULL) at (0,0);
    \coordinate (B) at (5,5);
    \coordinate (C) at ($0.5*(B)$);
    \coordinate (D1) at ($0.125*(B)$);
    \coordinate (D2) at ($0.71*(B)$);

    \coordinate (D11) at ($0.1*(B)$);

    \coordinate (L) at ($1.1*(B|-NULL)+0.96*(B-|NULL)$);

    % fill
    \filldraw[fill=black!10!white, line width=0.1pt] (NULL) -- (B|-NULL) -- ($(B|-NULL) + 0.16*(B-|NULL)$) -- ($(B|-NULL) + 0.16*(B-|NULL)$) arc (33.5:56.4:15) -- (B-|NULL) -- (NULL);

    % grid 
    \draw[line width=1pt] (0,0) rectangle (B);
    \draw[line width=0.5pt] (C |- NULL) -- (C |- B);
    \draw[line width=0.5pt] (C -| NULL) -- (C -| B);
    \draw[fill=TUMorange, draw=TUMorange] (C) circle (0.1);

    \draw[dashed] ($(C)+(C-|NULL)$) -- ($(C)+(B-|NULL)$);
    \draw[dashed] ($(C)+(C|-NULL)$) -- ($(C)+(B|-NULL)$);
    \draw[dashed] ($(C)-(C-|NULL)$) -- ($(C)-(B-|NULL)$);
    \draw[dashed] ($(C)-(C|-NULL)$) -- ($(C)-(B|-NULL)$);
    \draw[dashed] ($3*(C|-NULL)-(NULL|-C)$) -- ($-1*(C)$) -- ($3*(C-|NULL)-(NULL-|C)$) -- ($3*(C)$) -- cycle;

    % coordinates
    \draw[fill=black, draw=black] (C) circle (0.1) node[below left] {$(i,j,k)$};
    \draw[fill=black, draw=black] ($(C)+(B-|NULL)$) circle (0.1) node[below right] {$(i,j+1,k)$};
    \draw[fill=black, draw=black] ($(C)+(B|-NULL)$) circle (0.1) node[below right] {$(i+1,j,k)$};
    \draw[fill=black, draw=black] ($(C)-(B|-NULL)$) circle (0.1) node[below left] {$(i-1,j,k)$};
    \draw[fill=black, draw=black] ($(C)-(B-|NULL)$) circle (0.1) node[below right] {$(i,j-1,k)$};
    \draw[fill=black, draw=black] ($-1*(C)$) circle (0.1) node[below left] {$(i-1,j-1,k)$};
    \draw[fill=black, draw=black] ($3*(C)$) circle (0.1) node[below right] {$(i+1,j+1,k)$};
    \draw[fill=black, draw=black] ($3*(C-|NULL)-(NULL-|C)$) circle (0.1) node[below left] {$(i-1,j+1,k)$};
    \draw[fill=black, draw=black] ($3*(C|-NULL)-(NULL|-C)$) circle (0.1) node[below right] {$(i+1,j-1,k)$};

    % interface
    \draw[TUMblue, line width=1pt] ($1.1*(B|-NULL)$) arc (30:60:15); \label{tikz:interface}

    % reconstruction
    \draw[TUMorange, line width=1pt] (B -| D11) -- (D11 -| B) node[at start, below, yshift=-0.5cm, xshift=0.1cm] {$\Delta\Gamma_{i,j,k}$}; \label{tikz:interface_reconstruction}
    
    % apertures
    \coordinate (DIST) at (0.5,0.5);
    \dimline[extension start length=1cm, extension end length=1cm,extension style={black}, label style={above=0.5ex}] {(-0.5,0)}{($(NULL|-B) - (0.5,0)$)}{$A_{i-\frac{1}{2},j,k}=1.0$};
    \dimline[extension start length=-1cm, extension end length=-1cm,extension style={black}, label style={below=0.5ex}] {(0.0,-0.5)}{($(NULL-|B) - (0.0,0.5)$)}{$A_{i,j-\frac{1}{2},k}=1.0$};
    \dimline[extension start length=0.5cm, extension end length=0.5cm,extension style={black}, label style={above=0.5ex}] {($(NULL|-B) + (DIST -| NULL)$)}{($(NULL|-B) + (DIST -| NULL) + (D11|-NULL)$)}{$A_{i,j+\frac{1}{2},k}$};
    \dimline[extension start length=-0.5cm, extension end length=-0.5cm,extension style={black}, label style={below=0.5ex}] {($(NULL-|B) + (DIST |- NULL)$)}{($(NULL-|B) + (DIST|- NULL) + (D11-|NULL)$)}{$A_{i+\frac{1}{2},j,k}$};

  % coordinate system
    \draw[->, line width=1pt] (-1,-1) -- (-1,0) node[left] {$y$};
    \draw[->, line width=1pt] (-1,-1) -- (0,-1) node[below] {$x$};
    \node[inner sep=2, circle, draw=black, line width=1pt] at (-1,-1) {};
    \node[inner sep=1, circle, draw=none, fill=black] at (-1,-1) {};
    \node[below left] at (-1,-1) {$z$};

\end{tikzpicture}